\documentclass[twocolumn,pra,superscriptaddress,10pt]{revtex4}
\usepackage{graphicx,color}
\usepackage{bm}
\usepackage{hyperref}
\usepackage{amsmath,braket}
\usepackage{siunitx}
\usepackage{physics}
\usepackage{verbatim}

\definecolor{battleshipgrey}{rgb}{0.52, 0.52, 0.51}
\definecolor{cadet}{rgb}{0.33, 0.41, 0.47}
\definecolor{charcoal}{rgb}{0.21, 0.27, 0.31}
\hypersetup{
colorlinks   = true, 
urlcolor     = battleshipgrey, 
linkcolor    = cadet, 
citecolor   = charcoal }

\begin{document}

\title{Twisted quantum interference in photoelectron holography with elliptically polarized fields}

\author{G. Kim}
\email{gyeonghun.kim.phys@gmail.com}

\affiliation{Department of Physics \& Astronomy, University College London \\
Gower Street London WC1E 6BT, United Kingdom }
\affiliation{Department of Physics \& Astronomy, Seoul National University, Seoul, 08826, Republic of Korea}

\author{C. Hofmann}
\email{c.hofmann@ucl.ac.uk}
\affiliation{Department of Physics \& Astronomy, University College London \\
Gower Street London WC1E 6BT, United Kingdom }

\author{A. S. Maxwell}
\email{andrew.maxwell@phys.au.dk}
\affiliation{Department of Physics and Astronomy, Aarhus University, DK-8000 Aarhus C, Denmark}
\affiliation{Department of Physics \& Astronomy, University College London \\
Gower Street London WC1E 6BT, United Kingdom }

\author{C. Figueira de Morisson Faria}
\email{c.faria@ucl.ac.uk}
\affiliation{Department of Physics \& Astronomy, University College London \\
Gower Street London WC1E 6BT, United Kingdom }
\date{\today}

\begin{abstract}
We perform a systematic analysis of how ultrafast photoelectron holography is influenced by an elliptically polarized field, with emphasis on quantum interference effects. We find that the interplay of the external field and the binding potential leads to twisted holographic patterns for low ellipticities and recover well-known angular offsets for high ellipticities. Using the Coulomb quantum-orbit strong-field approximation (CQSFA), we assess how the field ellipticity affects specific holographic patterns, such as the fan and the spider. The interplay of the external field and the binding potential leads to twisted holographic patterns in the fan, and to loss of contrast in the spider. This behavior can be traced back to interfering electron trajectories, and unequal changes in tunneling probability due to non-vanishing ellipticity.  We also derive tunneling times analytically using the strong-field approximation (SFA), provide estimates for ellipticy ranges for which interference is expected to be prominent, and discuss how to construct continuous electron momentum distributions exploring the rotation symmetry around the origin. 
\end{abstract}

\maketitle


\section{Introduction}

When matter interacts with a laser field whose intensity is of the order of \SI[per-mode = symbol]{e14}{W\per\cm\tothe{2}}, valence electrons can absorb more photons than necessary for ionization. This phenomenon is called above-threshold ionization (ATI)\cite{agostini1979free}, and has been researched intensely in theory and applications \cite{Becker2002Review,MilosReviewATI}.  It is explained by a physical picture of an electron being 
released through strong-field quantum tunneling or multiphoton ionization, accelerated by the field in the continuum, and finally captured by the detector \cite{Corkum1993}.
Since it propagates to the detector via many possible pathways after ionization, an interference pattern is present in the photoelectron momentum distribution (PMD). These interference patterns are related to a wealth of information, such as the molecular structure and orbital geometries, essential for ultrafast imaging \cite{Lein2007,Krausz2009,Salieres2012R,Gallmann2012,Lepine2014}. This has led to the inception of photoelectron holography \cite{HuismansScience2011,Bian2011,Bian2012}, in which phase differences between distinct electron pathways lead to several types of structures (for a review see \cite{Faria2020}). Photoelectron holography has been widely explored in linearly polarized fields, although there are studies in orthogonally polarized two-color \cite{Xie2015,Gong2017,Han2017,Han2018} or elliptical fields \cite{Han2017,Xie2018}.

In particular, tailored fields are a powerful tool for controlling both electron ionization and continuum propagation, and thus phenomena such as ATI (for reviews see \cite{Brabec2000,Ehlotzky2001,Milos2006}).
Complicated electron dynamics induced by a tailored field provide intricate interference patterns that could be used for revealing detailed internal orbital structures, or focusing on specific electronic wavefunction evolution paths. Various tailored fields, such as orthogonal two-color fields (OTC) \cite{Shafir2012,Xie2015,Richter2015,Das2015,Henkel2015,Li2016,Gong2017,Han2017,Zhang2014,Xie2017,Han2018,Tulsky2018}, bicircular fields \cite{Milos2000,Smirnova2015JPhysB,Milos2015,Milos2016,Mancuso2016,Hoang2017,ALmajid2017,Busuladvic2017,Eckart2018,Milos2018,Ayuso2018,Ayuso2018II,Baykusheva2018,Eicke2019,Yue2020,Maxwell2021}, parallel two-color fields (PTC) \cite{Dudovich2006,Xie2013,Arbo2014,Skruszewicz2015,Xie2016b,Luo2017,Porat2018}, and elliptically polarized fields 
(see for example 
\cite{Nubbemeyer2008,Abu-samha2011,Das2013,Hofmann2016,Danek2018}),
have been studied extensively. In particular,  ATI with elliptically polarized fields has been widely investigated in circular streaking approaches, using the numerical solution of the time-dependent Schrödinger equation (TDSE) \cite{Ivanov2014,Torlina2015,Eicke2018,Sainadh2019}, classical orbit theories 
\cite{Kulander1993,Corkum1993,Pfeiffer2012,Landsman2014b}, the strong-field approximation (SFA) \cite{Paulus1998}, and the Coulomb-eikonal approximation \cite{Torlina2013}. These studies considered, and sometimes even required, the field to be almost circularly polarized. In contrast, the low-ellipticity regime is comparatively less studied. 
Thereby, a key question is how to detangle and interpret the holographic patterns that appear in the photoelectron spectra in terms of interfering electron orbits. This information is difficult to extract using the TDSE, in which specific quantum pathways cannot be switched on and off as one wishes. Furthermore, in its standard form, the SFA does not include the residual Coulomb potential in the electron's continuum propagation (for reviews see, e.g.,  \cite{Popruzhenko2014a,Amini2019}). This potential will influence ionization, continuum propagation and consequently the shapes of the photoelectron momentum distributions. 

Substantial progress in this direction has been made using the Coulomb Quantum-orbit Strong-Field Approximation (CQSFA) for linearly polarized fields \cite{Faria2020}. The CQSFA has allowed explicit investigation into how holographic patterns form, through the isolation of interfering pairs of orbits. This includes the fan-shaped pattern close to the ionization threshold \cite{Rudenko2004,Maharjan2006,Gopal2009}, spider-like fringes along and near the field-polarization axis \cite{HuismansScience2011,Bian2011,Huismans2012,Marchenko2011,Hickstein2012}, and a spiral-like structure recently identified in experiments \cite{Maxwell2020,Qin2021}. Multipath interference \cite{Werby2021} and phase differences that can be used to probe orbital parity \cite{Kang2020} have also been explored in conjunction with experiments. The CQSFA, however, has not yet been applied to fields with elliptical or circular polarization. As it is a non-adiabatic, fully Coulomb-distorted orbit based method, the CQSFA is a powerful tool to assess photoelectron holography in this context. 

Therefore, in this work, we perform an analysis on the effect of the Coulomb potential in ATI with an elliptically polarized field based on quantum-orbit methods. We provide the fully analytic form of the SFA solutions with an elliptically polarized field and classify the orbits based on these solutions. Subsequently, we focus on the low-ellipticity regime, with emphasis on photoelectron holography. We show that the field ellipticity, together with the residual potential, modify holographic structures, leading to changes in contrast or twisted, spiral-like interference patterns. These features can be traced back to the quantum interference of specific pairs of orbits, which are  affected in different ways.  In the high ellipticity regime, we recover the angular offsets known from previous angular streaking studies \cite{Pfeiffer2012,Landsman2014b,Torlina2015}. 

This paper is organised as follows: Section \ref{sec:background} offers and overview of the theoretical background upon which the work presented in this paper is based. Section \ref{sec:SaddlePointSolutions} introduces saddle-point solutions for the ionization times, the formal extension of the CQSFA method for arbitrary ellipticity, and estimates for the maxima of the distributions and the ellipticity range for which interference is prominent. Subsequently, in Sec.~\ref{sec:PMD}, we present the photoelectron momentum distributions computed for several field ellipticities, which are analyzed in terms of interfering electron trajectories. Finally, in Sec.~\ref{sec:conclusions} we state the main conclusions to be drawn from this work. 
We use atomic units throughout, unless otherwise stated. 

\section{Background}
\label{sec:background}
The Hamiltonian $H$ in strong-field ionization can be split into the atomic Hamiltonian $H_a$ and the interaction Hamiltonian $H_I$ as $H = H_a + H_I$ where
\begin{equation}
H_a = \frac{{\mathbf{\hat{p}}}^2}{2}+ V(r) ,
    \label{eqn: atomic hamiltonian}
\end{equation}
\begin{equation}
   V(r)= - \frac{2}{\sqrt{\mathbf{\hat{r}} \cdot \mathbf{\hat{r}}}} f(r,r_0)
\end{equation}
and
\begin{equation}
H_I(t) = -\mathbf{\hat{r}} \cdot \mathbf{E}(t)
    \label{eqn: interaction hamiltonian}
\end{equation}
in the length gauge, and assuming the dipole approximation \cite{Reiss2008}.  Unless otherwise stated, $f(r,r_0)=1$ throughout, so that the binding potential is of Coulomb type representing a helium atom, but excludes multielectron effects. 
However, in Fig.~\ref{fig:QpropTruncated} we consider it to be of the form \cite{DeMorissonFaria2002}
\begin{equation}
   f(r,r_0)= \left\{  
   					\begin{array}{ll}
   					1	&	\mbox{for }r < r_0, \\
   					\cos^7\left(\pi \frac{r - r_0}{2(L-r_0)}\right)	& \mbox{for }r_0 \leq r < L, \\
   					0	& \mbox{for }r \geq L,
   					\end{array}
   			 \right.
   \label{eq:truncation}
\end{equation}
truncating the Coulomb potential smoothly starting at $r_0$ and leaving only the Coulomb-free laser potential outside $L$. In the present work, the distance $r_0$ is chosen as a multiple of the radius defined by the tunnel exit, which is the coordinate at which the electron reaches the continuum by tunneling through the potential barrier. It is defined as in \cite{maxwell2017coulomb}, while $L$ is defined as $r_0+$ half an excursion amplitude.  

The parameters used in this article are for helium, which is a widely used target in attosecond angular streaking studies (see, e.g., the reviews \cite{Landsman2015,Hofmann2019}), although we work within the single active electron approximation.
From the Schr\"odinger equation with the Hamiltonian above, we can calculate the transition amplitude of an electron from the bound state $\left| \psi_0 \right>$ to a final continuum state $\left| \psi_{\mathbf{p}} \right>$ with momentum $\mathbf{p}$. The transition amplitude is defined as 
\begin{equation}
    M(\mathbf{p}) = -i \lim_{t\to\infty} \int^t_{-\infty} dt' \left< \psi_{\mathbf{p}}(t) \left| U(t, t') H_I(t') e^{iI_pt'} \right| \psi_0 \right>,
    \label{eqn: transition amplitude ATI}
\end{equation}
where $I_p$ is the ionization potential, and $U(t, t')$ is the time evolution operator associated with the full Hamiltonian $H_a+H_I(t)$. 
This integral equation is a general formal solution, and a good starting point for developing quantum orbit-based approaches. 

\subsection{Strong-field approximation}
The strong-field approximation (SFA) is a useful and often applied way to evaluate \eqref{eqn: transition amplitude ATI} analytically. The SFA consists in approximating the continuum by field-dressed plane waves, and in neglecting the influence of the external laser field when the electron is bound, although continuum-to-continuum contributions may be incorporated perturbatively (for a recent review see \cite{Amini2019}). In its standard form, it also neglects bound-to-bound transitions and considers only the ground state and the continuum, although one may also modify it to incorporate excitation \cite{Shaaran2010,Shaaran2010a}. 

In the SFA computations performed in this work, we will focus on the direct electrons, which reach the detector after tunnel ionization without further interacting with the core. This approximation corresponds to replacing the full time-evolution operator $U(t,t')$ by the Volkov time-evolution operator $U^{(V)}(t,t')$ in Eq.~(\ref{eqn: transition amplitude ATI}). This is also known as the Keldysh-Faisal-Reiss (KFR) approximation \cite{Keldysh1965,Faisal1973,Reiss1980}; for the specific formulation used here see also \cite{Lohr1997}.  
Then the semi-classical action corresponding to the propagation after tunnel ionization time $t'$ can be calculated analytically as
\begin{equation}
    S(\mathbf{p}, t') = -\frac{1}{2} \int^{\infty}_{t'} \left[ \mathbf{p} + \mathbf{A}(\tau) \right]^2 d\tau + I_p t'.
    \label{eqn: SFA action}
\end{equation}
Here, $I_p$ is the ionization potential and $\mathbf{A}$ denotes the vector potential. The SFA transition amplitude may be associated to the coherent superposition of electron orbits in the continuum using saddle-point methods. Therefore, we seek values of $t'$ for which Eq.~\eqref{eqn: SFA action} is stationary. This gives the saddle-point equation
\begin{equation}
    \frac{\partial S(t')}{\partial t'} = \frac{[\mathbf{p} + \mathbf{A}(t')]^2}{2} + I_p =0 ,
    \label{eqn: SFA SPE}
\end{equation} 
and \eqref{eqn: transition amplitude ATI} can be approximated by the coherent sum of orbits
\begin{equation}
    M(\mathbf{p}) \sim \sum_{s} \mathcal{C}(t'_s) e^{iS(\mathbf{p}, t'_s)},
    \label{eqn: transition amplitude SPA SFA}
\end{equation}
where the prefactor $\mathcal{C}(t'_s)$ is given as
\begin{equation}
    \mathcal{C}(t'_s) = \sqrt{\frac{2\pi i}{\partial^2 S(\mathbf{p}, t'_s) / \partial {t'_s}^2}} \left< \mathbf{p} + \mathbf{A}(t'_s) \left| H_I(t'_s) \right| \psi_0 \right>
    \label{eqn: prefactor SFA}
\end{equation}
and $t'_s$ are the saddle-point solutions. 
Since more than one orbit are related to a single final momentum, interference patterns will appear in the photoelectron momentum distributions. Due to the residual binding potential being neglected in the continuum propagation, the momentum $\mathbf{p}$ is conserved throughout. 

\subsection{Coulomb quantum-orbit strong-field approximation}

The Coulomb quantum-orbit strong-field approximation (CQSFA) also starts from Eq.~(\ref{eqn: transition amplitude ATI}), but instead of approximating the full time-evolution operator by its Volkov counterpart, time-slicing techniques and path-integral methods are used. Correspondingly, the Coulomb potential and the external field are treated on equal footing. 
We use the CQSFA action integrated over a two-pronged  contour,  first along the imaginary time axis from $t'$ to its real part, and subsequently along the real time axis from $\mathrm{Re}[t']$ up to $t \rightarrow \infty$ \cite{maxwell2017coulomb}, and make the further approximation that the orbits are real in the continuum. A full treatment requires complex coordinates throughout and will lead to branch cuts, and has been discussed in \cite{Maxwell2018}. Within the CQSFA, the Coulomb-distorted transition amplitude within the saddle-point approximation reads
\begin{equation}
\begin{split}
    M(\mathbf{p_f}) \propto -i &\lim_{t\to\infty} \sum_{s} \left\{ \det\left[\frac{\partial\mathbf{p_s}(t)}{\partial\mathbf{r_s}(t'_s)}\right]\right\}^{-1/2} \\ 
    &\times \mathcal{C}(t'_s) e^{iS(\mathbf{p_s}, \mathbf{r_s},t, t'_s)},
\end{split}
    \label{eqn: transition amplitude SPA CQSFA}
\end{equation}
where the
semi-classical action is given by
\begin{equation}\label{eq:stilde}
    S(\mathbf{p},\textbf{r},t,t')=I_p t'-\int^{t}_{t'}[
    \dot{\mathbf{p}}(\tau)\cdot \mathbf{r}(\tau)
    +H(\mathbf{r}(\tau),\mathbf{p}(\tau),\tau)]d\tau.
\end{equation}
The full Hamiltonian reads as
\begin{equation}
H(\mathbf{r}(\tau),\mathbf{p}(\tau),\tau)=(1/2)\left[\mathbf{p}(\tau)+\mathbf{A}(\tau)\right]^2+V\left(\mathbf{r}(\tau)\right).
\end{equation}
The variables $t'_s$, $\mathbf{p}_s$ and $\mathbf{r}_s$ are the solutions of the saddle-point equations
\begin{eqnarray}
[\mathbf{p}(t')+\mathbf{A}(t')]^2 = -2I_p \label{eq:SPEt}\\
\mathbf{\dot{r}}(\tau) = \mathbf{p}(\tau) + \mathbf{A}(\tau) \label{eq:SPEp}\\
\mathbf{\dot{p}}(\tau) = -\nabla_rV(\mathbf{r}(\tau) \label{eq:SPEr}
\end{eqnarray}
for energy conservation at tunnel ionization and the electron's intermediate momentum and position, respectively. One should note that, in Eq.~(\ref{eq:SPEt}), an additional approximation was made, namely that the momentum in the first part of the contour is constant and equal to $\mathbf{p}_0=\mathbf{p}(t')$, and that, in contrast to the SFA transition amplitude, one must take into consideration the intermediate variables $\mathbf{r}(\tau)$ and $\mathbf{p}(\tau)$, $t'<\tau<t$ in the  the continuum propagation equations (\ref{eq:SPEp}) and (\ref{eq:SPEr}). The momentum at the detector is $\mathbf{p}(t)=\mathbf{p}_f$.  The term $\mathcal{C}(t'_s)$ is given by Eq.~(\ref{eqn: prefactor SFA}), but with $\mathbf{p}$ replaced by the initial momentum $\mathbf{p}_0$. For details about the CQSFA, we refer to \cite{maxwell2017coulomb, lai2015influence}. 

In the present work, we consider ionization times within up to four cycles, and the continuum propagation extends to roughly 20 cycles of the field. Since the laser field is periodic and no pulse envelope is considered in the CQSFA method, restricting ionization times to a single cycle leads to some ambiguity with regard to where the cycle starts and finishes. This ambiguity will influence the intra-cycle interference patterns, and could be removed by considering distributions incoherently averaged over the offset phases marking the start of these unit cells, but this method will not be employed here. For details see \cite{Werby2021}.

\section{Analytical estimates and ionization times}
\label{sec:SaddlePointSolutions}

In the following, we consider an elliptically polarized field approximated by two orthogonally polarized monochromatic waves of frequency $\omega$, 
so that the vector potential and  corresponding electric field are given by \begin{equation}
    \begin{split}
        \mathbf{A}(t) &= \frac{2\sqrt{U_p}}{\sqrt{1 + \epsilon^2}}[\cos(\omega t) \mathbf{\hat{e}_z} + \epsilon \sin(\omega t) \mathbf{\hat{e}_x}] \\
        \mathbf{E}(t) &= \frac{2\omega\sqrt{U_p}}{\sqrt{1 + \epsilon^2}}[\sin(\omega t) \mathbf{\hat{e}_z} - \epsilon \cos(\omega t) \mathbf{\hat{e}_x}],
    \end{split}
    \label{eqn: elliptically polarized field}
\end{equation}
 where $\epsilon$ is the field ellipticity, and we keep the ponderomotive energy $U_p$ constant for varying ellipticity. This approximation holds for long enough pulses. For simplicity, we will restrict the dynamics to the polarization plane.  Note that the field major axis is $\hat{z}$ and its minor axis is $\hat{x}$. Eq.~\eqref{eqn: elliptically polarized field} implicitly states that we define a unit cell starting at a phase $\phi=0$. Other unit cells could be chosen by setting $\omega t \rightarrow \omega t + \phi$, where $\phi$ is an offset phase used to define the beginning of the unit cell. For a coherent sum of ionization times over many cycles, this will not play a role, but for a single-cycle photoelectron momentum distribution this will lead to some ambiguity in the patterns \cite{Werby2021}.

Next, we will use the action associated with the direct SFA transition amplitude to provide analytic estimates for the centers of the electron momentum distributions, as well as the parameter range for which quantum interference is expected to be significant. We will also employ the tunnel ionization equation \eqref{eqn: SFA SPE} to derive analytic solutions for the ionization times. Although such estimates are approximate in the presence of residual potentials, they give valuable insight and can also be used as initial guesses for the CQSFA.

The SFA action for the elliptically polarized fields \eqref{eqn: elliptically polarized field} reads 
\begin{eqnarray}
S_d(\mathbf{p},t')&=&\left(
\frac{p_z^2+p^2_x}{2}+ I_p+U_p\right)t'\\&&+\frac{U_p}{2\omega(1+\epsilon^2)}\left[ \sin 2 \omega t' -\epsilon^2\sin 2( \omega t')
\right]\notag \\&&+\frac{2\sqrt{U_p}}{\omega \sqrt{1+\epsilon^2}}\left[p_z\sin\omega t'-\epsilon p_x \cos (\omega t')\right].\notag
\label{eq:action2}
\end{eqnarray}
The corresponding tunnel ionization equation \eqref{eqn: SFA SPE} then reads
\begin{widetext}

\begin{equation}
		\left(p_z + \frac{2\sqrt{U_p}}{\sqrt{1 + \epsilon^2}}\cos(\omega t')\right)^2
		+ \left(p_x + \frac{2\epsilon\sqrt{U_p}}{\sqrt{1 + \epsilon^2}}\sin(\omega t')\right)^2 + 2I_p = 0.
	\label{eqn: expanded saddle point equation}
\end{equation}
Eq.~\eqref{eqn: expanded saddle point equation} is 
a superposition of circles of complex radii centered at
\begin{equation}
    (p^{(c)}_z(t'),p^{(c)}_x(t'))=(-\frac{2\sqrt{U_p}}{\sqrt{(1+\epsilon^2)}}\cos\omega t',-\epsilon\frac{2\sqrt{U_p}}{\sqrt{(1+\epsilon^2)}}\sin\omega t'), \label{eqn: radii centers}
\end{equation}
which can be used to estimate the maxima of the distributions and the region for which quantum interference is significant. 
\end{widetext}

\subsection{Widths and maxima of the distributions}

The previous section dealt with the centers of momentum distributions (or most probable final photoelectron momenta).
However, in order to see interference patterns in the PMD we must consider the widths of photoelectron wave packets, and whether or not they overlap. Here we provide estimates for the ellipticity range for which prominent intra-cycle interference patterns are expected. The estimates below assume that ionization is  most probable at the peak of the field, which lies along the z-axis, and is valid for small or medium ellipticities. They have also been performed within the SFA, for which the field-dressed momentum is conserved in the continuum.
For linear and elliptical polarization, $p^{(c)}_z=0$ is expected since $\omega t' = (2n+1)\pi/2 $ are the peaks in the electric field $\mathbf{E}(t')$, which points along the major axis and $A_z(t')=0$ for those specific times  (see equations \eqref{eqn: elliptically polarized field} and \eqref{eqn: radii centers})
\begin{equation}
	(p^{(c)}_z,p^{(c)}_x)=\left(0,-\epsilon\frac{2\sqrt{U_p}}{\sqrt{1+\epsilon^2}}(-1)^n\right).
\end{equation}
For events displaced by half a cycle, $p^{(c)}_x$ will always have opposite signs, so that the distance between the center of the distributions yields
\begin{equation}
 p^{(c)}_{x1}- p^{(c)}_{x2}=\epsilon\frac{4\sqrt{U_p}}{\sqrt{1+\epsilon^2}}.
\end{equation}

In this work we are investigating interference patterns, and hence require wavepackets ionized at opposite half cycles to still overlap to some degree in the final momentum distribution. 
We can find an estimate of the interference width, starting from the Ammosov, Delone, Krainov (ADK) \cite{Ammosov1986,Delone1991} description of the width of the wavepackets approximated to Gaussian shapes
\begin{equation}
\sigma_{\perp} = \sqrt{ \frac{\omega \sqrt{U_p}}{\sqrt{1+\epsilon^2}\sqrt{2 I_p}}}.
\end{equation}
Requiring that the two centers are a maximum of $5\sigma_{\perp}$ apart from each other, such that there is still a significant enough overlap between them to show interference patterns originating from different half cycle orbits, we find
\begin{eqnarray}
p^{(c)}_{x1}- p^{(c)}_{x2} &=& 5 \sigma_{\perp} \notag \\
\epsilon\frac{4\sqrt{U_p}}{\sqrt{1+\epsilon^2}} &=&5 \sqrt{ \frac{\omega \sqrt{U_p}}{\sqrt{1+\epsilon^2}\sqrt{2 I_p}}}.
\end{eqnarray}
Although somewhat arbitrary, this limiting shift of $5\sigma_{\perp}$ has been chosen as an educated guess, by assuming that an overlap of at least $1\sigma_{\perp}$ may occur within a $3\sigma_{\perp}$ range from each of the peaks such that some interference pattern is still visible (on a log scale such as presented in figure \ref{fig:QpropvsCQSFAlow}).
Solving for $\epsilon$ yields
\begin{equation}
\epsilon_c = \frac{5 \sqrt{\omega}}{32 \sqrt{I_p} \sqrt{U_p}}\sqrt{5^2 \omega + \sqrt{2048 I_p U_p + 5^4 \omega^2} }, \label{eq:criticalEllipticity}
\end{equation}
an estimate for a critical value of ellipticity, beyond which interferences are expected to vanish. 
This expression scales $\propto \sqrt{\omega} + \order{\sqrt{\omega}^3}$ for constant ponderomotive energy $U_p$ and all other parameters. This scaling is consistent with the parameter range employed in this work. 
Evaluating \eqref{eq:criticalEllipticity} for the laser parameters typically used in our study (see for example figure \ref{fig:QpropvsCQSFAlow}), we find a critical value of around $\epsilon_c = 0.33$, comparable to the results of our calculations in that same figure. 

\subsection{SFA Ionization times}
With the elliptically polarized field as given in equation \eqref{eqn: elliptically polarized field}, we obtained the analytic form of the ionization time by solving the SFA saddle point equation \eqref{eqn: SFA SPE} for tunnel ionization in the whole $p_zp_x$ plane. This goes beyond the work in \cite{Javsarevic2020}, which proved that the ionization time with elliptically polarized fields has two solutions, but only provided analytic expressions along the major polarization axis. In the present paper, we calculated the ionization times analytically and determined the regions in the momentum plane for which they are valid.

Since the ionization times' analytic form is complicated, we introduced $t_{11, n}$, $t_{12, n}$, $t_{21, n}$, and $t_{22, n}$ in Eq.~\eqref{eqn: four solutions}, where n denotes  the cycle number and the variables $\zeta$ and $\eta$ are defined in Appendix 1. These expressions are the candidates for the ionization time. The solutions $t_{11, n}$, $t_{21, n}$ are both valid in the first and third quadrant in the momentum plane, and $t_{12, n}$, $t_{22, n}$ both hold in the second and fourth quadrant. We grouped $t_{11, n}$, $t_{12, n}$, and $t_{21, n}$, $t_{22, n}$ as $t_{1, n}$ and $t_{2, n}$ to render a solution valid in the full $p_zp_x$ plane (equation \eqref{eqn: grouped ionization time}). We will refer to the ionization quantum trajectory associated with the ionization time $t_{1, n}$ as an orbit $a$ and one with $t_{2, n}$ as an orbit $b$. 
Please see the next section \ref{sec:classification} for more details on orbit classification. 
The necessity of specifying domains in the  $p_zp_x$  plane stems from the fact that Eq.~\eqref{eqn: four solutions} was obtained from a quartic equation. This implies that two solutions become spurious in specific domains. This derivation is sketched in Appendix 1. 
\begin{widetext}
\begin{equation}
    \begin{split}
        t_{11, n} &= \frac{2\pi(n+1)}{\omega}  - \frac{1}{\omega}
        \arccos{\left[\frac{
            -p_z + \zeta + i\sqrt{
            (1-\epsilon^2)(2I_p + p_x^2) + \epsilon^2 p_z^2  + 4\zeta^2+\eta
        }}{2(1-\epsilon^2)\sqrt{U_p}}\right]}\\
        t_{12, n} &= \frac{2\pi(n+1)}{\omega}  - \frac{1}{\omega}
        \arccos{\left[\frac{
            -p_z - \zeta + i\sqrt{
            (1-\epsilon^2)(2I_p + p_x^2) + \epsilon^2 p_z^2  + 4\zeta^2-\eta
        }}{2(1-\epsilon^2)\sqrt{U_p}}\right]}\\
        t_{21, n} &= \frac{2\pi(n)}{\omega}  + \frac{1}{\omega}
        \arccos{\left[\frac{
            -p_z + \zeta - i\sqrt{
            (1-\epsilon^2)(2I_p + p_x^2) + \epsilon^2 p_z^2  + 4\zeta^2+\eta
        }}{2(1-\epsilon^2)\sqrt{U_p}}\right]}\\
        t_{22, n} &= \frac{2\pi(n)}{\omega}  + \frac{1}{\omega}
        \arccos{\left[\frac{
            -p_z - \zeta - i\sqrt{
            (1-\epsilon^2)(2I_p + p_x^2) + \epsilon^2 p_z^2  + 4\zeta^2-\eta
        }}{2(1-\epsilon^2)\sqrt{U_p}}\right]}\\
   \end{split}
   \label{eqn: four solutions}
\end{equation}
\end{widetext}

\begin{equation}
\openup 2\jot
\begin{split}
    t_{1, n} =
        \left\{
        	\begin{array}{ll}
        		t_{11, n}  & \mbox{if } p_{z}p_{x} \geq 0 \\
        		t_{12, n} & \mbox{if } p_{z}p_{x} < 0
        	\end{array}
        \right. \\
    t_{2, n} =
        \left\{
        	\begin{array}{ll}
        		t_{21, n}  & \mbox{if } p_{z}p_{x} \geq 0 \\
        		t_{22, n} & \mbox{if } p_{z}p_{x} < 0
        	\end{array}
        \right.
\end{split}
    \label{eqn: grouped ionization time}
\end{equation}

Further to that, the two orbits given by the times in \eqref{eqn: grouped ionization time} are not always physically significant. Depending on the parameter range, one of the saddle-point solutions may become inaccurate and lead to divergencies in the PMDs. This is due to the presence of Stokes transitions, which are directly related to topological changes in the contours introduced by nonzero ellipticity. For high ellipticities, a \textit{single} ionization time will be associated with a specific final momentum, or angle. This behavior makes the attosecond angular streaking method, `the attoclock', possible \cite{Eckle2008,Eckle2008a}. These Stokes transitions have been first identified in \cite{Paulus1998} for a restricted parameter range, and in \cite{Javsarevic2020} it was stressed that for elliptically polarized fields there is \textit{always} a Stokes transition. However, for low ellipticity they are outside the physically relevant parameter range. Details about Stokes transitions and the high ellipticity limit for $\Re[t']$ are provided in Appendices 2 and 3, respectively. 

\subsection{CQSFA ionization times and orbit classification}
\label{sec:classification}

\begin{figure*}[h!tb]
    \centering
   \includegraphics[width=0.24\textwidth] {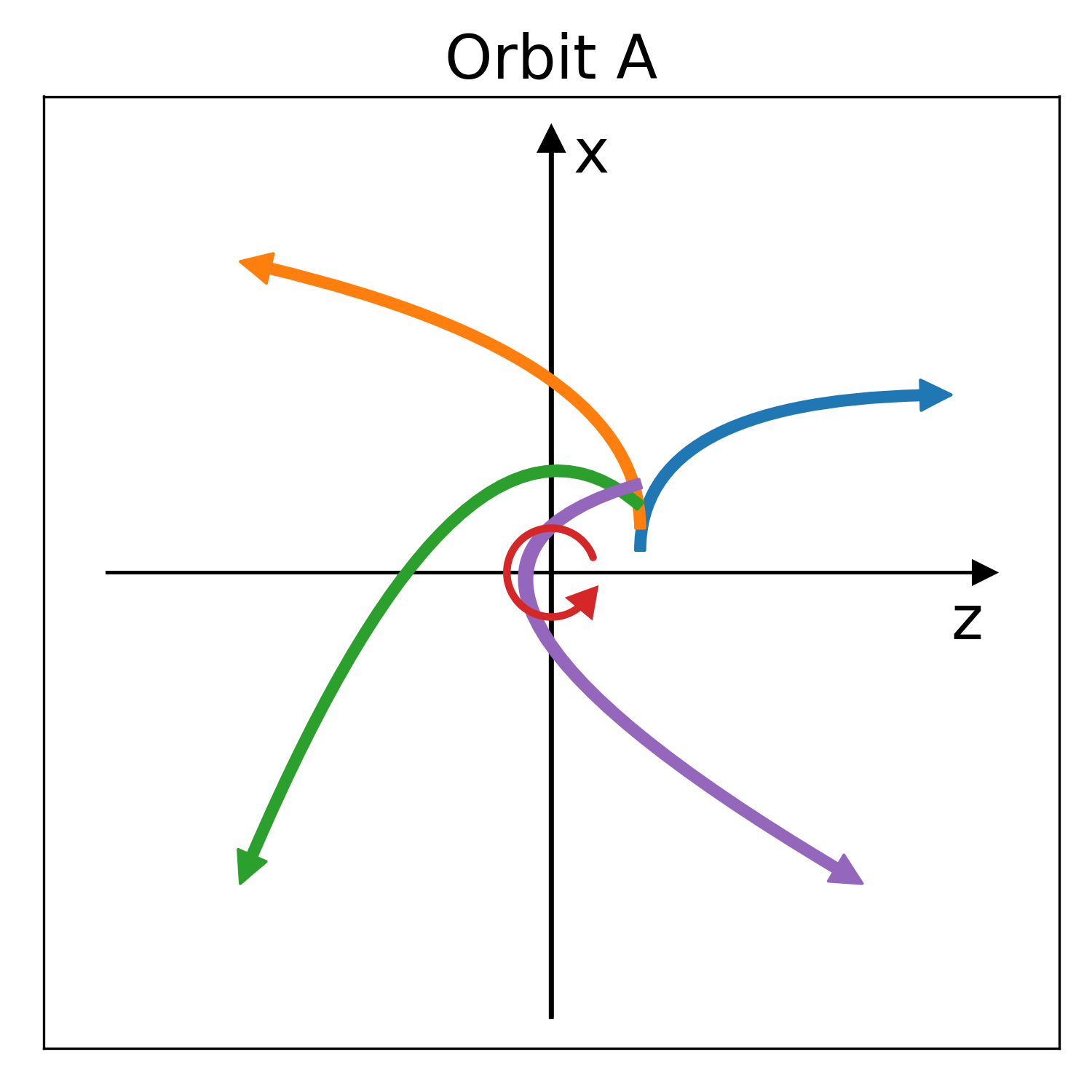}
   \includegraphics[width=0.24\textwidth] {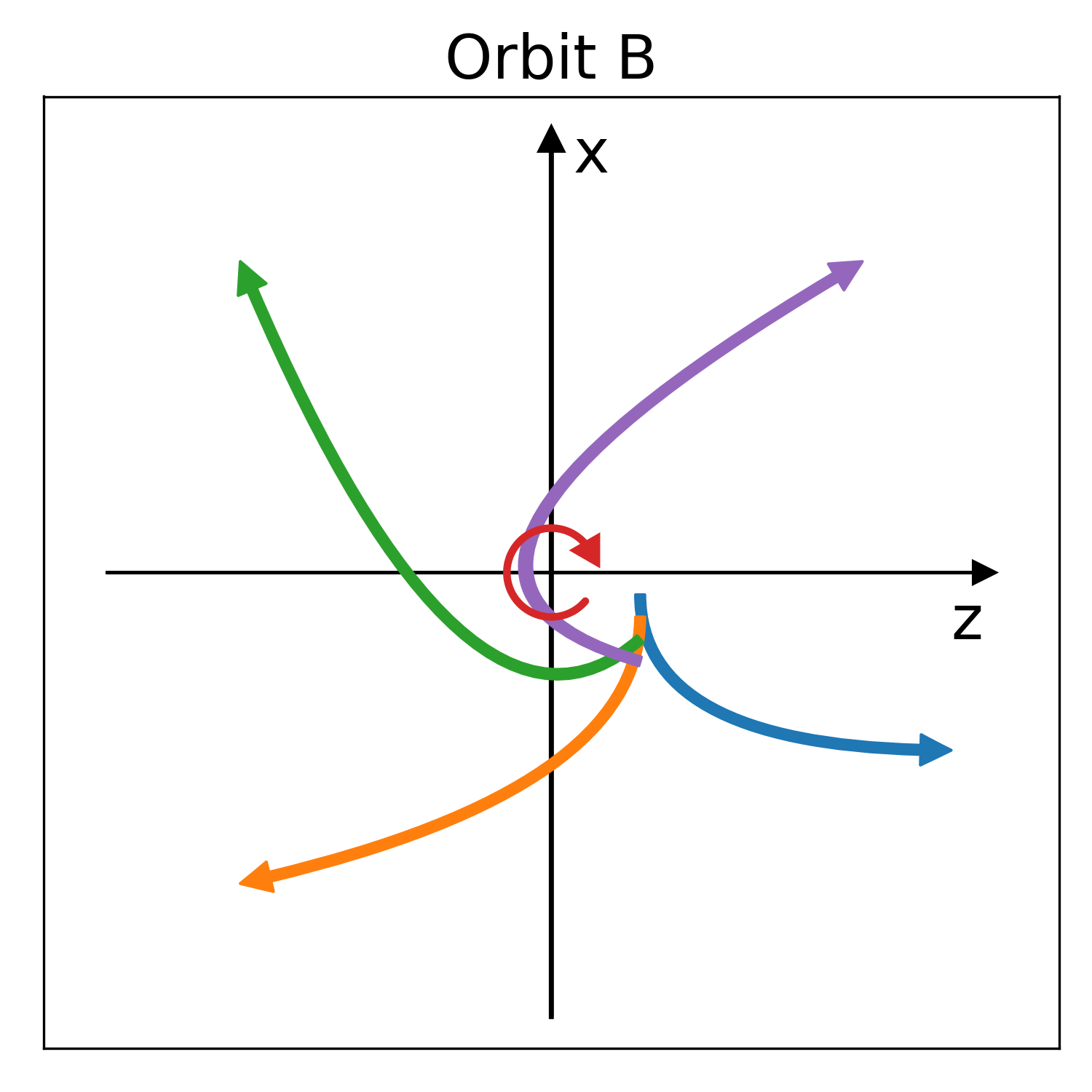}
   \includegraphics[width=0.24\textwidth] {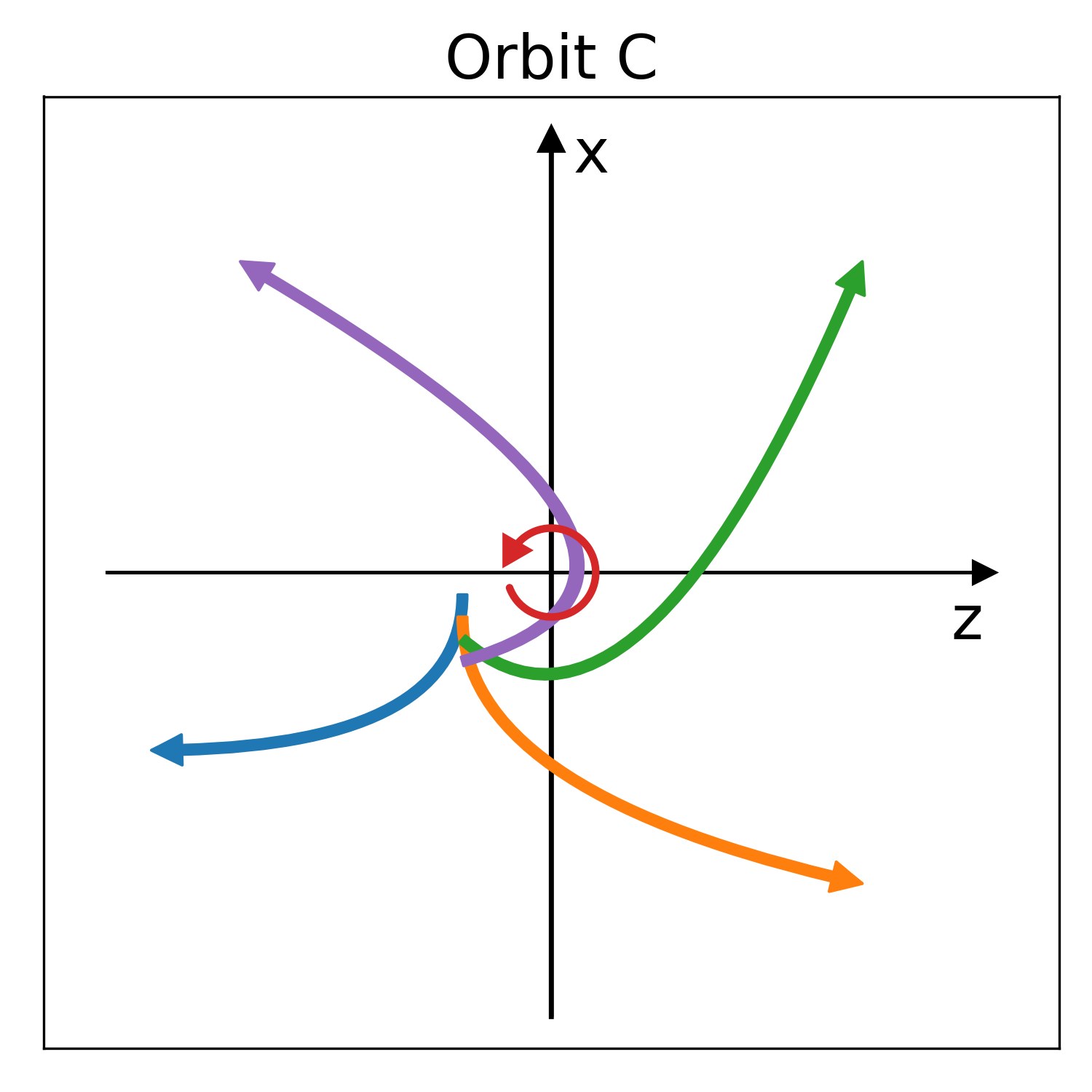}
   \includegraphics[width=0.24\textwidth] {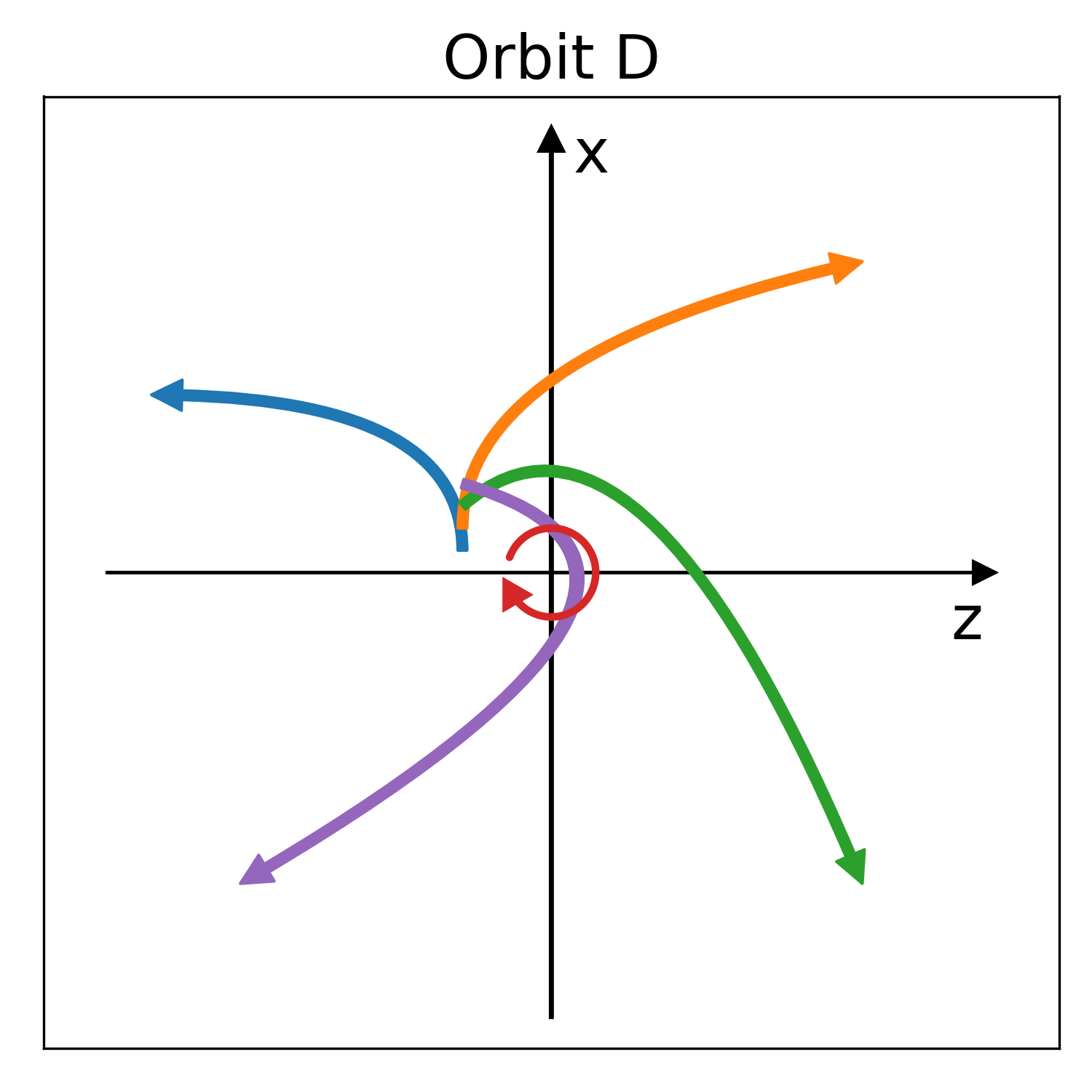} \\
   \includegraphics[width=0.24\textwidth] {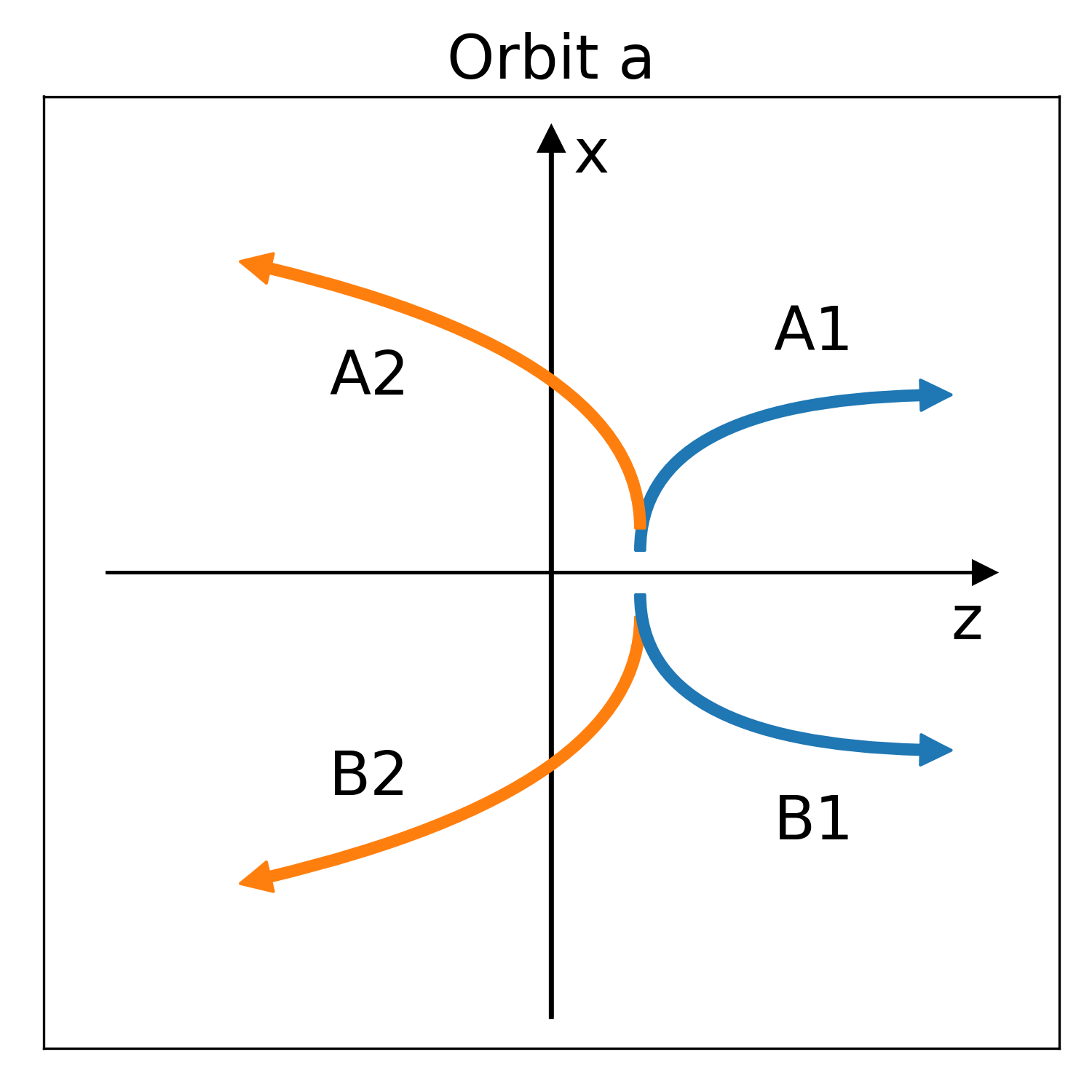}
   \includegraphics[width=0.24\textwidth] {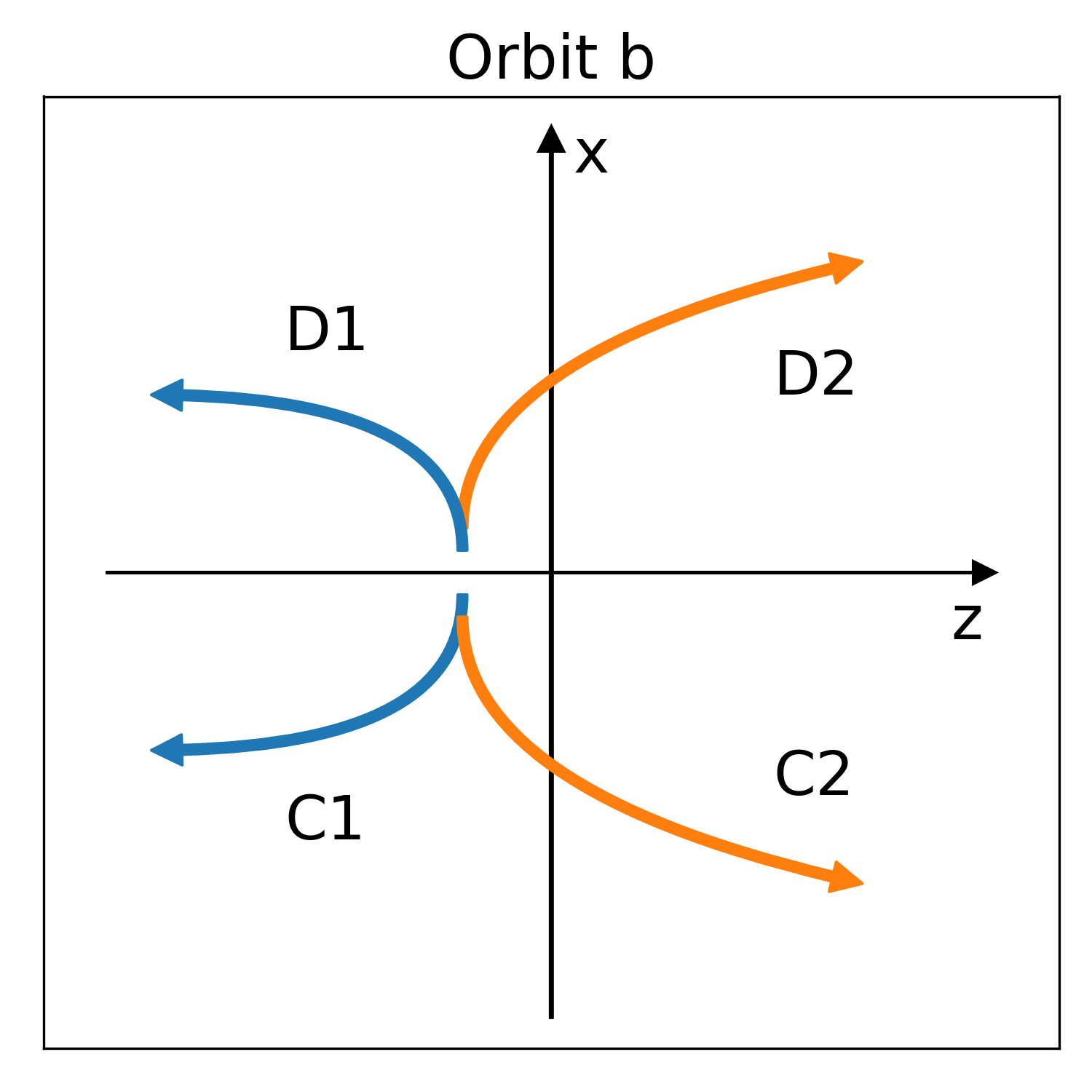}
   \includegraphics[width=0.24\textwidth] {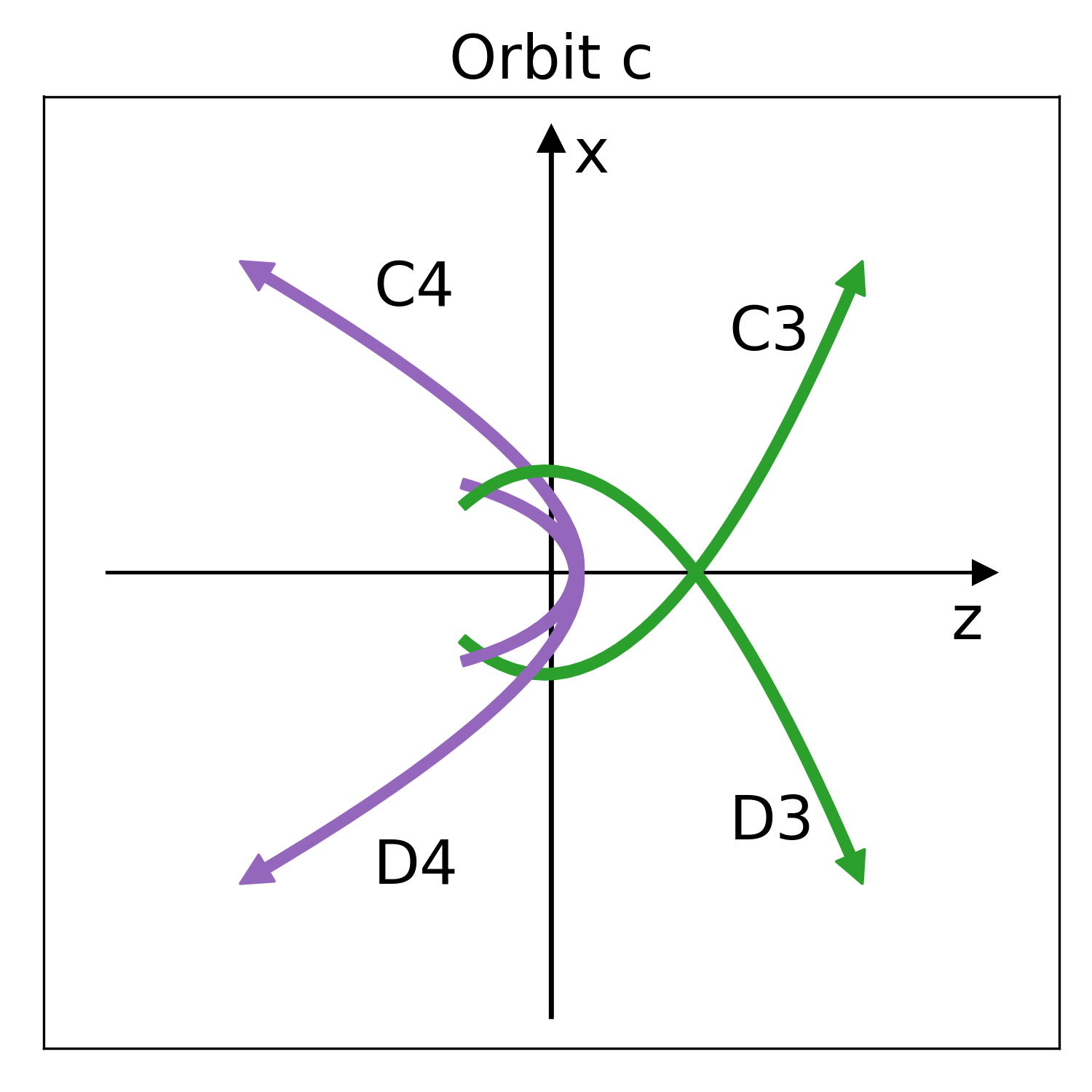}
   \includegraphics[width=0.24\textwidth] {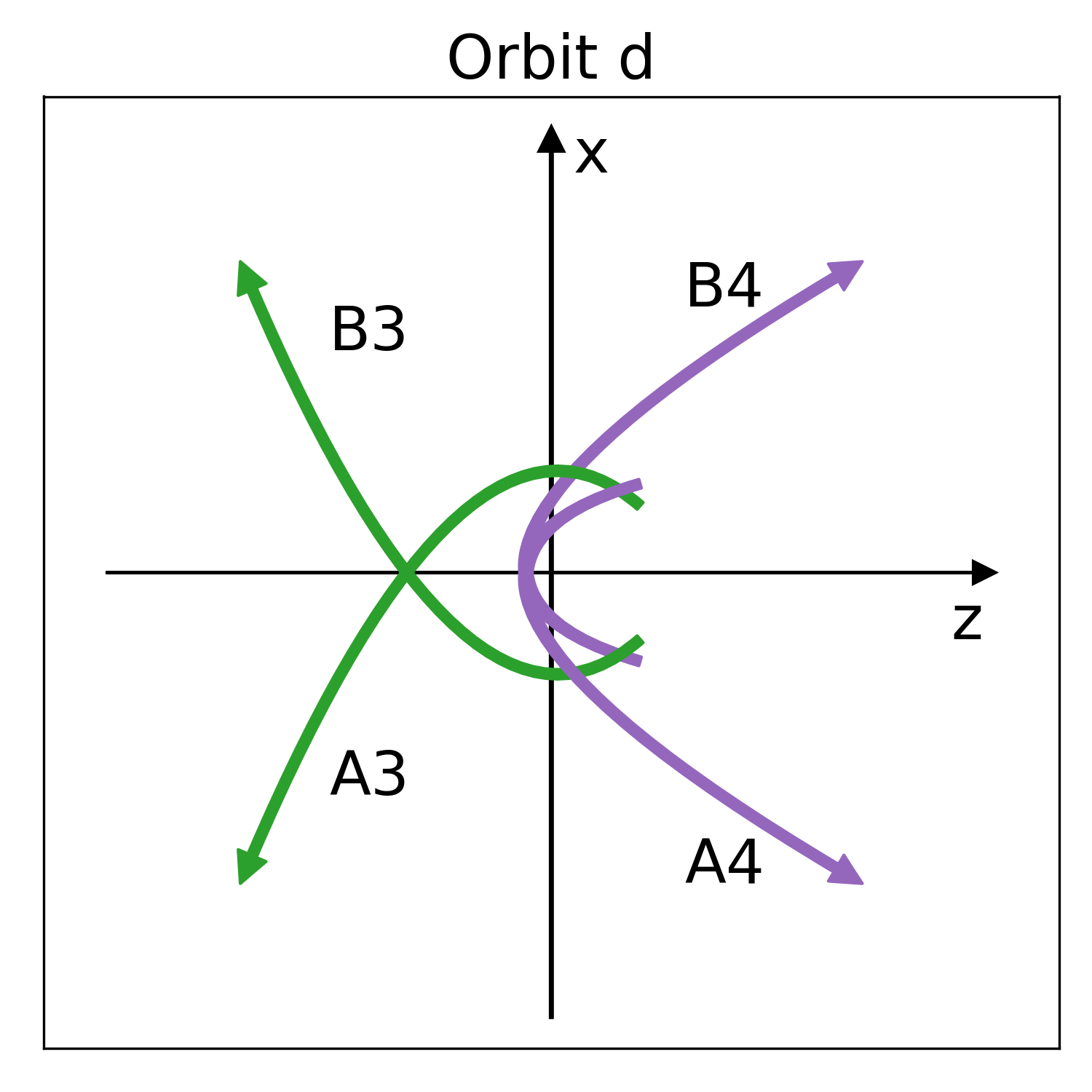}
   \\
   \includegraphics[width=0.96\textwidth] {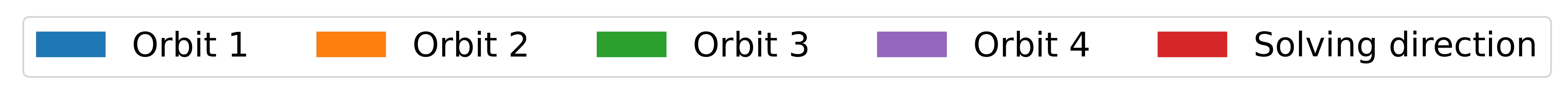}
    \caption{\textbf{Schematic representation of the classification of orbits used in this paper.}  The colors of the orbits give orbits 1 to 4, according to the classification criteria introduced in \cite{Yan2010,lai2015influence} of the CQSFA with a linearly polarized field. Orbits A and B (C and D) correspond to the times t1,n (t2,n) in Eq.~\ref{eqn: grouped ionization time}, solved asymptotically in the anticlockwise and clockwise direction, which are marked as red circular arrows. Classification of orbits $a-d$ are used for calculating PMDs.  The curves indicated in the figure are not actual electron trajectories, but schematic illustrations of the orbit's overall  behavior. \label{fig:Classification}}
\end{figure*}

\begin{table}
\centering

\vspace{5pt}
\begin{tabular}{ c c c c c c c}
  \hline
  Orbit & $z_0 p_{fz}$ & $p_{fx}p_{0z}$ & $1^{st} \rm{quad.}$ & $2^{nd} \rm{quad.}$ & $3^{rd} \rm{quad.}$ & $4^{th} \rm{quad.}$ \\ 
  \hline\hline
  1 & + & + & a & b & b & a \\ 
  2 & - & + & b & a & a & b \\ 
  3 & - & - & c & d & d & c \\ 
  4 & + & - & d & c & c & d \\ 
  \hline
\end{tabular}
\caption{\textbf{Orbit classification compared to the linearly polarized case.} In the CQSFA calculation with linearly polarization field, orbit classification 1-4 classifies the orbit with two different conditions, the sign of $z_0p_{fz}$ and $p_{fx}p_{0z}$. However, with an elliptically polarized field, because of the broken reflection symmetry, orbits $1-4$ could not directly be matched with orbits $a-d$. For convenience, we provide a relation between orbits $1-4$ and $a-d$ here. The first column in the table gives the orbit classification $1-4$ used in our previous work for linearly polarized fields, the second column provides the conditions upon the tunnel exit and momentum components for each orbit, and the remainining columns yield the classification a-d in each quadrant of the $p_zp_x$ plane. }
\label{tab:OrbitClassification}
\end{table}

In contrast to the SFA, the saddle point solutions for the CQSFA cannot be directly expressed analytically; thus, we have to solve the saddle point equations \eqref{eq:SPEt}-\eqref{eq:SPEr} numerically. Nevertheless, we can expect that the SFA and CQSFA solution dynamics are most similar in those orbits whose final momentum is the same with its ionization direction. Therefore, we can use the SFA solution as a first guess at some point in the momentum space. Then we can asymptotically solve other points by using a solved neighbor point as an initial guess. In this method, because of the core in the center, clockwise and counterclockwise solved solutions are different; therefore, we can obtain two CQSFA solutions from one SFA solution. As shown in Fig.~\ref{fig:Classification}, we named the counterclockwise (clockwise) solved solution derived from $t_{1, c}$ as solution class $A(B)$, and that derived from $t_{2, c}$ as $C(D)$. 

Each solution class $A$ to $D$ contains a single orbit $1$ to $4$, whose criteria are its ionization direction and final momentum. 
This classification into orbits $1$ to $4$ is robust for the linearly polarized case because orbits in each classification $A$ to $D$ show similar dynamics.  For linear polarization, orbits $1$ reach the detector directly, orbits $2$ and $3$ are field-dressed Kepler hyperbolae and orbits $4$ go around the core. This classification was introduced in \cite{Yan2010} and employed by us in previous publications (see, e.g.,  \cite{lai2015influence,maxwell2017coulomb}), and one can understand how such orbits interfere by piecing them together. For instance, orbit $A1$ reaches the detector directly on the first quadrant of the figure, and interferes with orbits $C3$ and $D2$, which both start on the ``wrong" side half a cycle later, but differ in their transverse momentum behavior. Finally, orbit $B4$ starts on the same side as $A1$ (meaning it is born in the same half cycle), but goes around the core before eventually reaching the detector. The orbits in this example form interference patterns in the first quadrant of the momentum plane. 

For an elliptically polarized field, on the other hand, the reflection symmetry about both major and minor polarization axis breaks down. Therefore, this symmetry breaking makes the PMD of each orbit with the classification employed in the linear case discontinuous in both the major and minor polarization axis. 
Instead, the final PMD typically shows a point-symmetry about the origin. 
Therefore we introduce a new generalized orbit classification $a-d$, according to the lower panels in Fig.~\ref{fig:Classification}.
Orbits whose tunnel exits are located at $z>0$ ($z<0$) and whose momentum components $p_x$ do not change sign during continuum propagation are designated as $a$ ($b$), while orbits $c$ ($d$) start at $z<0$ ($z>0$) and switch half planes during continuum propagation. Note that this classification respects the fact that a non-vanishing field ellipticity introduces a preferential rotational direction, which must be taken into consideration. For clarity, in Table \ref{tab:OrbitClassification} we provide the correspondence between both labeling systems. For the first and fourth quadrant, the classifications $1-4$ and $a-d$ coincide, but this is not the case in the other two quadrants. 

In this paper, we will use this new heuristic classification for orbit $a$ to $d$ as outlined in Fig.~\ref{fig:Classification} and specified in Table \ref{tab:OrbitClassification}. Two main points considered in this classification are grouping the orbits with similar dynamical behaviors and minimizing discontinuity issues in the single-orbit PMDs. Since $a$ and $b$ are less affected by the Coulomb potential, their action is more similar to the SFA case than $c$ and $d$. This difference is more conspicuous with higher ellipticity. However, it is impossible to create fully continuous PMDs only with the orbit $a$ and $b$ ($c$ and $d$). Most of the significant interference patterns appear near the minor polarization axis, not the major axis; we choose the classification which makes PMD continuous on the minor axis. Note that the method of classification does not affect the complete PMD calculation, whose outcome is displayed in Fig.~\ref{fig:QpropvsCQSFAlow} and Fig.~\ref{fig:QpropvsCQSFAhigh}.

\section{Photoelectron momentum distributions}
\label{sec:PMD}

Below, we discuss photoelectron momentum distributions (PMDs) and provide interpretations for the features encountered in terms of the saddle-point solutions for the ionization time $t'$. Throughout, unless necessary, $\mathbf{p}$ refers to the \textit{final} momentum, measured at the detector. Due to the presence of the Coulomb potential in the continuum, for the CQSFA and TDSE computations this will not be the momentum $\mathbf{p}_0$ at the instant of ionization. For the SFA, the final and initial momentum are identical. 

\begin{figure*}[h!p]
	\centering
	\includegraphics[width=0.4\textwidth] {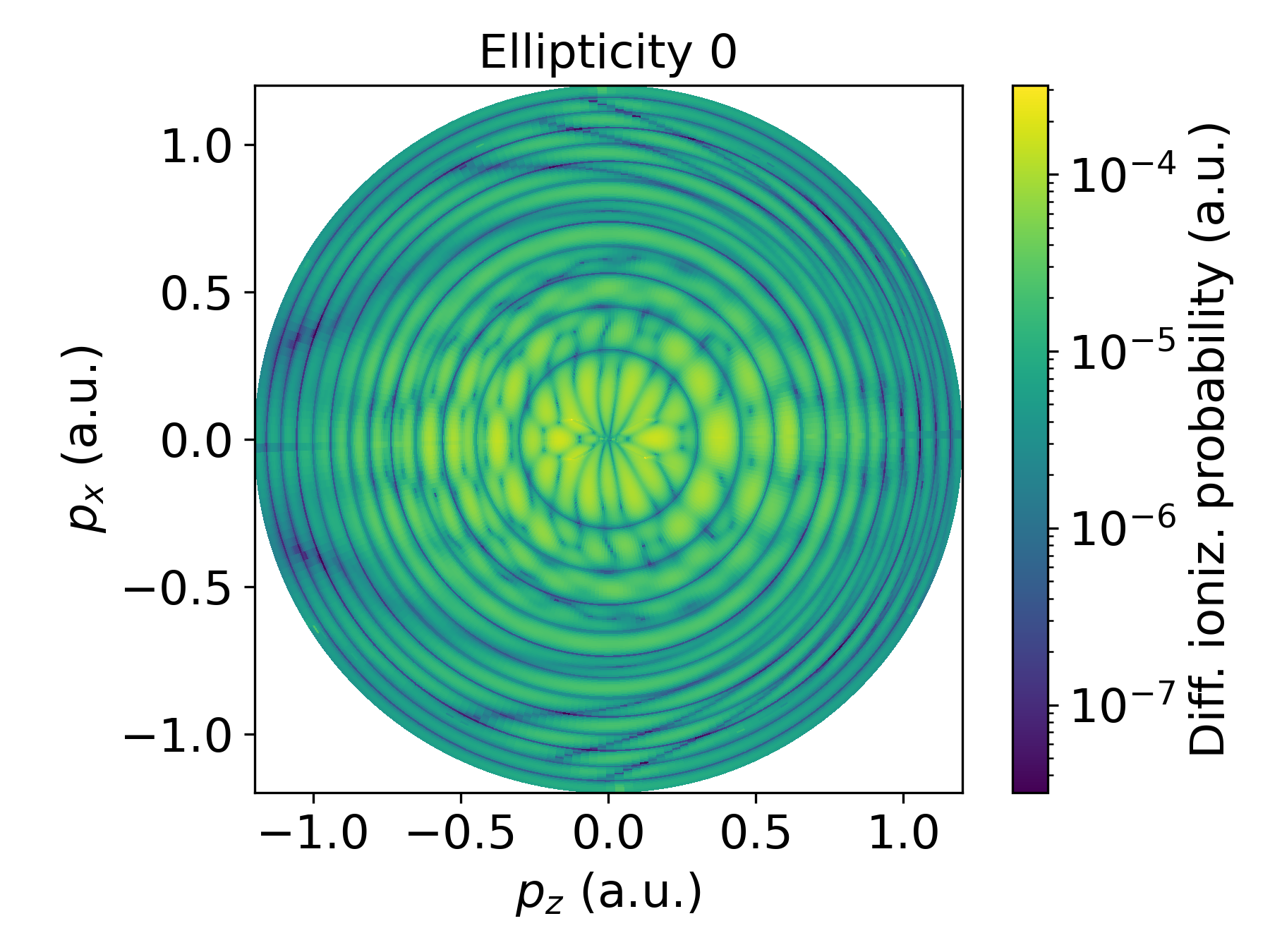}
	\includegraphics[width=0.4\textwidth] {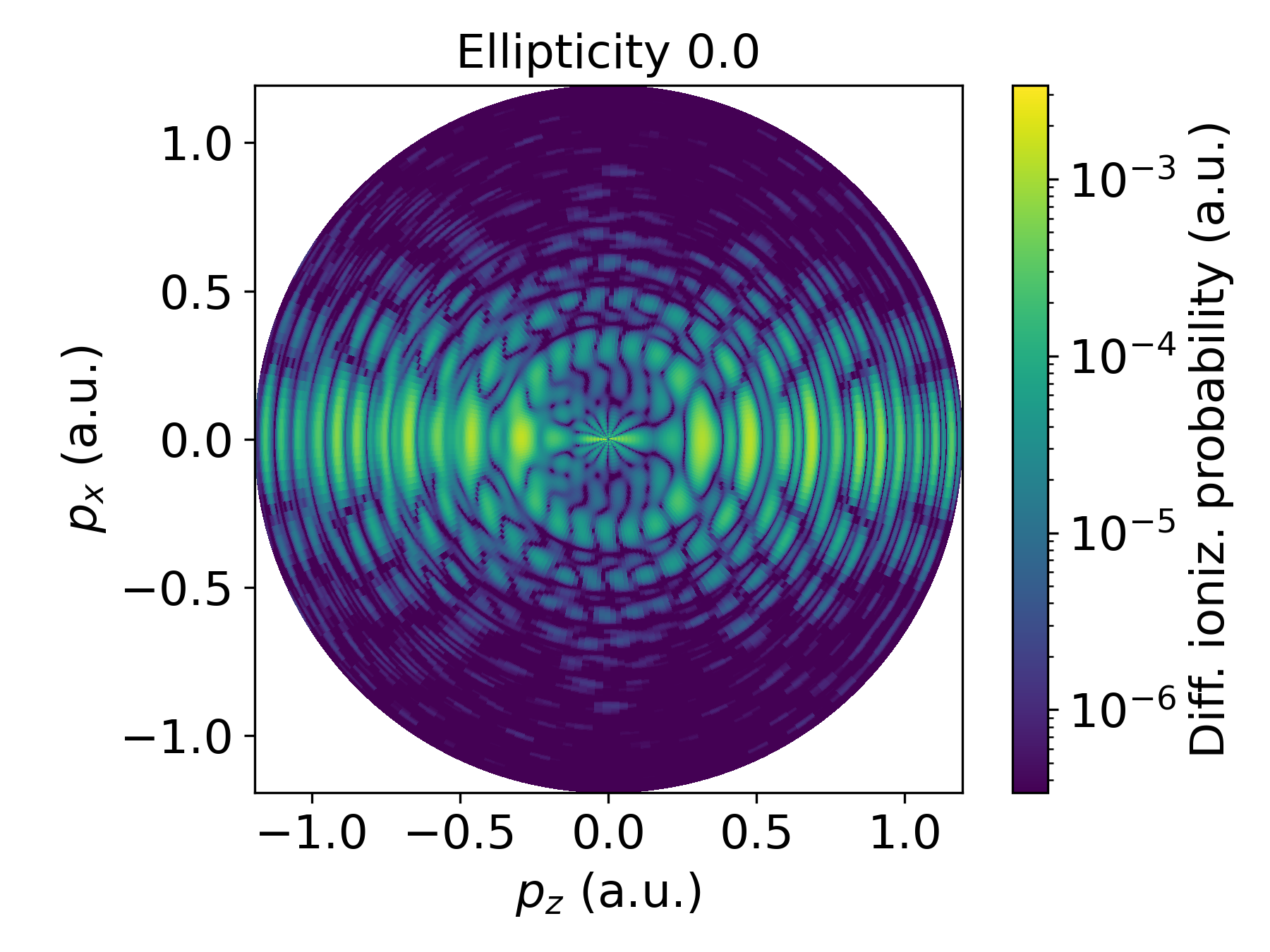} \\
	\includegraphics[width=0.4\textwidth]
	{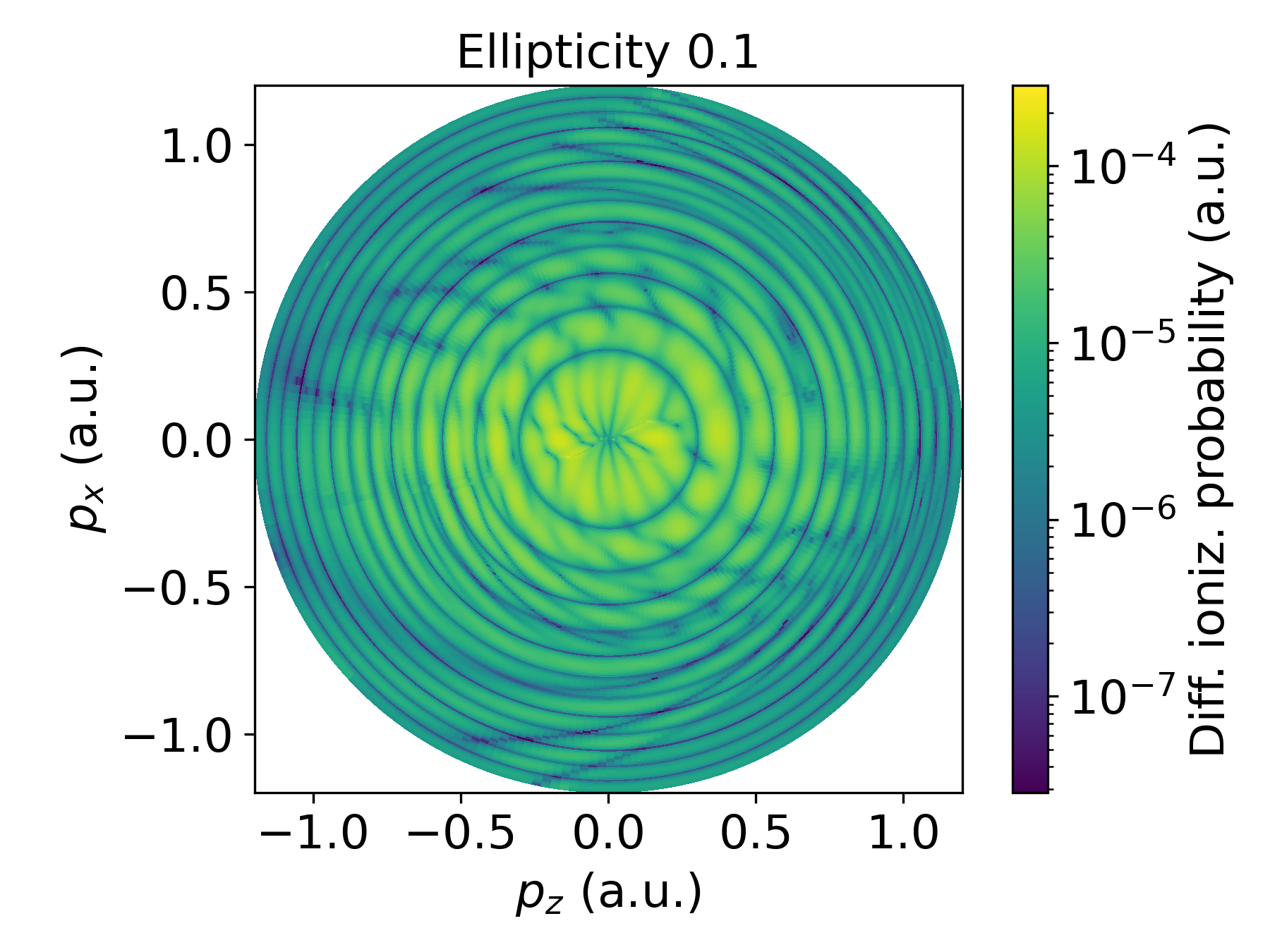}
	\includegraphics[width=0.4\textwidth]
	{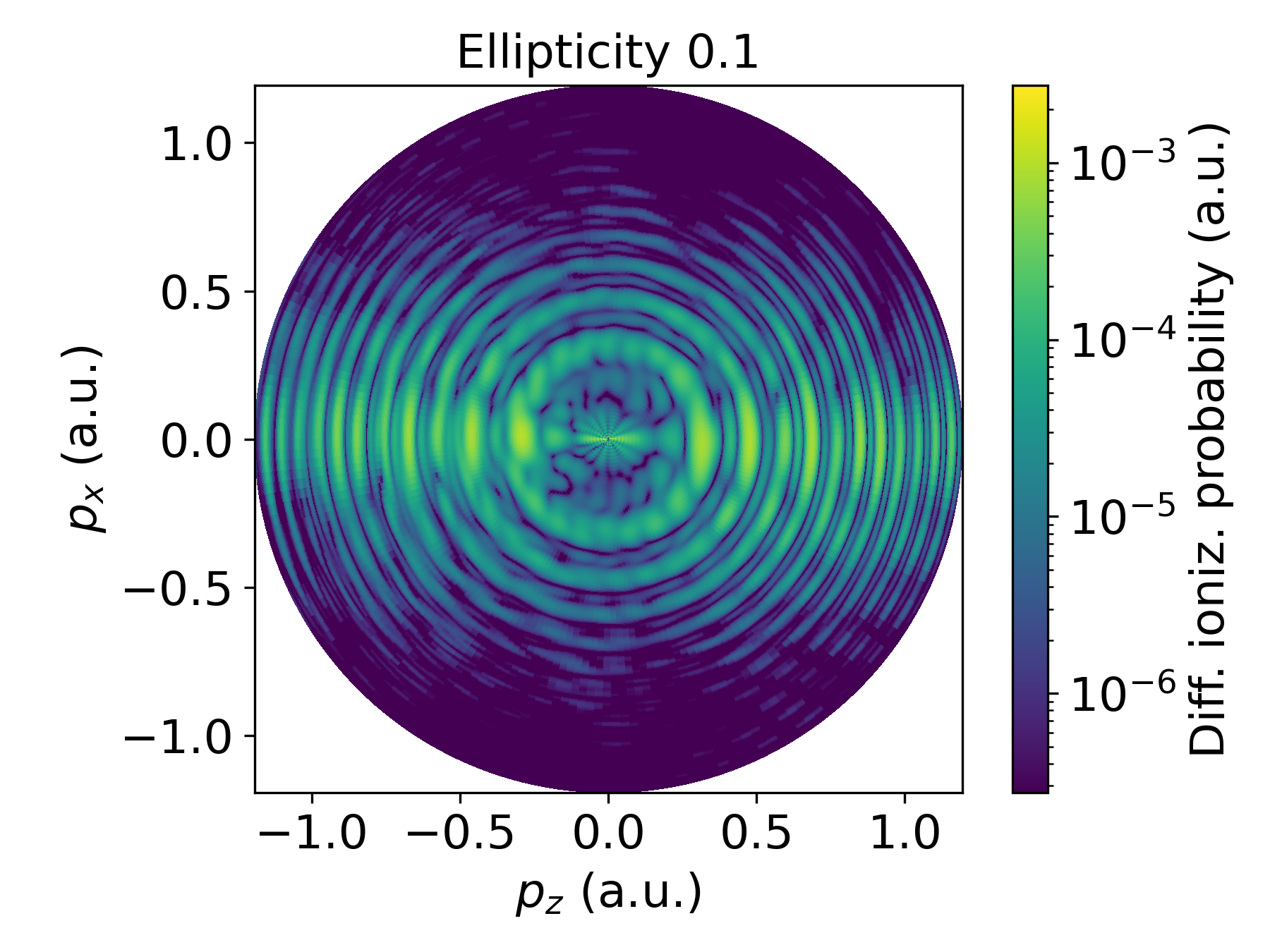} \\
	\includegraphics[width=0.4\textwidth]
	{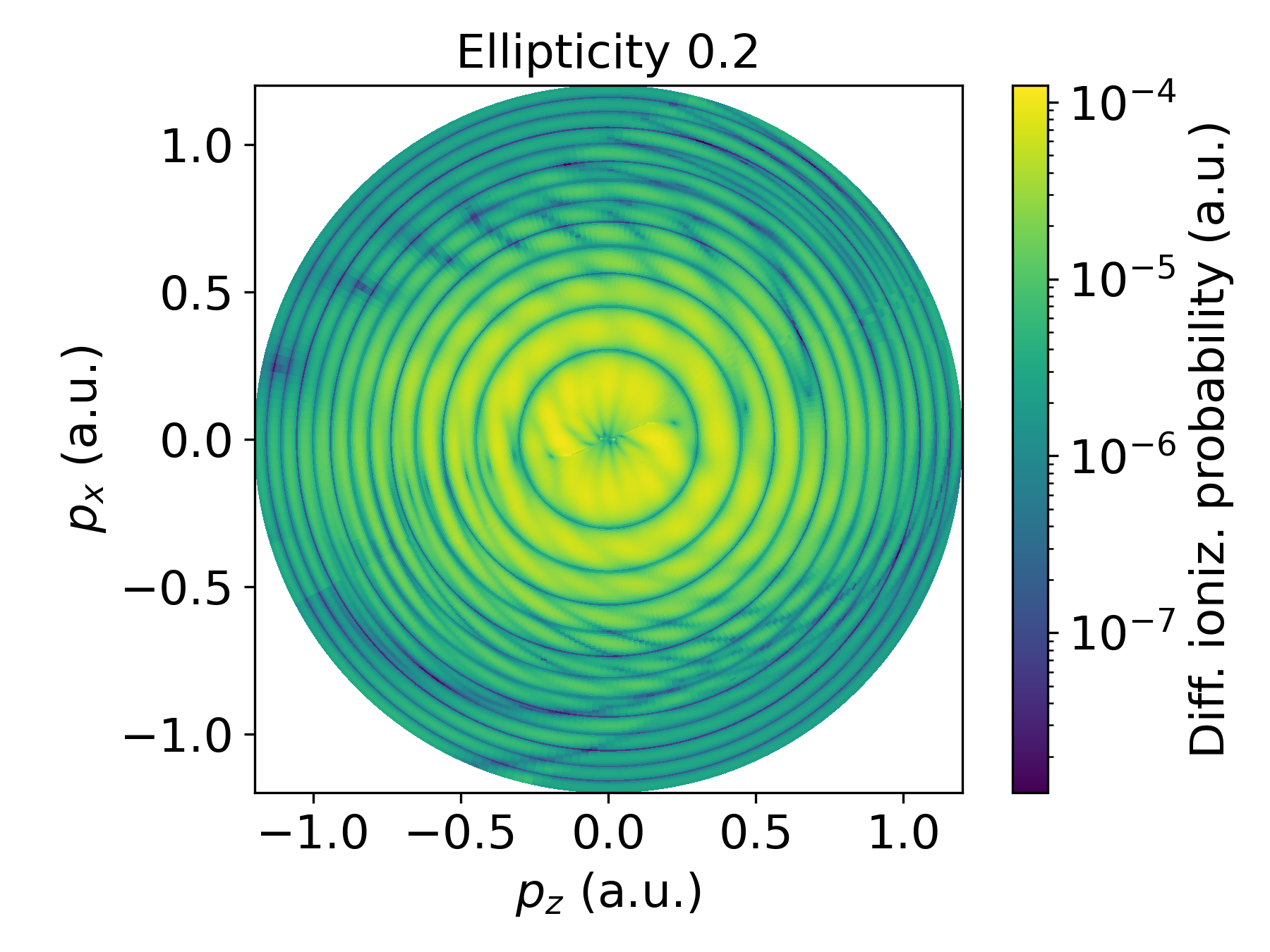}
	\includegraphics[width=0.4\textwidth]
	{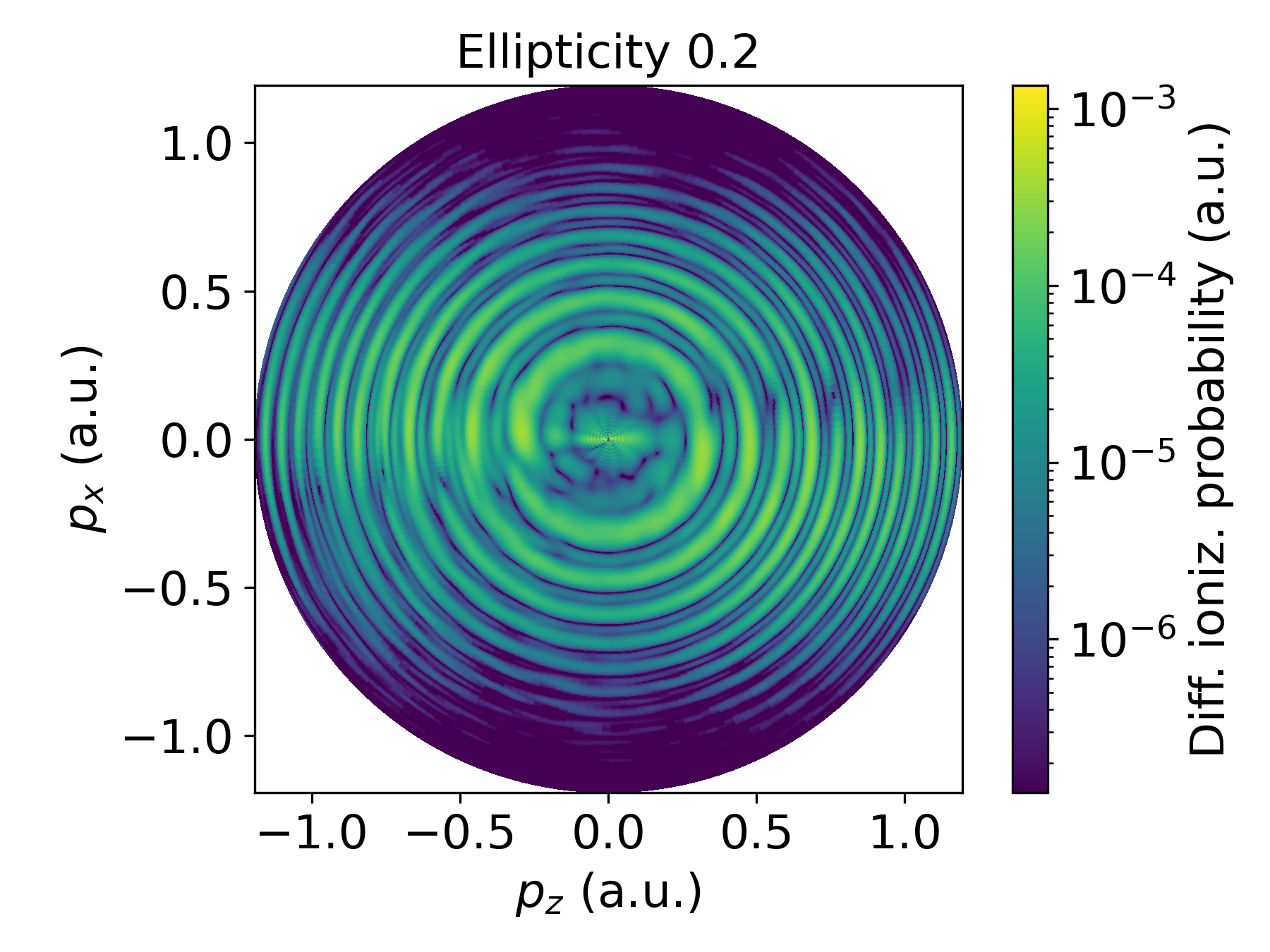} \\
	\includegraphics[width=0.4\textwidth]
	{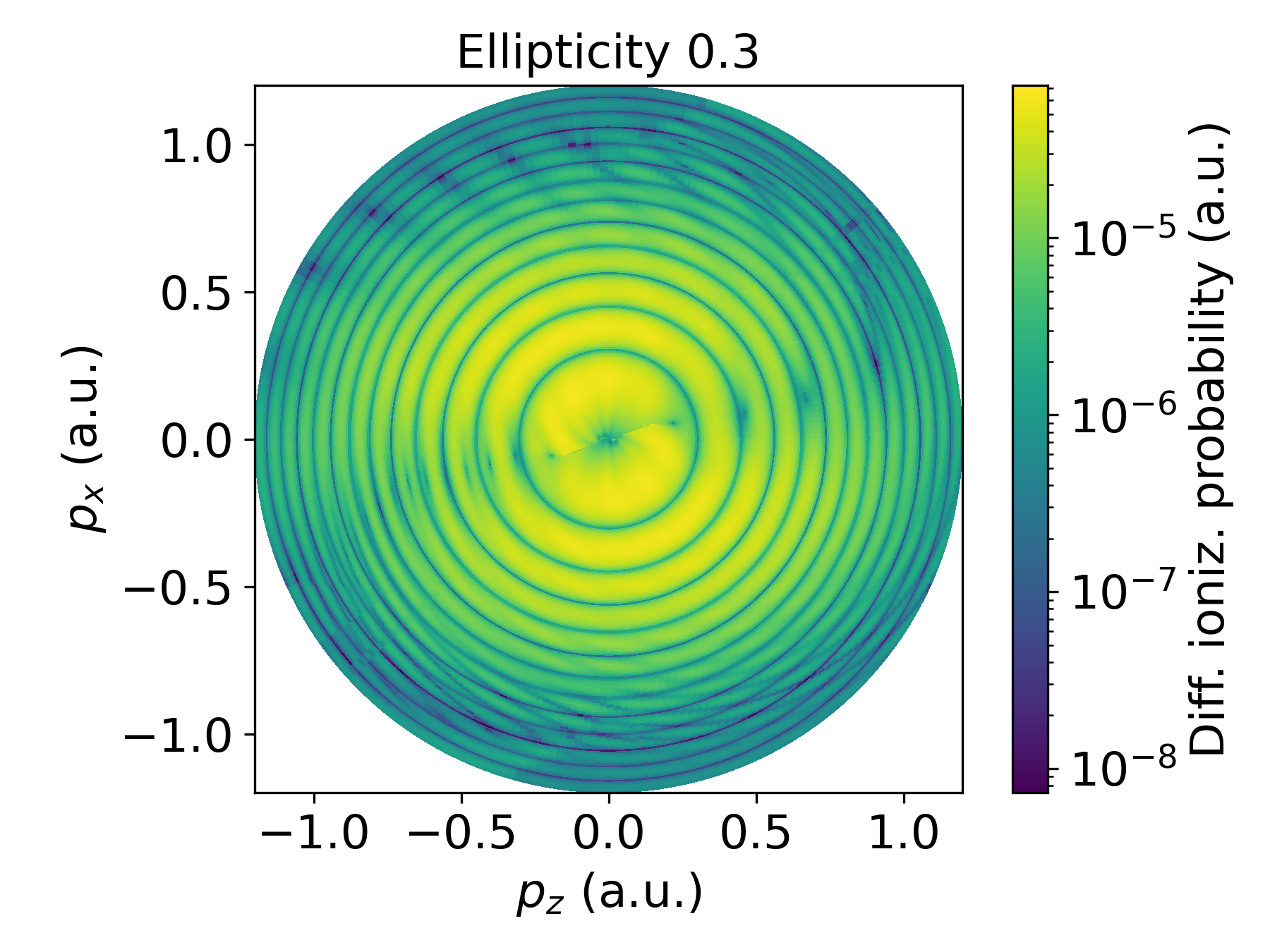}
	\includegraphics[width=0.4\textwidth]
	{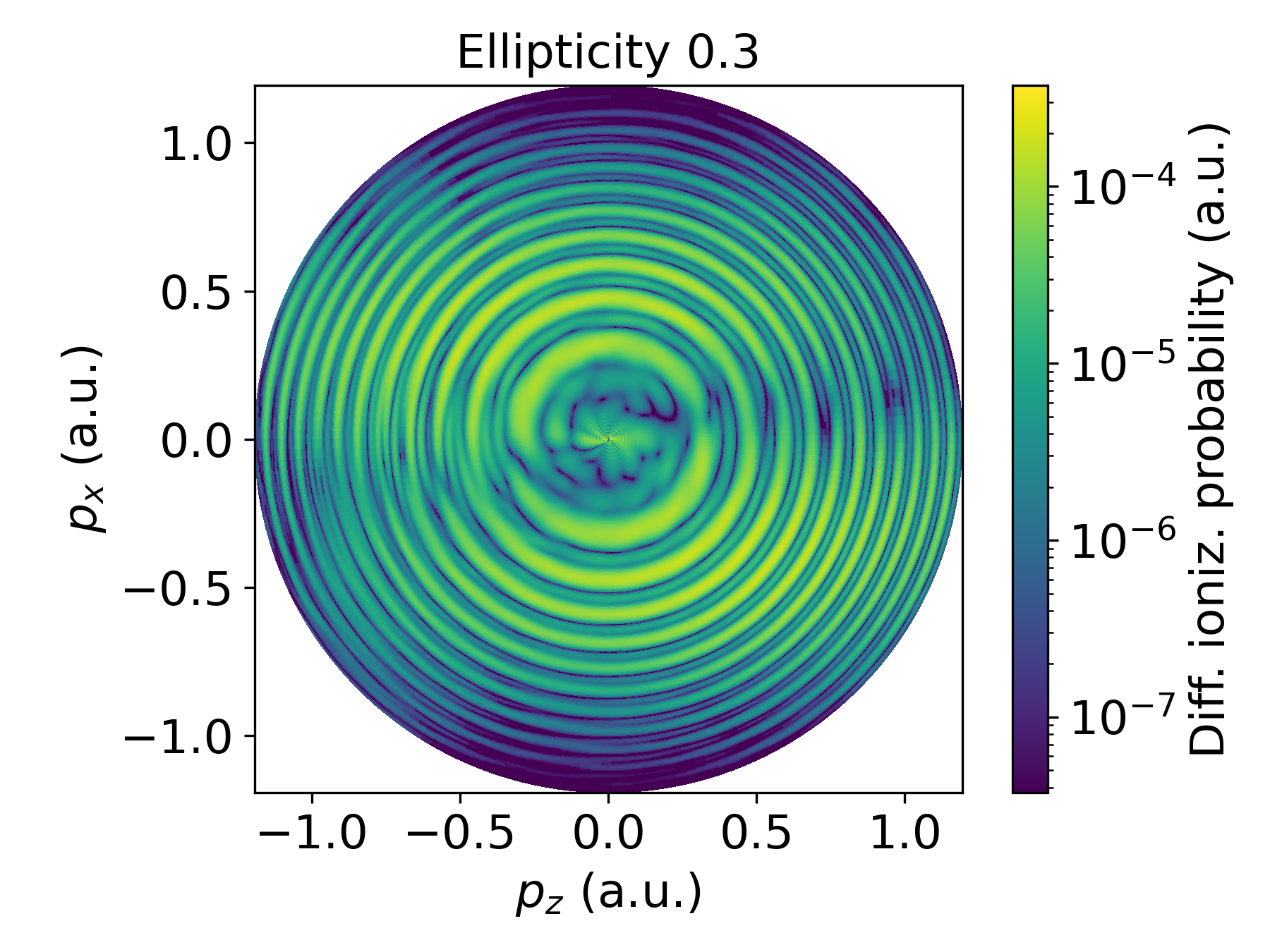} 
	\caption{Photoelectron momentum distributions calculated for  helium in a field of intensity $2.5 \times 10 ^{14}$ W/cm$^2$, wavelength $\lambda =$ 735 nm, whose ellipticity increases from $\epsilon=0$ to $\epsilon=0.3$. The left and the right columns have been calculated with the CQSFA and the Schr\"odinger equation solver Qprop \cite{Tulsky2020}, respectively. For Qprop, we have used an envelope with $(1-\cos^{16})$ shape for a pulse with four cycles total duration, creating a near-flattop pulse, and the final PMDs have been calculated with the isurfv option \cite{Tulsky2020}. Each panel has been normalized to its maximum value and plotted in a logarithmic scale.}
	\label{fig:QpropvsCQSFAlow}
\end{figure*}
Fig.~\ref{fig:QpropvsCQSFAlow} shows photoelectron-momentum distributions computed with the CQSFA and the one-electron time-dependent Schr\"odinger solver Qprop \cite{Mosert2016,Tulsky2020}, for a range of  driving-field ellipticities (left and right columns, respectively). The PMDs exhibit above-threshold ionization (ATI) rings stemming from inter-cycle interference, and, for vanishing and low ellipticity, holographic patterns resulting from interfering intra-cycle events. These patterns are clearly identifiable for linearly polarized fields (upper row in the figure), as the spider-like fringes near the polarization axis ($\theta=0$), a fan-shaped structure close to the ionization threshold and a carpet-like structure around the angle $\theta=90^{\circ}$. Aside minor issues associated with the finite pulse length, the patterns are symmetric with regard to reflections around $\theta=0^{\circ}$ and $\theta=90^{\circ}$. For non-vanishing ellipticity, these symmetries are lost and the patterns start to twist in the  anticlockwise direction, following the rotational sense of the driving field. Furthermore, the centers of the distributions start to split and the interference fringes become more and more blurred for increasing ellipticity. These features are observed throughout, although there are quantitative differences.

\begin{figure*}[h!t]
    \centering
   \includegraphics[width=0.4\textwidth] {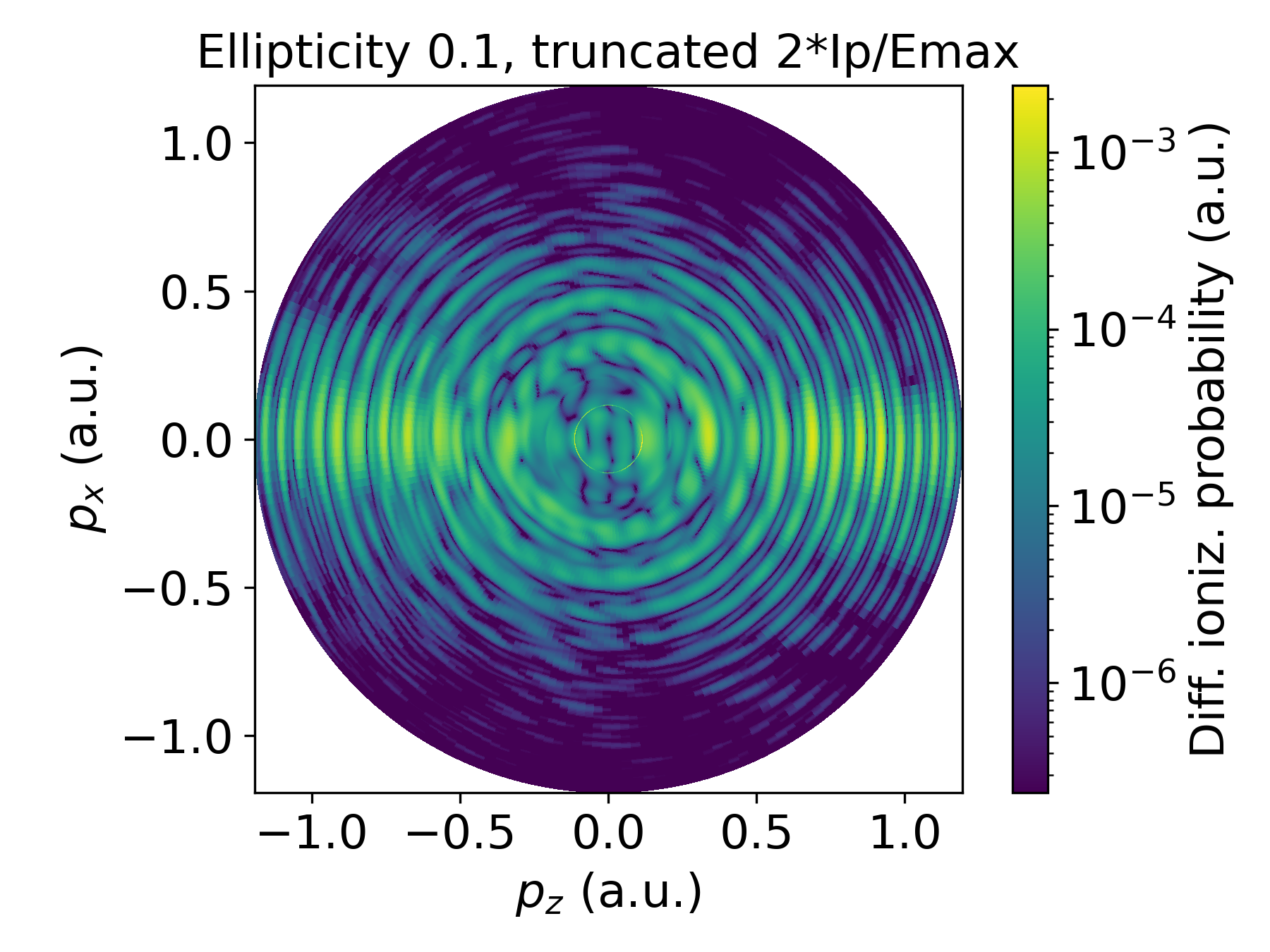}
   \includegraphics[width=0.4\textwidth] {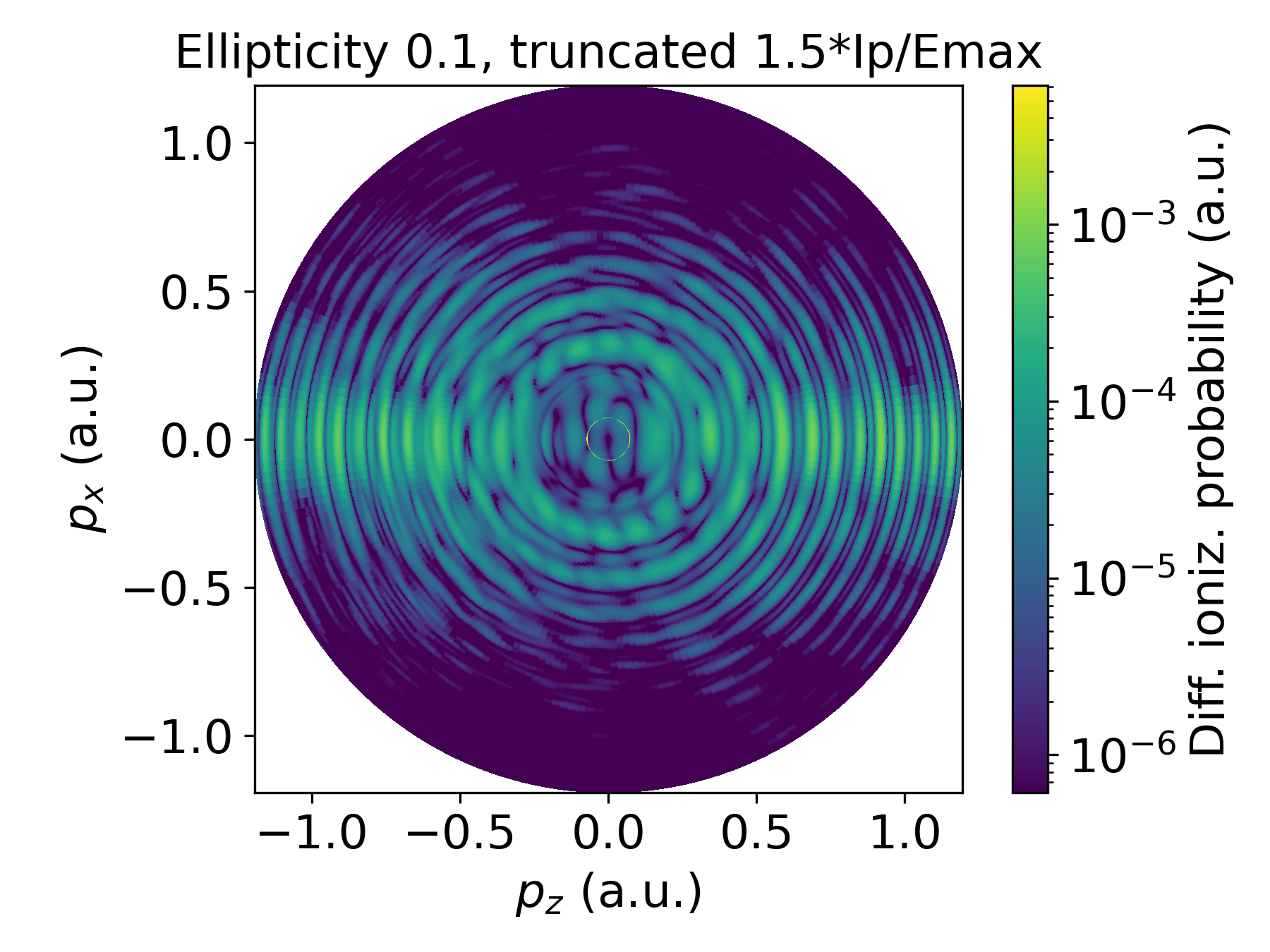} 
   \caption{Photoelectron momentum distributions calculated for  helium in a field of ellipciticity $\epsilon=0.1$ and the same parameters as in Fig.~\ref{fig:QpropvsCQSFAlow} using the Schr\"odinger solver Qprop, but considering a truncated potential according to Eq.~\eqref{eq:truncation}. We have used an envelope with $(1-\cos^{16})$ shape for a pulse with four cycles total duration, creating a near-flattop pulse, and the final PMDs have been calculated with the isurfv option \cite{Tulsky2020}. Each panel has been normalized to its maximum value and plotted in a logarithmic scale.}
   \label{fig:QpropTruncated}
   \end{figure*}
If one considers a binding potential such that its effective barrier remains the same, but its tail is truncated by multiplying a smooth function to the Coulomb potential \cite{DeMorissonFaria2002}, the rotational shifts in the holographic patterns are reduced. This is shown in Fig.~\ref{fig:QpropTruncated}, for which the potential started to be altered at 
\begin{equation}
\begin{split}
r_0 &= 2\cdot \frac{I_p}{E_{\mathrm{max}}} \qquad \text{or} \qquad 1.5\cdot \frac{I_p}{E_{\mathrm{max}}}, \\
L &= r_0 + \frac{E_{\mathrm{max}}}{2 \omega^2},
\end{split}
\label{eq:TruncationLimits}
\end{equation}
corresponding to $r_0 = 2\times$ or $r_0 = 1.5\times$ the approximate tunnel exit, respectively.
The width over which the Coulomb potential is smoothed out always corresponds to half an excursion amplitude of the electron, using  $E_{\mathrm{max}}= \frac{2 \omega \sqrt{U_p}}{\sqrt{1+\epsilon^2}}$.

Interestingly, the fan-shaped pattern near the ionization threshold is also considerably altered, which is expected due to it being caused by the Coulomb tail \cite{Lai2017,maxwell2017coulomb}. There is still some residual twisting, possibly associated with under-the-barrier dynamics.

\begin{figure*}[h!tb]
	\centering
	\includegraphics[width=0.3\textwidth]{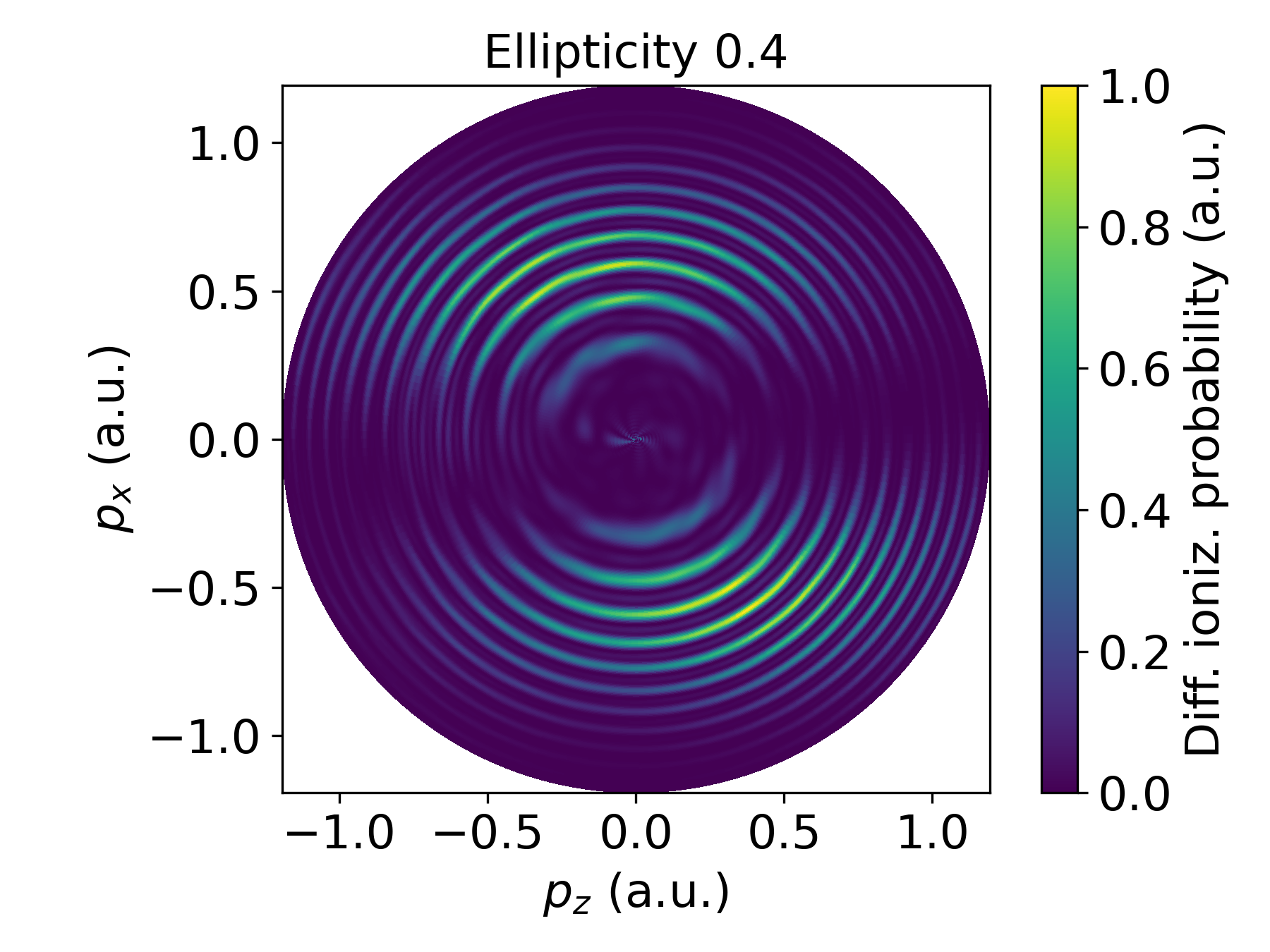}
	\includegraphics[width=0.3\textwidth]{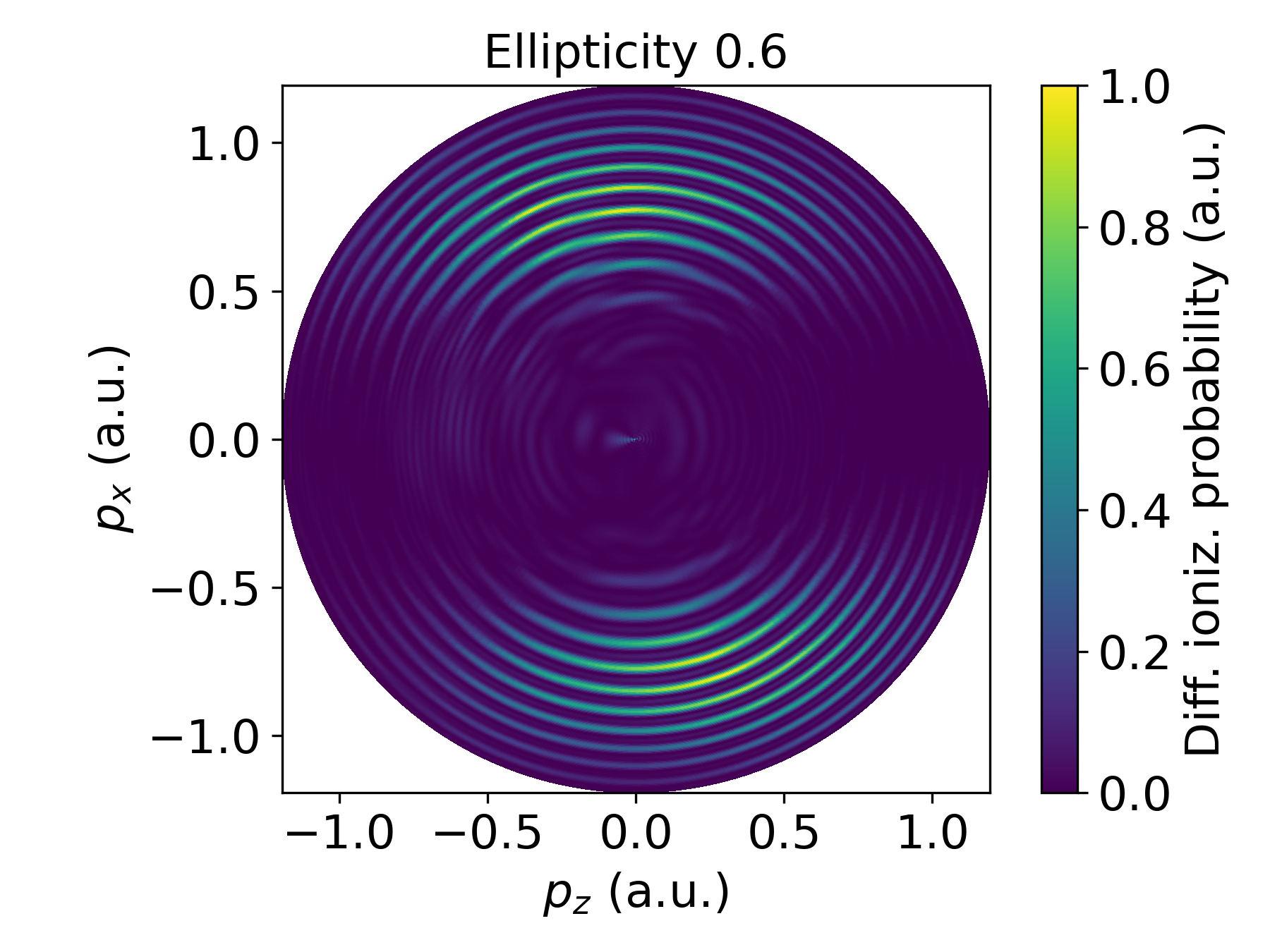}
	\includegraphics[width=0.3\textwidth]{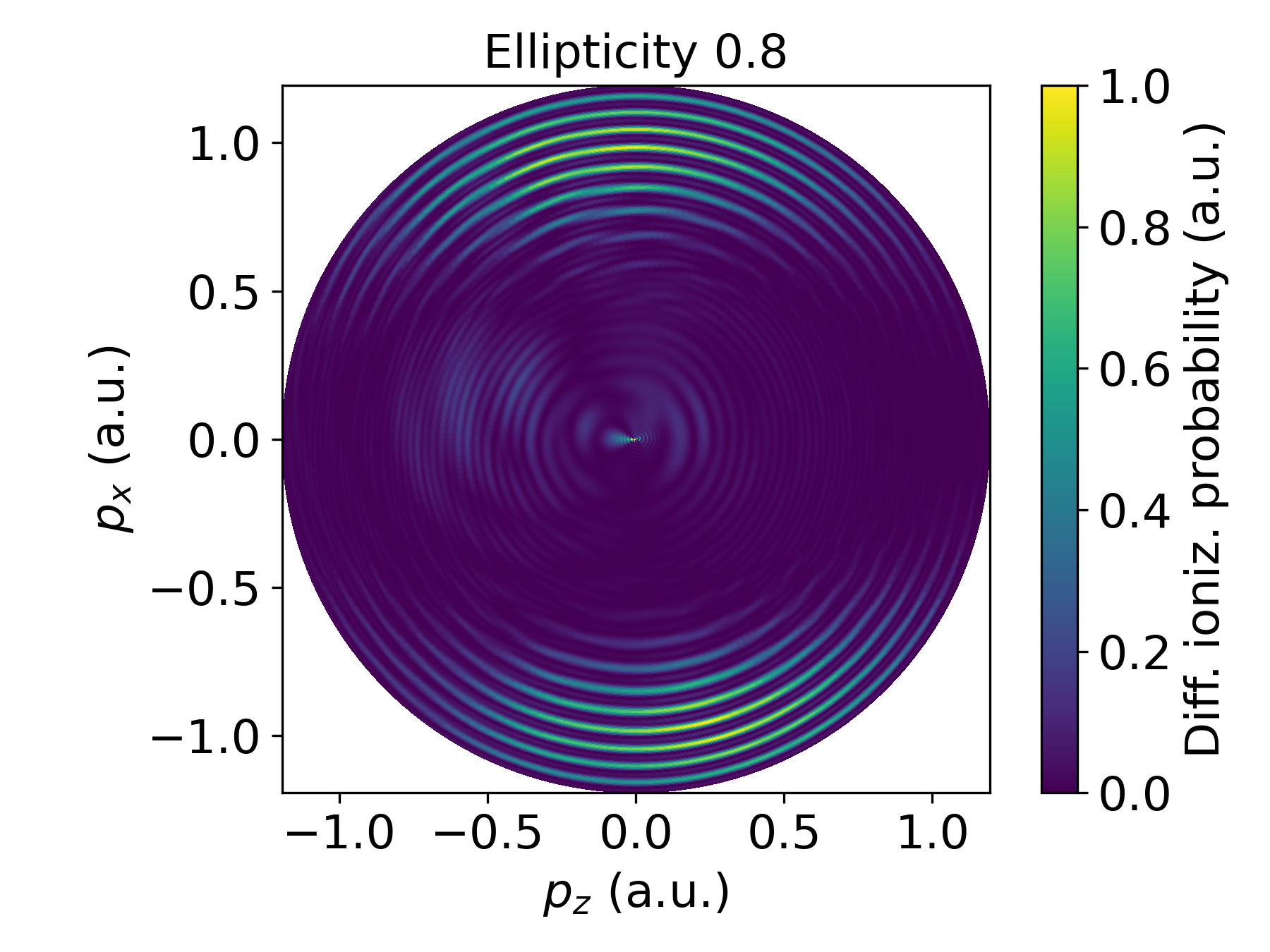} 
	
	\includegraphics[width=0.3\textwidth]{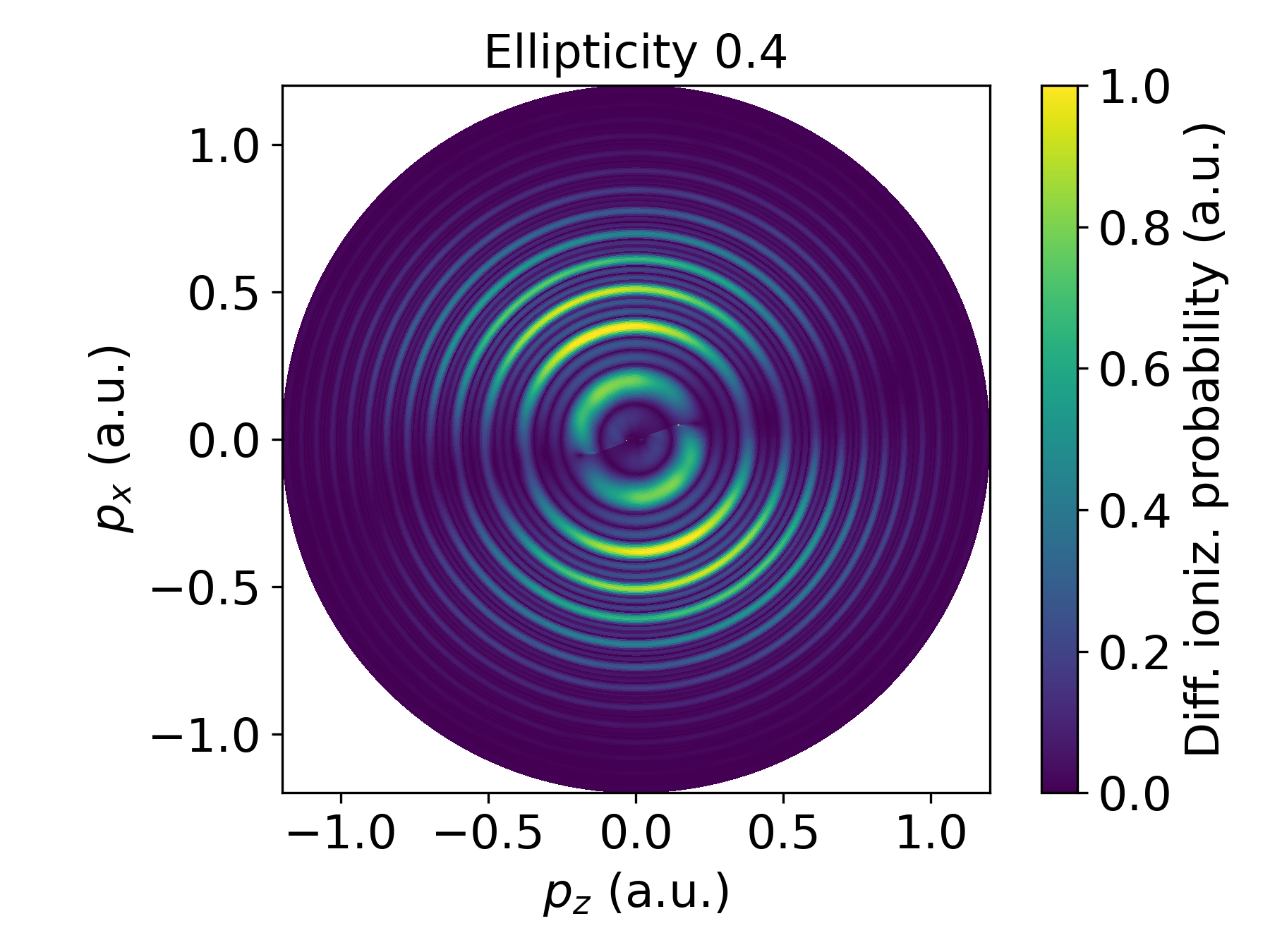}
	\includegraphics[width=0.3\textwidth]{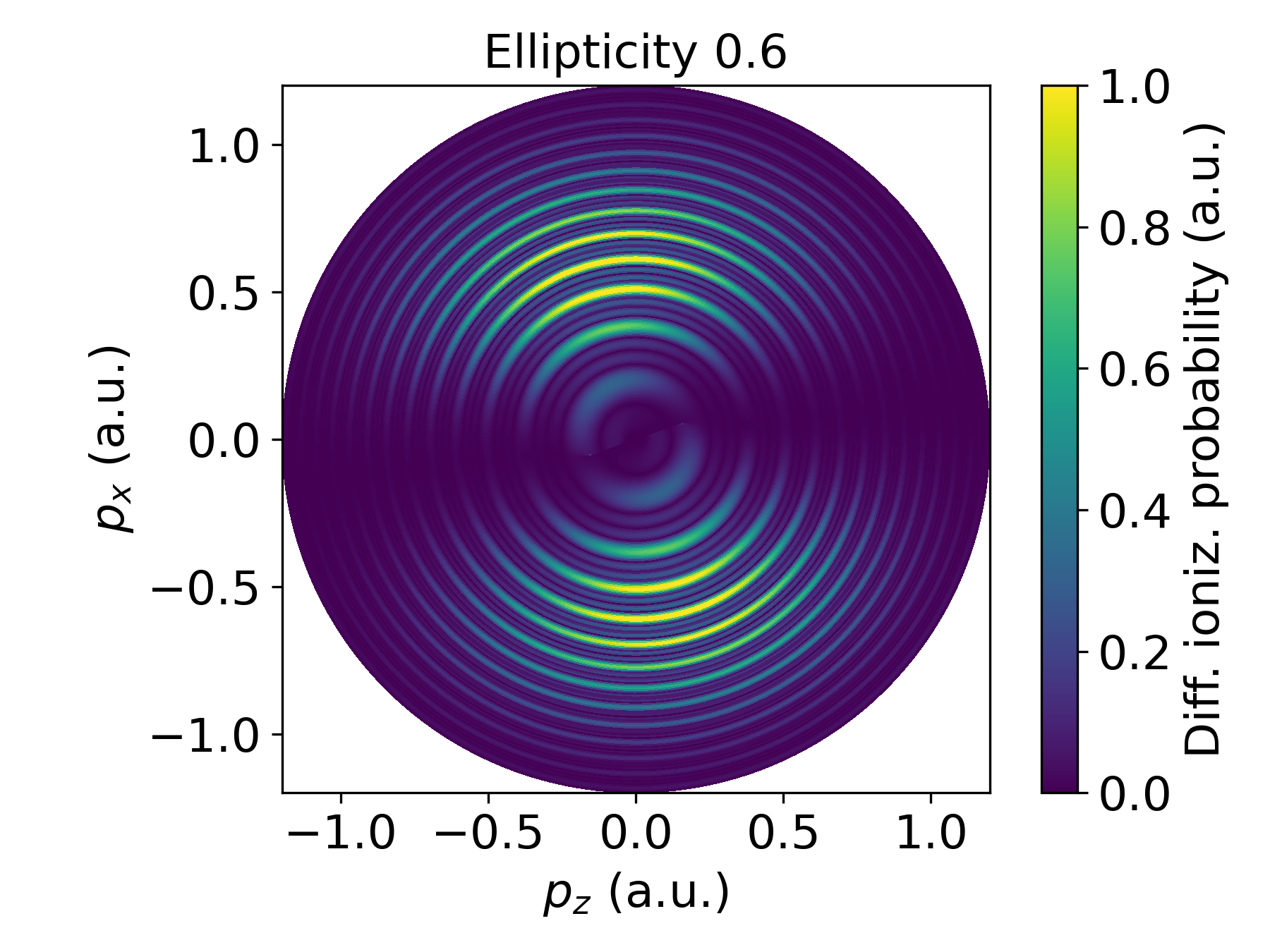}
	\includegraphics[width=0.3\textwidth]{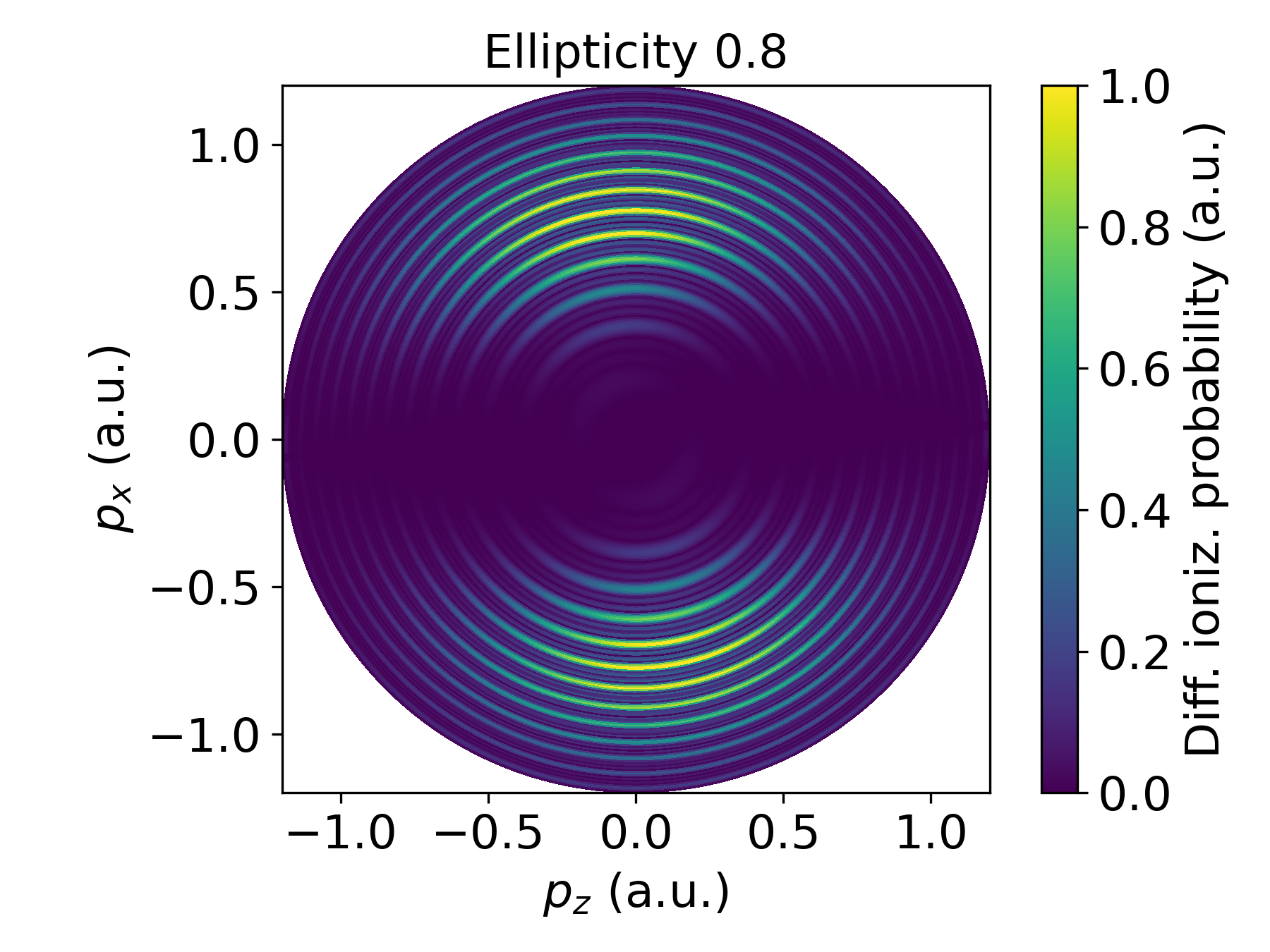}
	
	\includegraphics[width=0.3\textwidth]{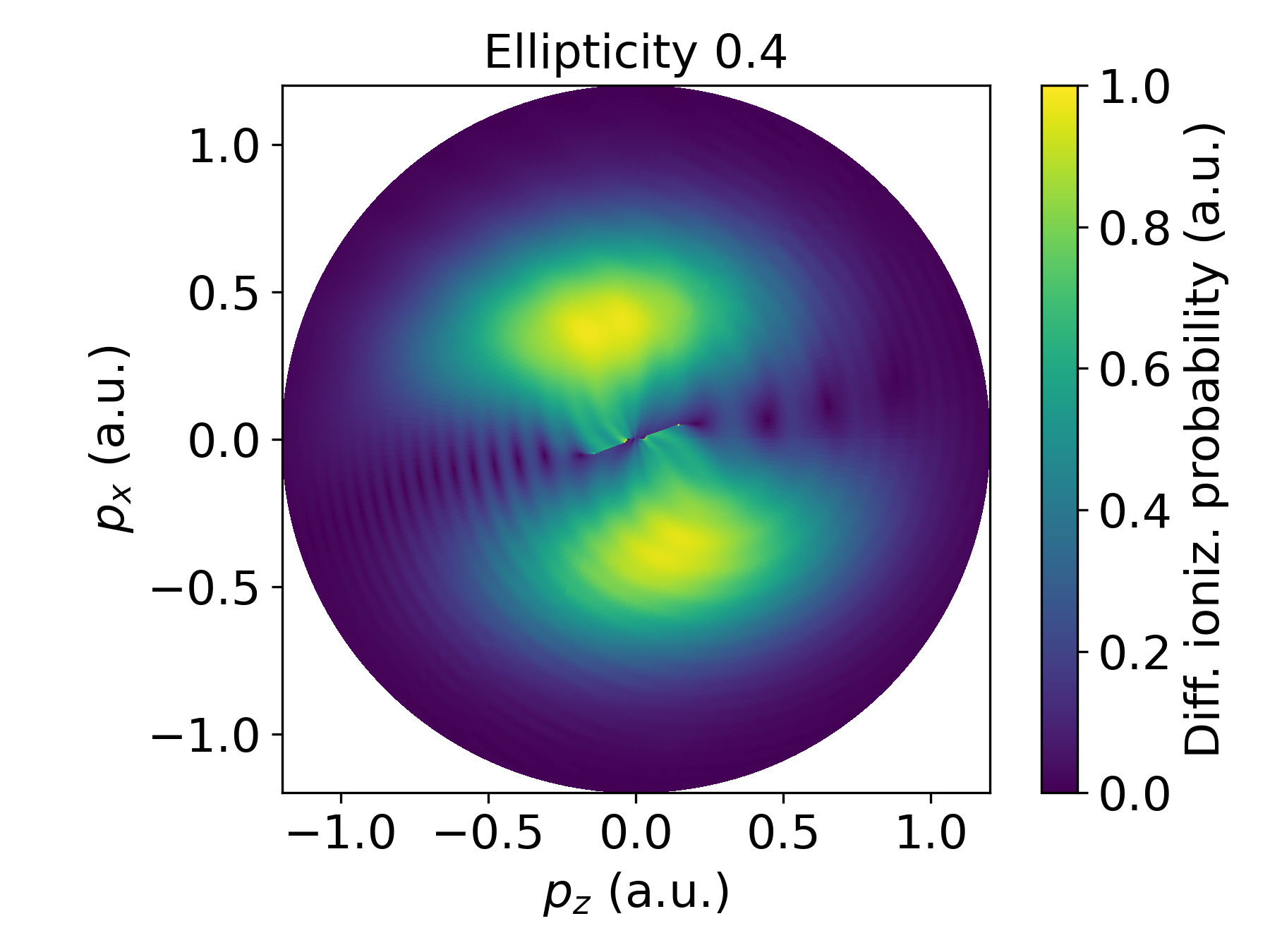}
	\includegraphics[width=0.3\textwidth]{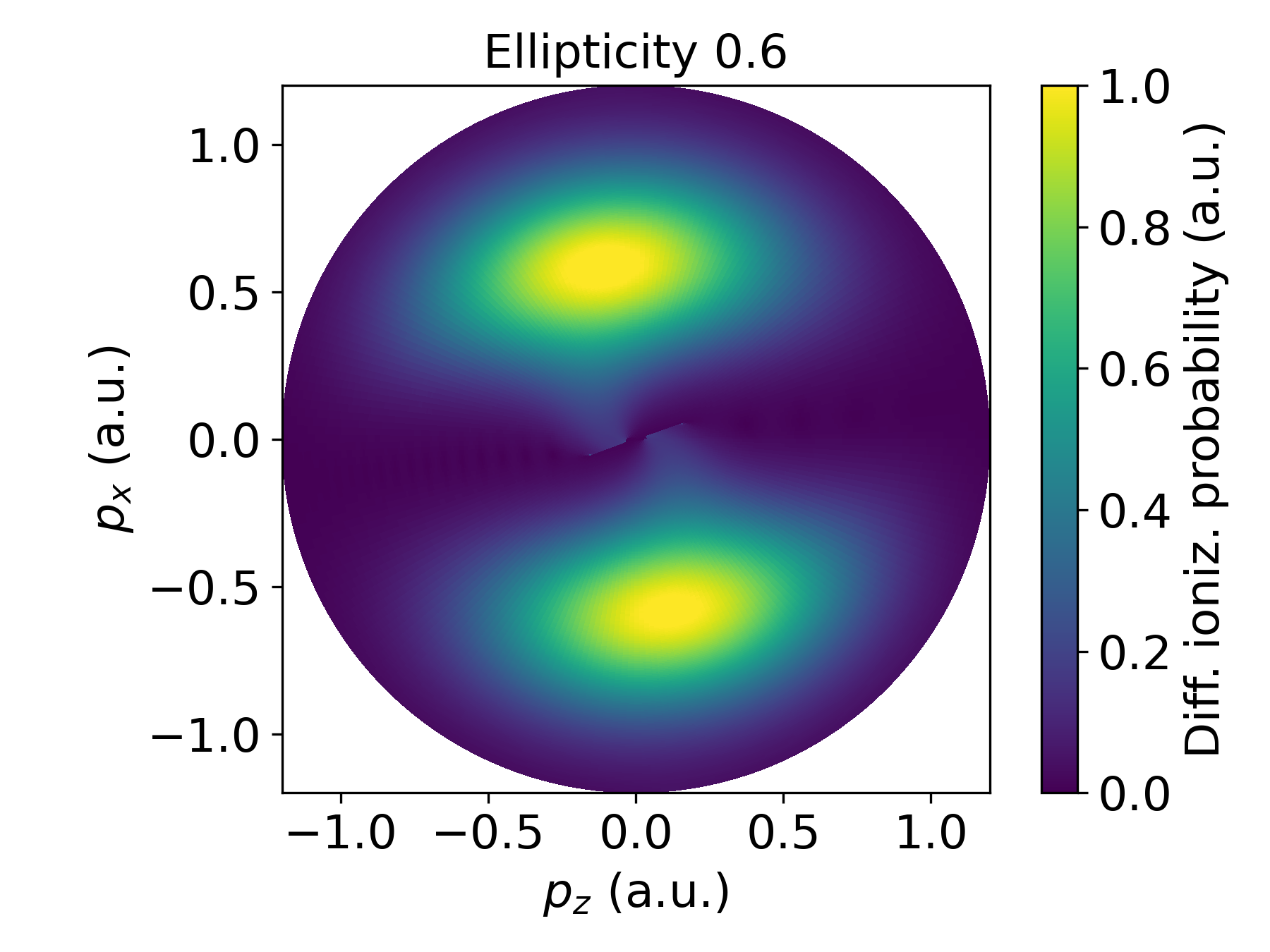}
	\includegraphics[width=0.3\textwidth]{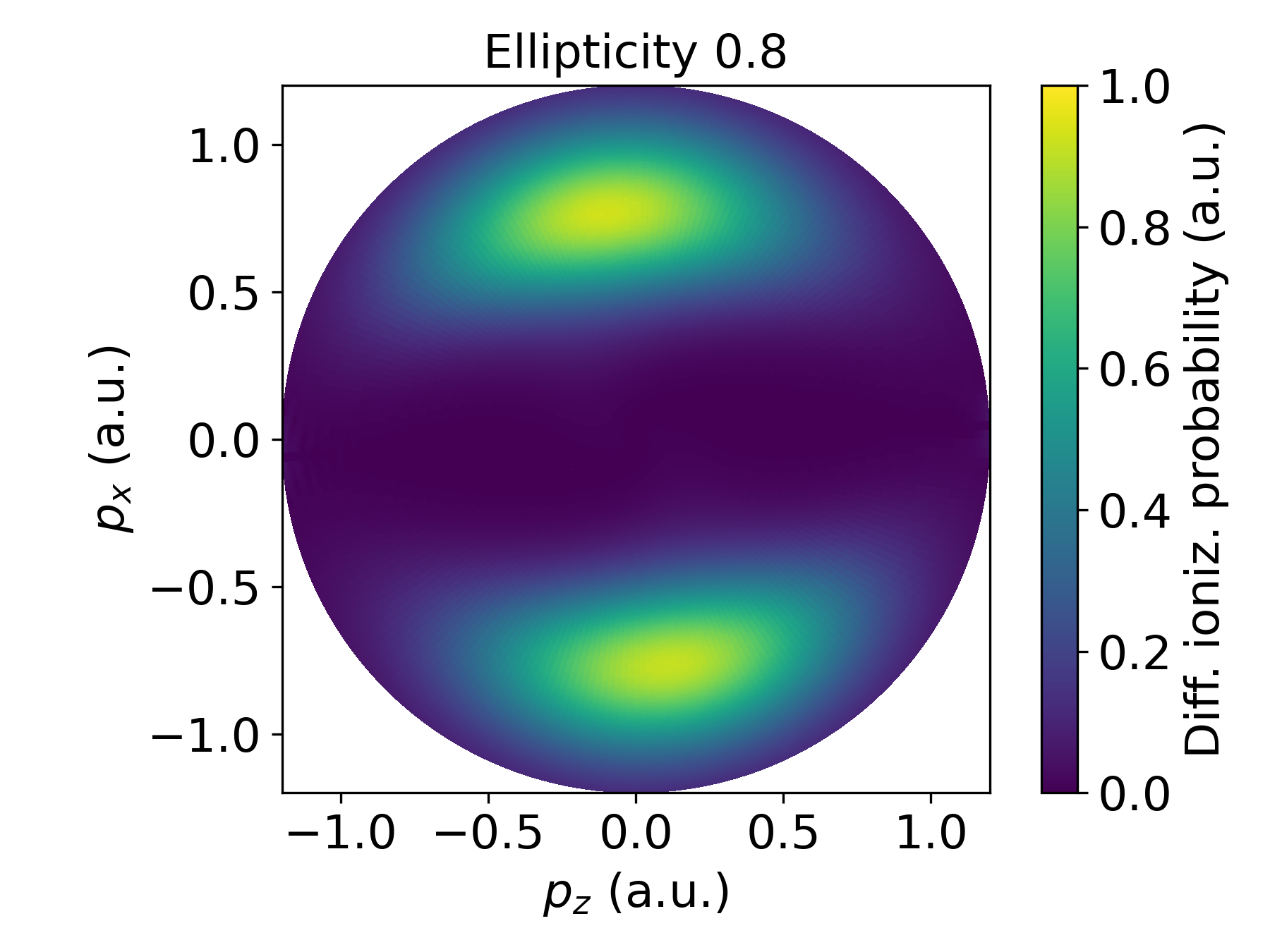}
	\caption{Photoelectron momentum distributions computed for ellipticity values higher than that specified in Sec.~\ref{sec:SaddlePointSolutions}, shown in linear scale. For the Qprop calculation in the top row and the CQSFA results in the middle row, all other field and atomic parameters are the same as in Fig.~\ref{fig:QpropvsCQSFAlow}. For the CQSFA in all cases, we have only employed orbits $a$ and $b$ as the remaining orbits are strongly suppressed in this ellipticity range (see appendices and discussion of $\mathrm{Im}[t']$). The bottom row shows the single-cycle CQSFA result, thus no inter-cycle ATI rings are visible anymore.}
	\label{fig:QpropvsCQSFAhigh}
\end{figure*}

The behavior described above is markedly different from that observed for high ellipticities, which we illustrate in Fig.~\ref{fig:QpropvsCQSFAhigh}. Both for the CQSFA and TDSE computations, the figure shows sickle-shaped distributions with an angular offset. This is the shape of the photoelectron distribution  typically used in angular streaking `attoclock' measurements \cite{Landsman2014b,Camus2017,Sainadh2019,Khan2020}. The holographic patterns are practically washed out and the only visible interference patterns are the ATI rings, resulting from inter-cycle rather than intra-cycle interference. The rings are quite prominent in the upper and middle row of the figure, which were computed for four-cycle-pulses. In order to highlight the shapes of the distributions and the angular offsets, in the lower row, we plot CQSFA results for a single cycle. The single-cycle distributions confirm that the intra-cycle patterns are either very faint (see plot for ellipticity 0.4, on the left hand side) or absent (see remaining plots). 

\begin{figure*}[h!p]
	\centering
	\includegraphics[width=0.4\textwidth] {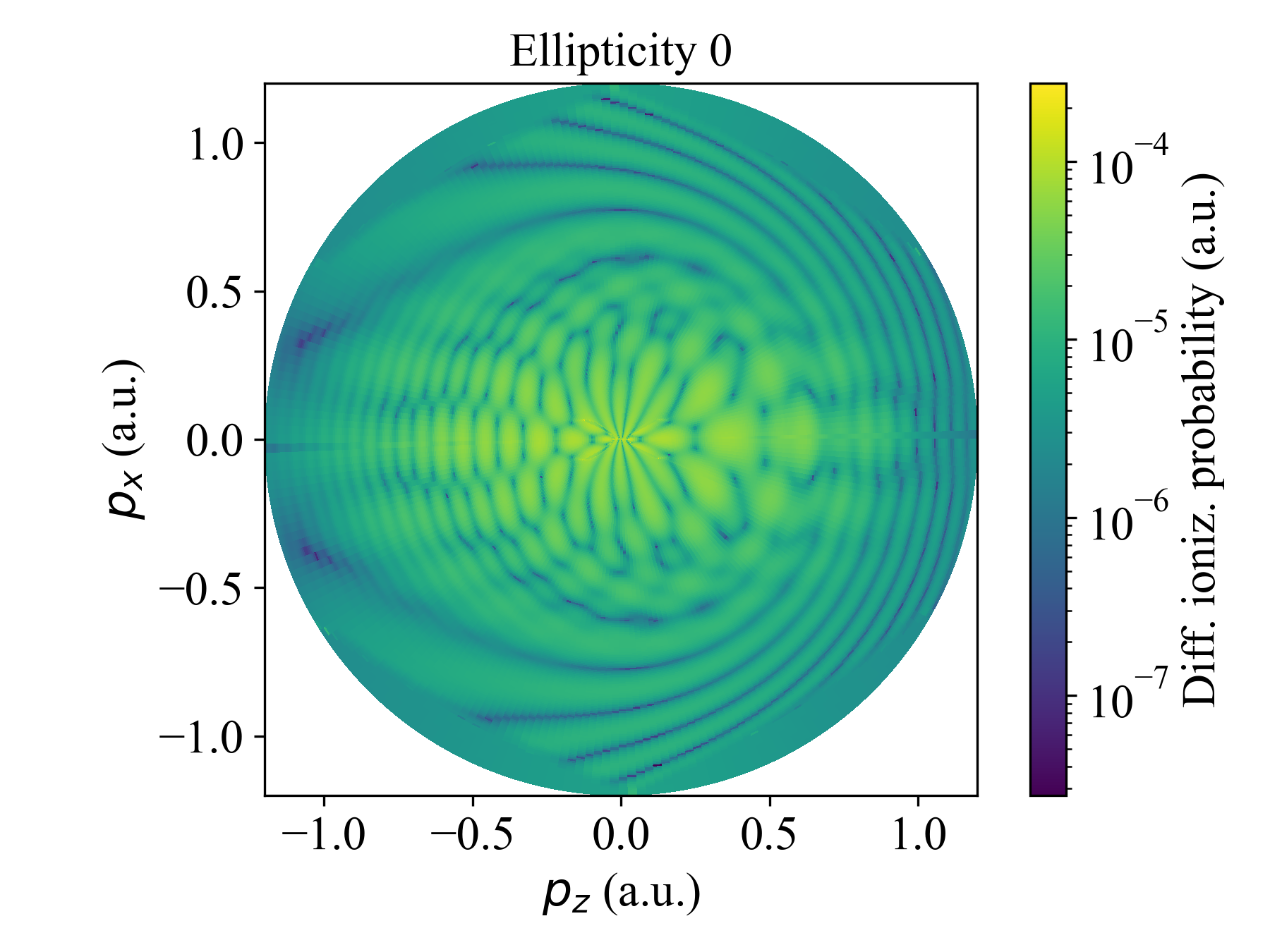}
	\includegraphics[width=0.4\textwidth] {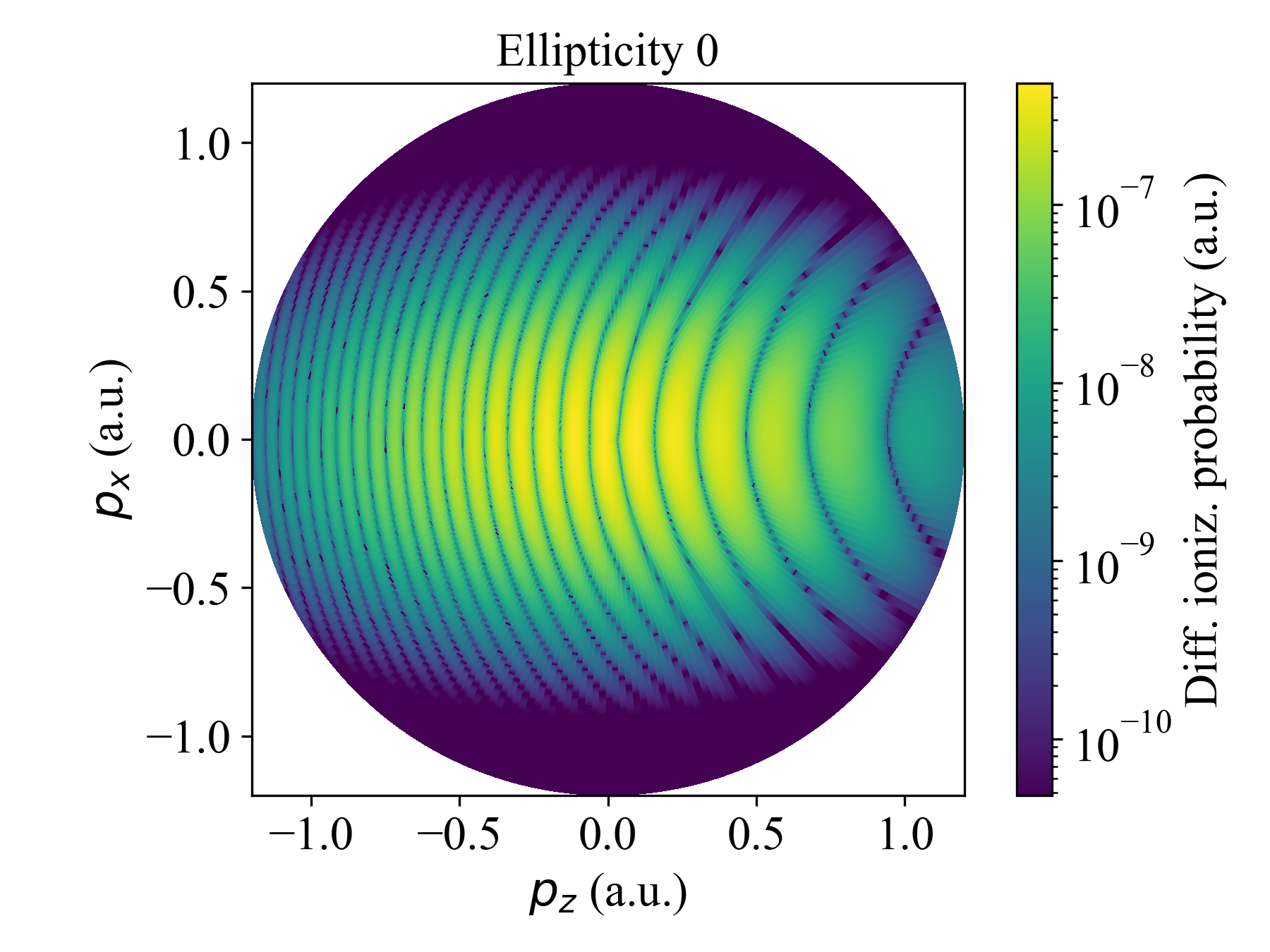} \\
	\includegraphics[width=0.4\textwidth] {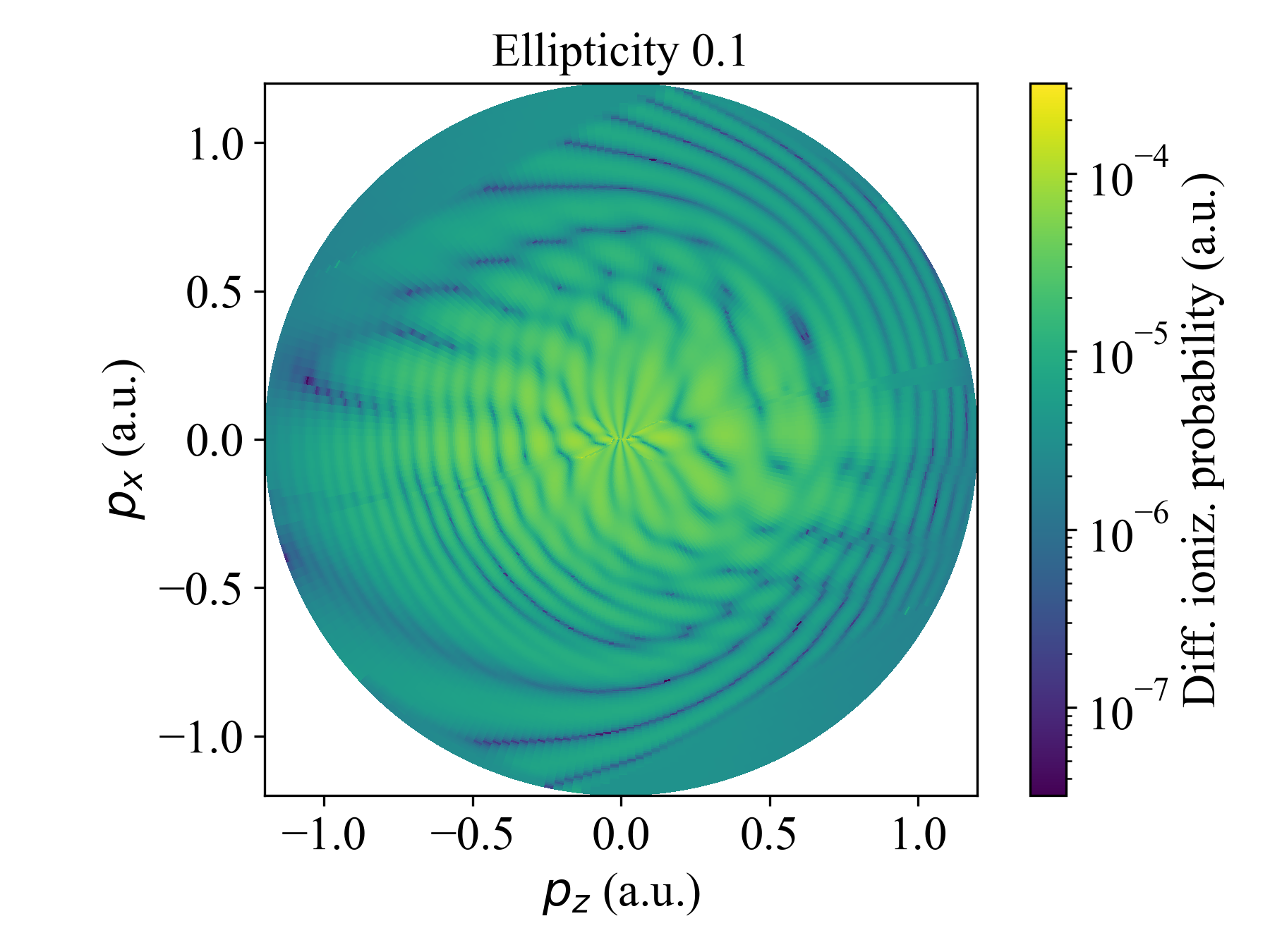}
	\includegraphics[width=0.4\textwidth] {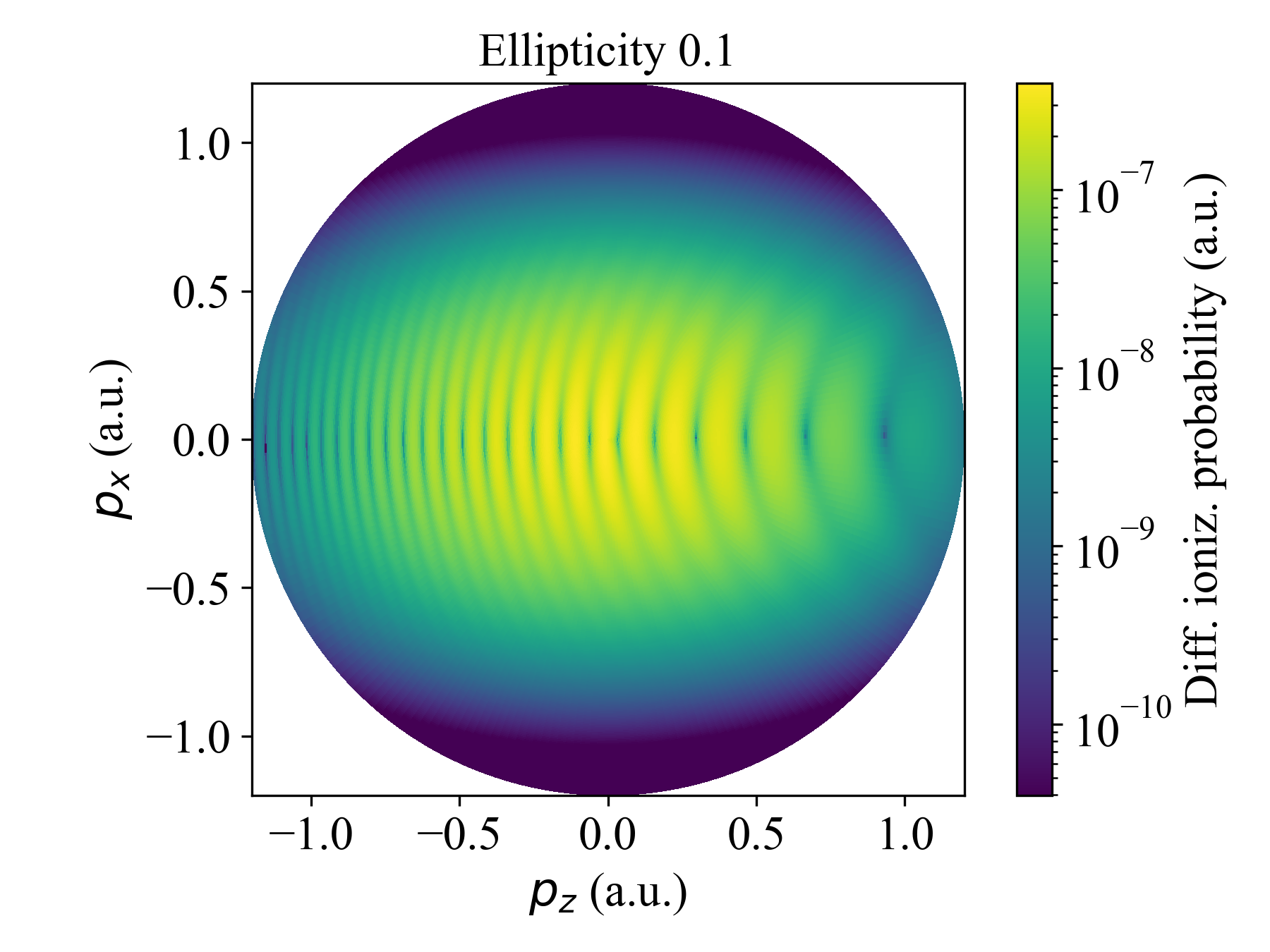} \\
	\includegraphics[width=0.4\textwidth] {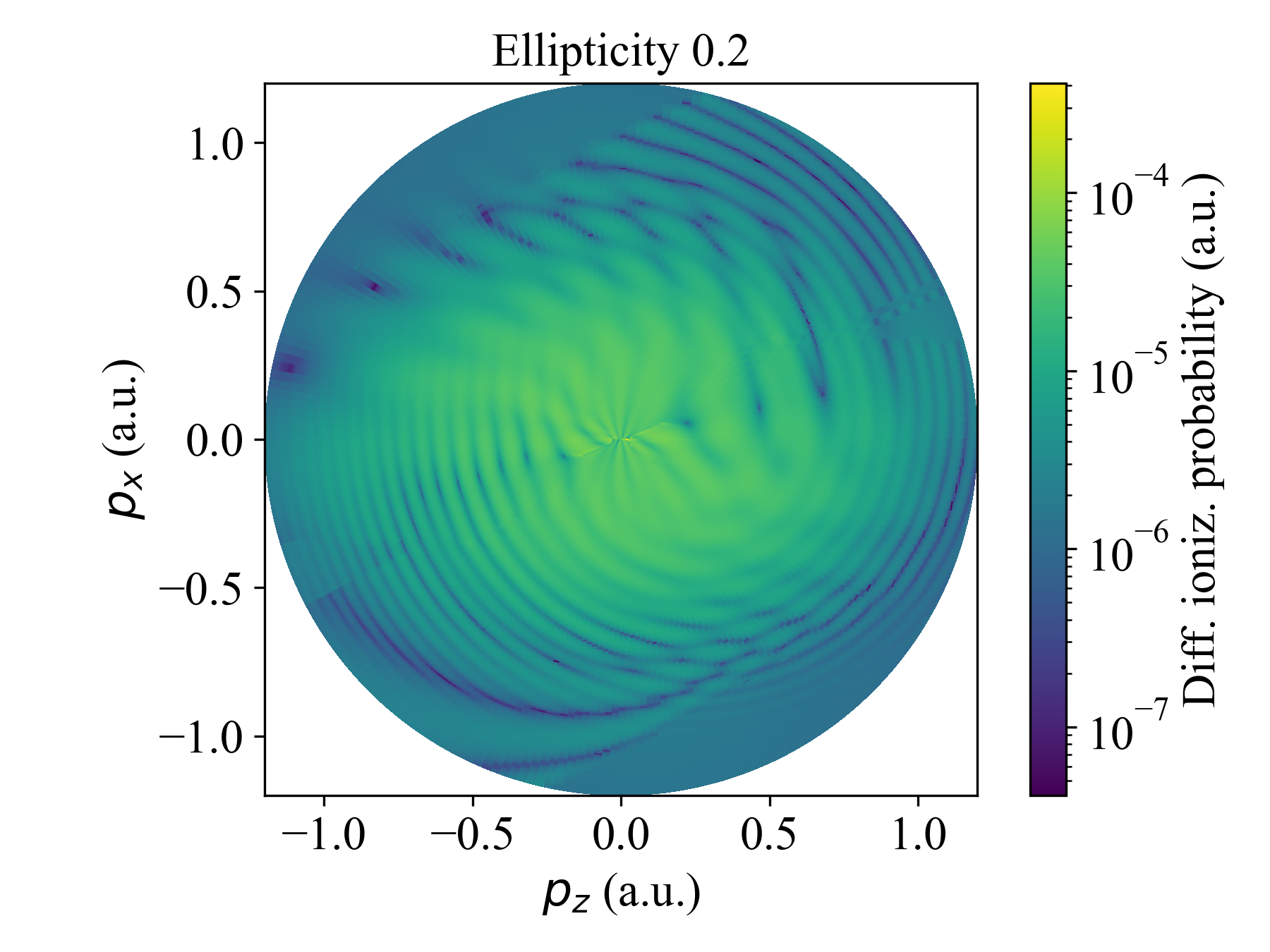}
	\includegraphics[width=0.4\textwidth] {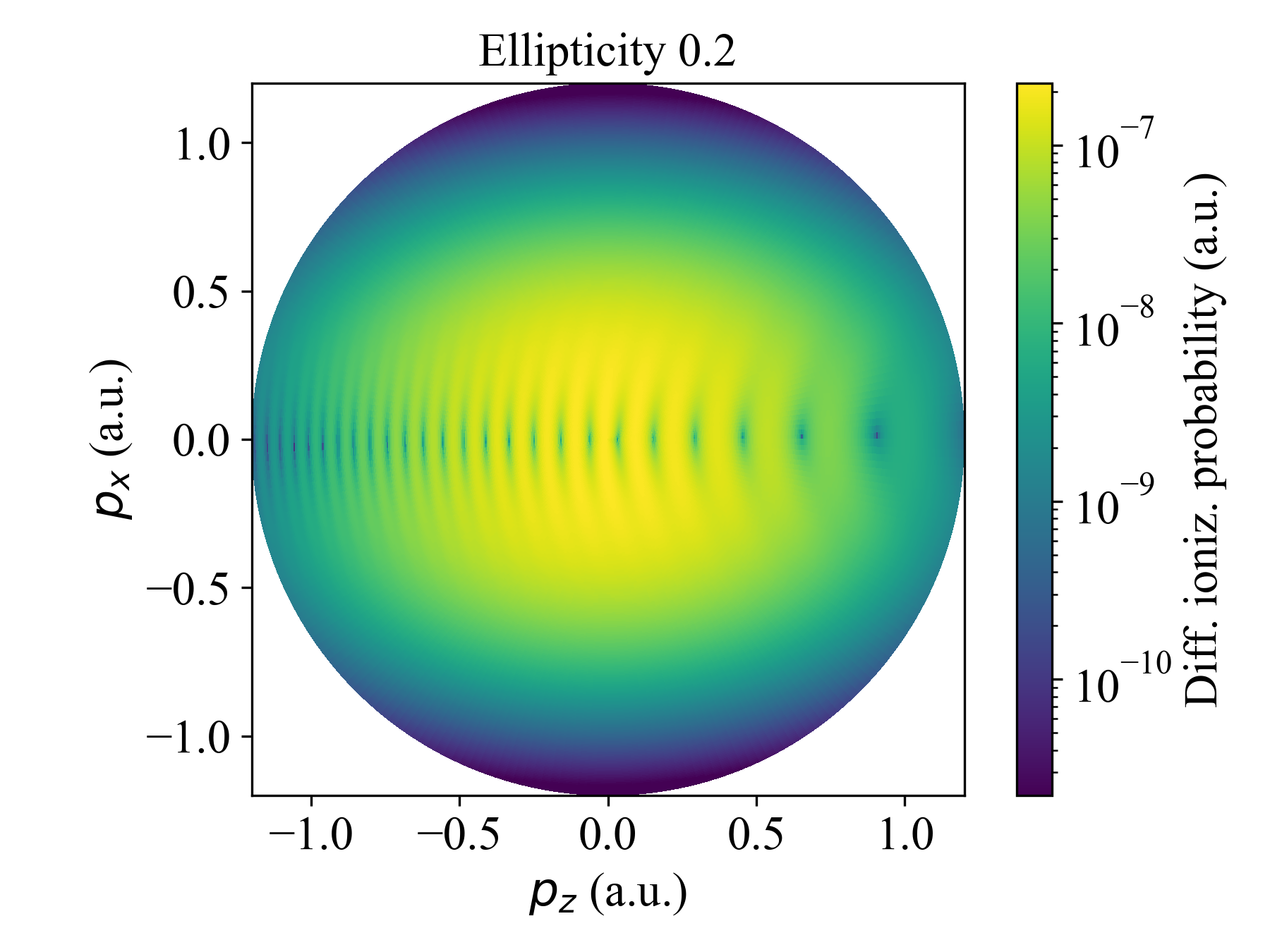} \\
	\includegraphics[width=0.4\textwidth] {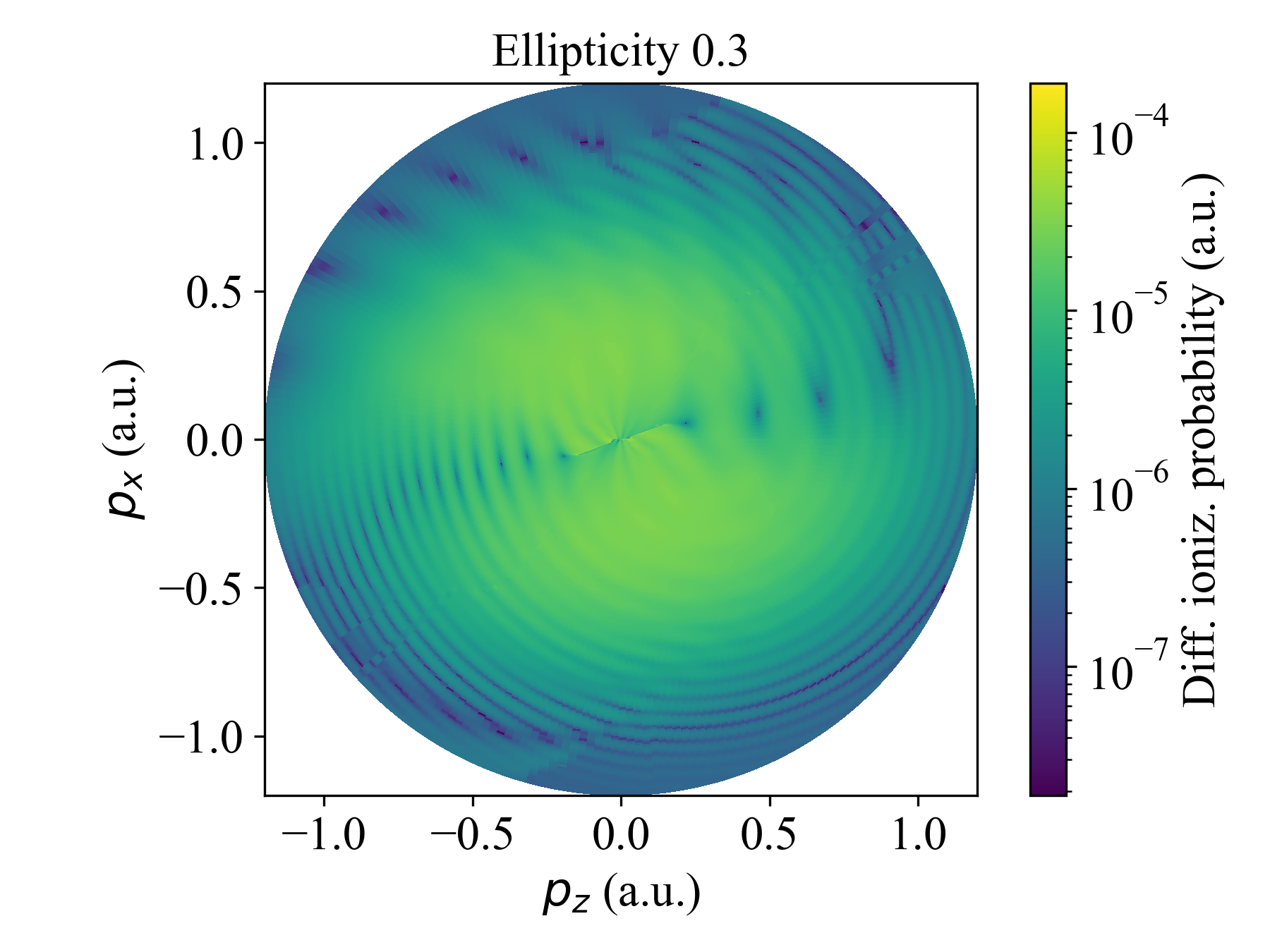}
	\includegraphics[width=0.4\textwidth] {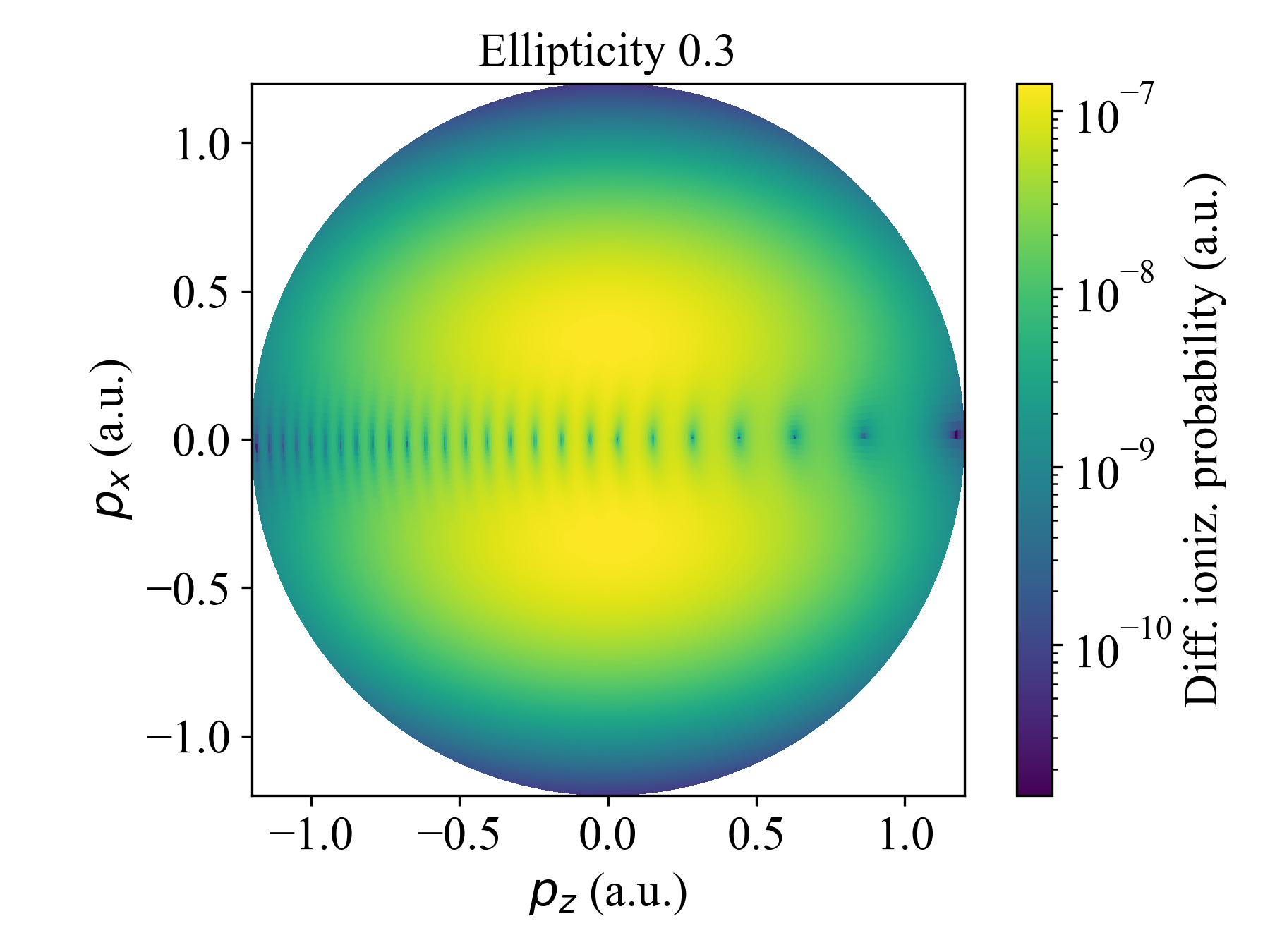}
	\caption{Photoelectron momentum distributions calculated for  helium in a field of intensity $2.5 \times 10 ^{14}$ W/cm$^2$, wavelength $\lambda =$ 735 nm, whose ellipticity increases from $\epsilon=0$ to $\epsilon=0.3$, considering a single cycle and a unit cell with $\phi=0$. The left and right panels display the outcome of the CQSFA and SFA, respectively. For the SFA, we have employed the direct orbits a and b, while for the CQSFA orbits a to d were included. All panels have been normalized to their maximum values and a logarithmic scale was used. }
	\label{fig:Allintracycle}
\end{figure*}

In Fig.~\ref{fig:Allintracycle}, we perform a direct comparison of the CQSFA with the interference from the direct SFA pathways, within a single cycle. This facilitates the study of holographic, intra-cycle interference due to the absence of ATI rings.  By making these choices, the role of the Coulomb potential becomes even clearer, and the CQSFA plots show superimposed sets of `twisted' fringes in the anti-clockwise direction for elliptical polarization. These fringes are more visible in the high-energy region, but are present in a broad range of momenta.  Another noteworthy feature is that the spider loses its dominance around ellipticity $\epsilon=0.2$ (left-hand side, second row from the bottom), and all patterns become increasingly blurred. For the largest ellipticity in the figure (left hand side, lowest row), the PMD exhibits a typical splitting, with offset phase shifts due to the presence of the Coulomb potential. 

This behavior is markedly different from that of the SFA PMDs, shown in the right column of the figure, which display near vertical fringes and no angular offset. For increasing ellipticity, the PMDs in the SFA split, but the fringes remain roughly the same.
The quantum interference fades around $\epsilon=0.3$, in agreement with our estimates in Sec.~\ref{sec:SaddlePointSolutions}.  Furthermore, for intermediate elliptiticies, the CQSFA maxima are closer to the major polarization axis than the SFA estimates, which is evidence of Coulomb focusing. 

\begin{figure*}[h!tb]
    \centering
   \includegraphics[width=0.32\textwidth] {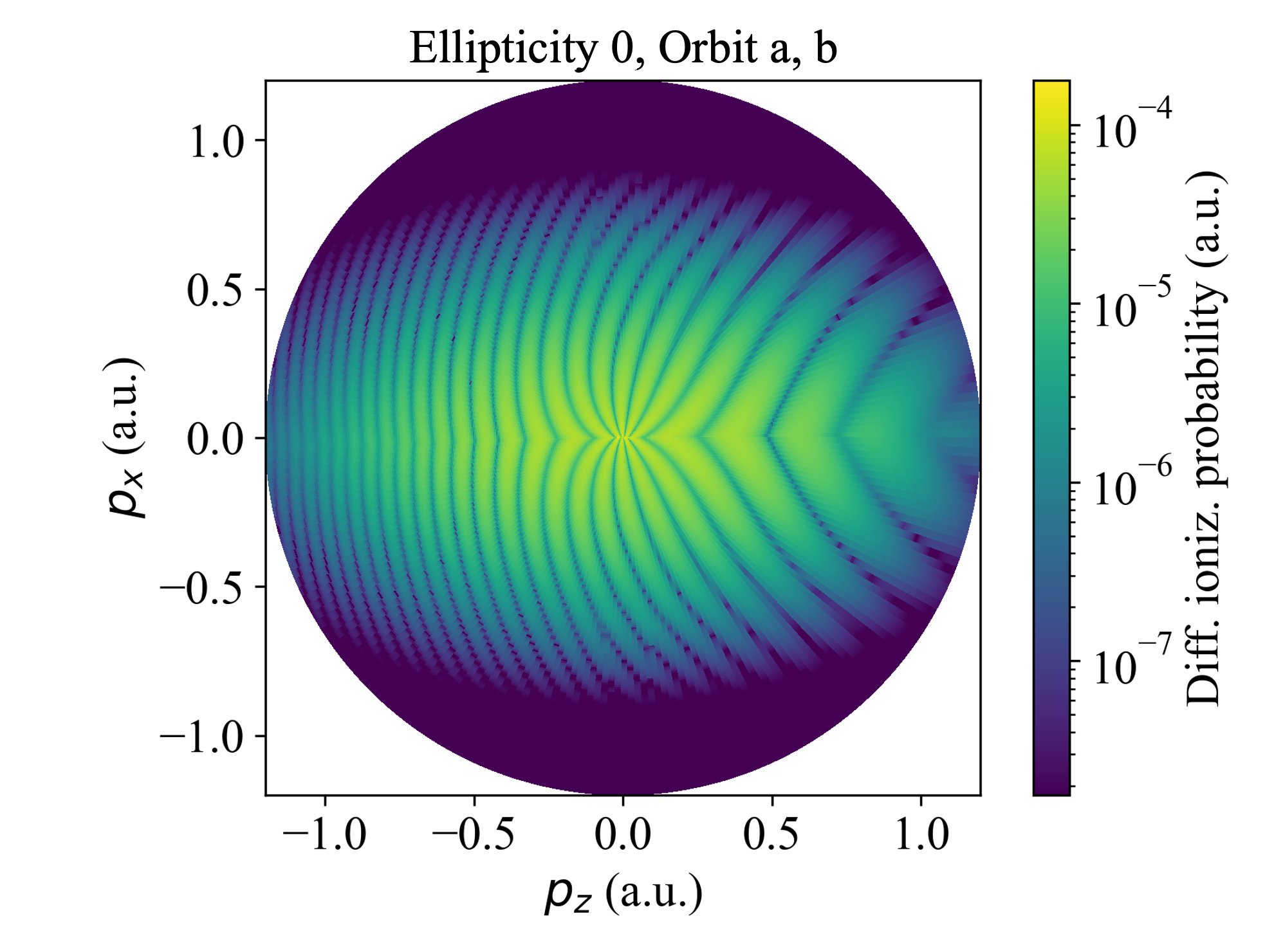}
   \includegraphics[width=0.32\textwidth] {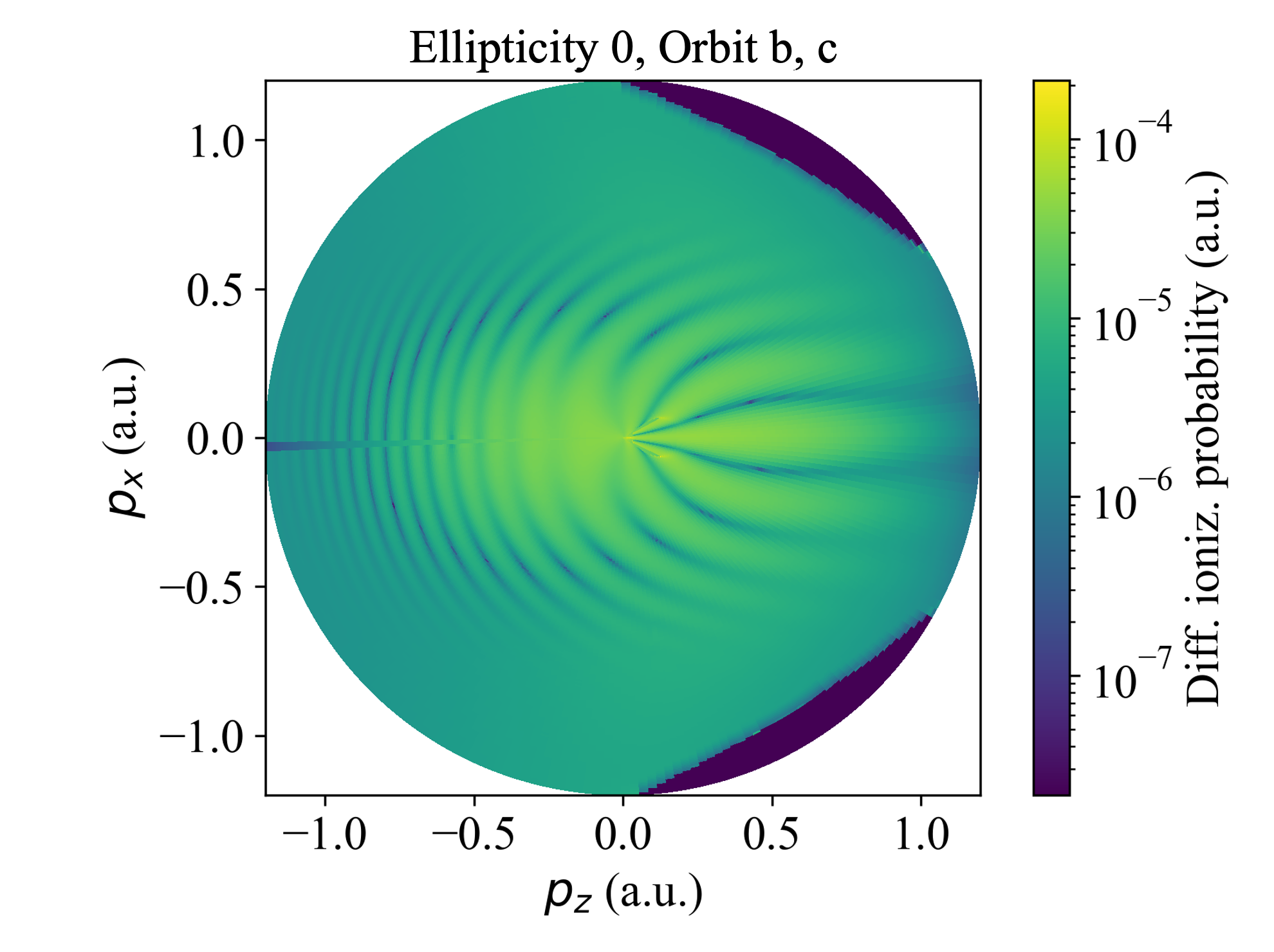} 
   \includegraphics[width=0.32\textwidth] {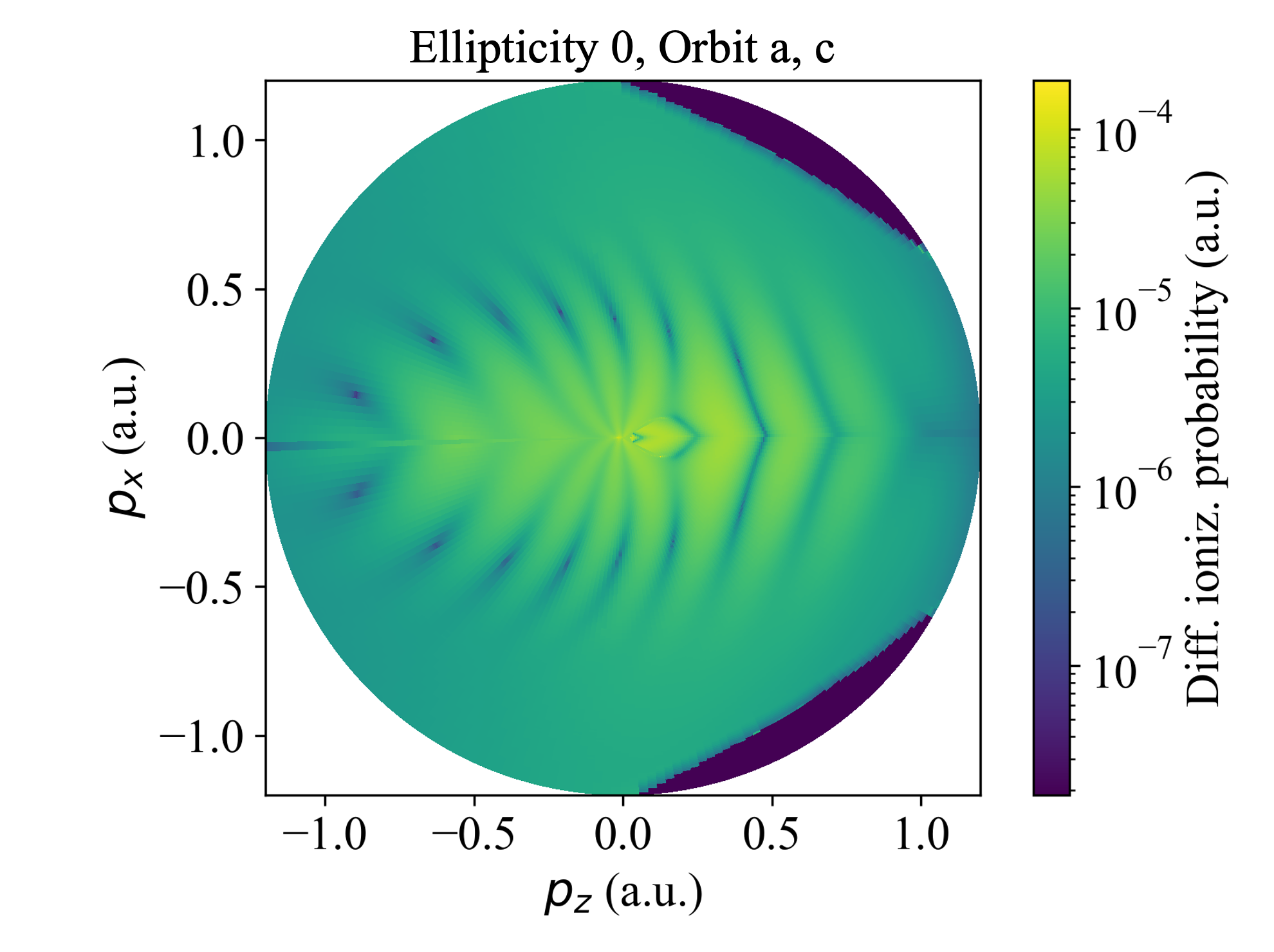}
  \\
   \includegraphics[width=0.32\textwidth] {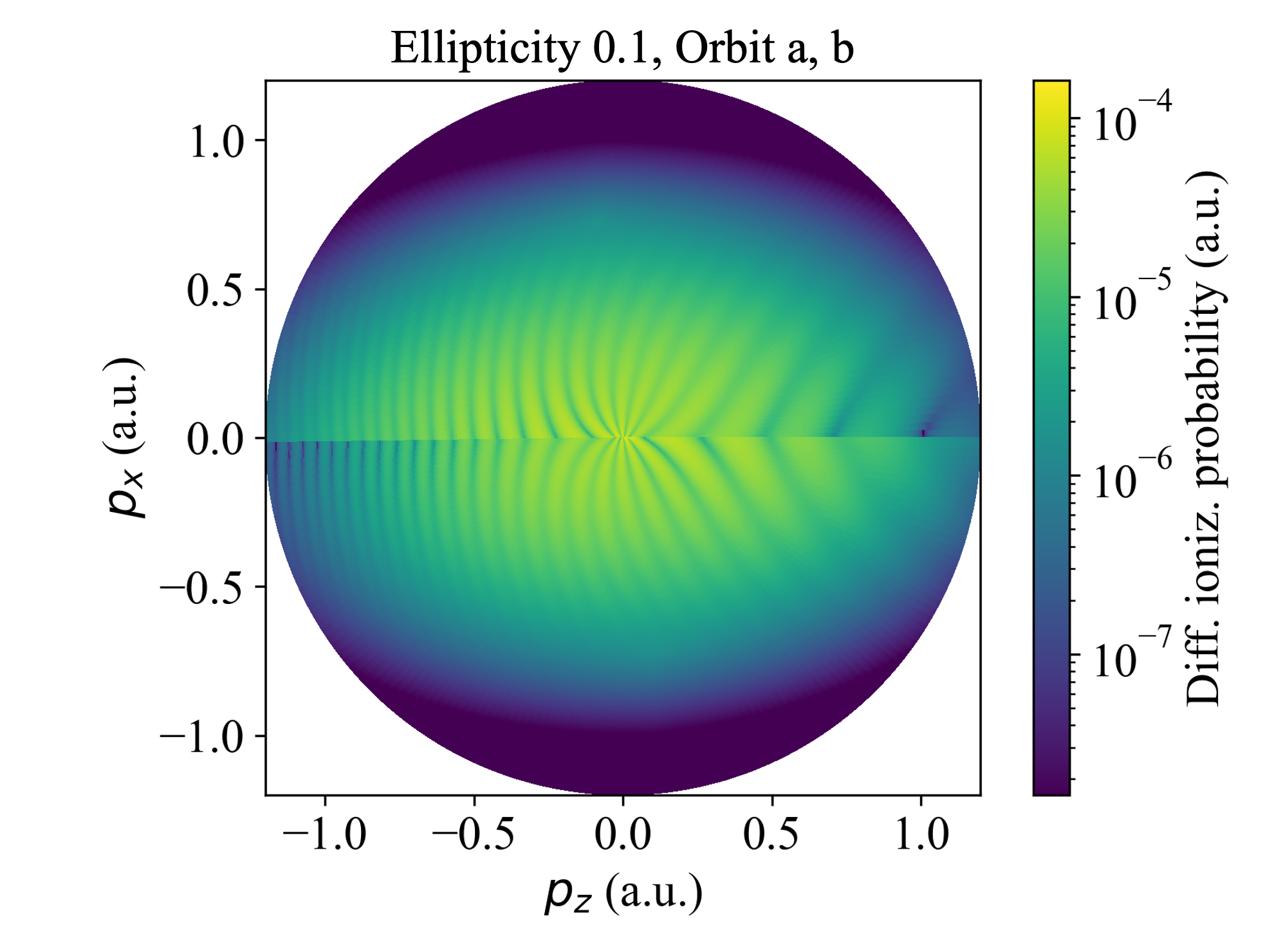}
   \includegraphics[width=0.32\textwidth] {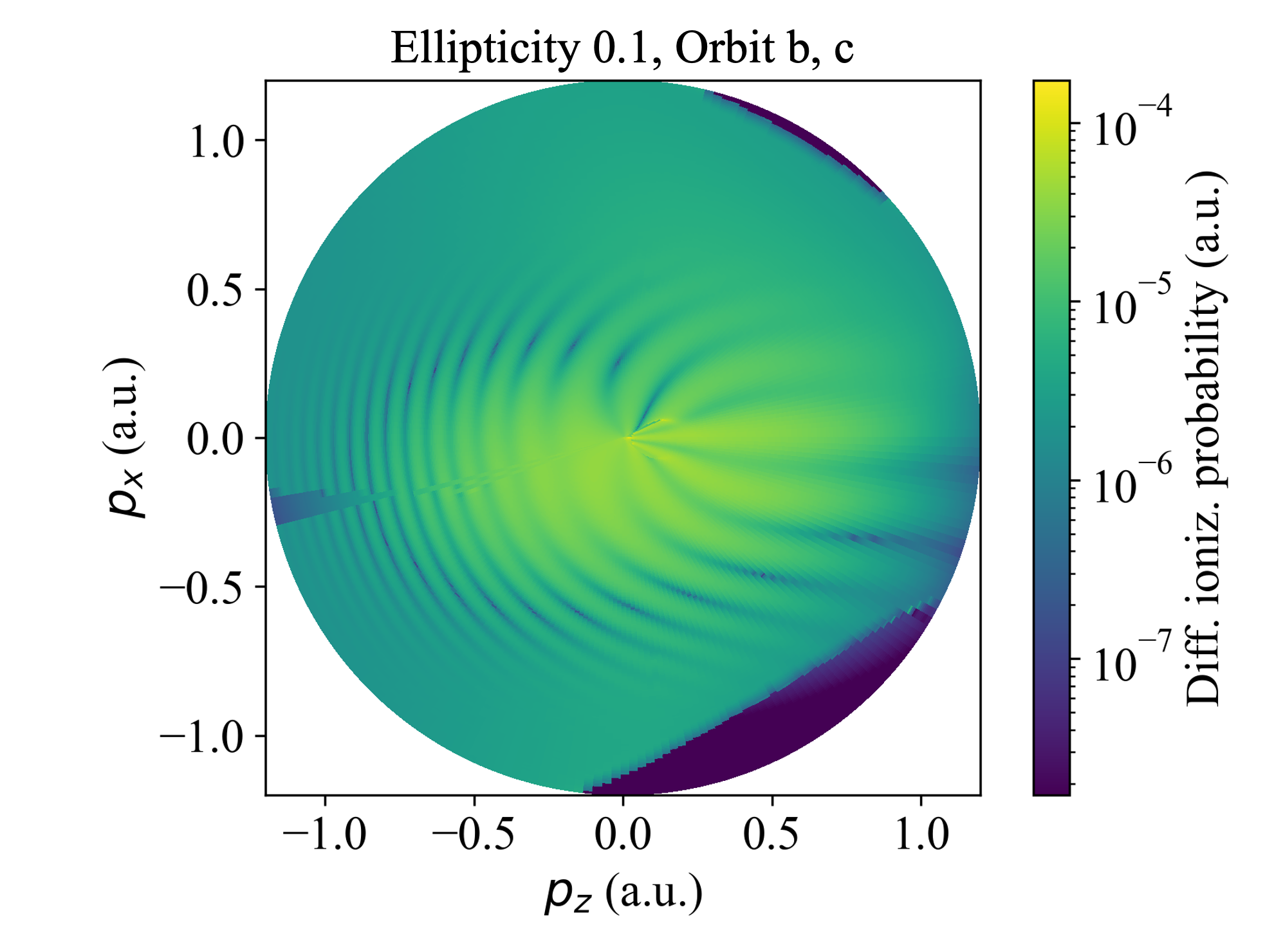} 
   \includegraphics[width=0.32\textwidth] {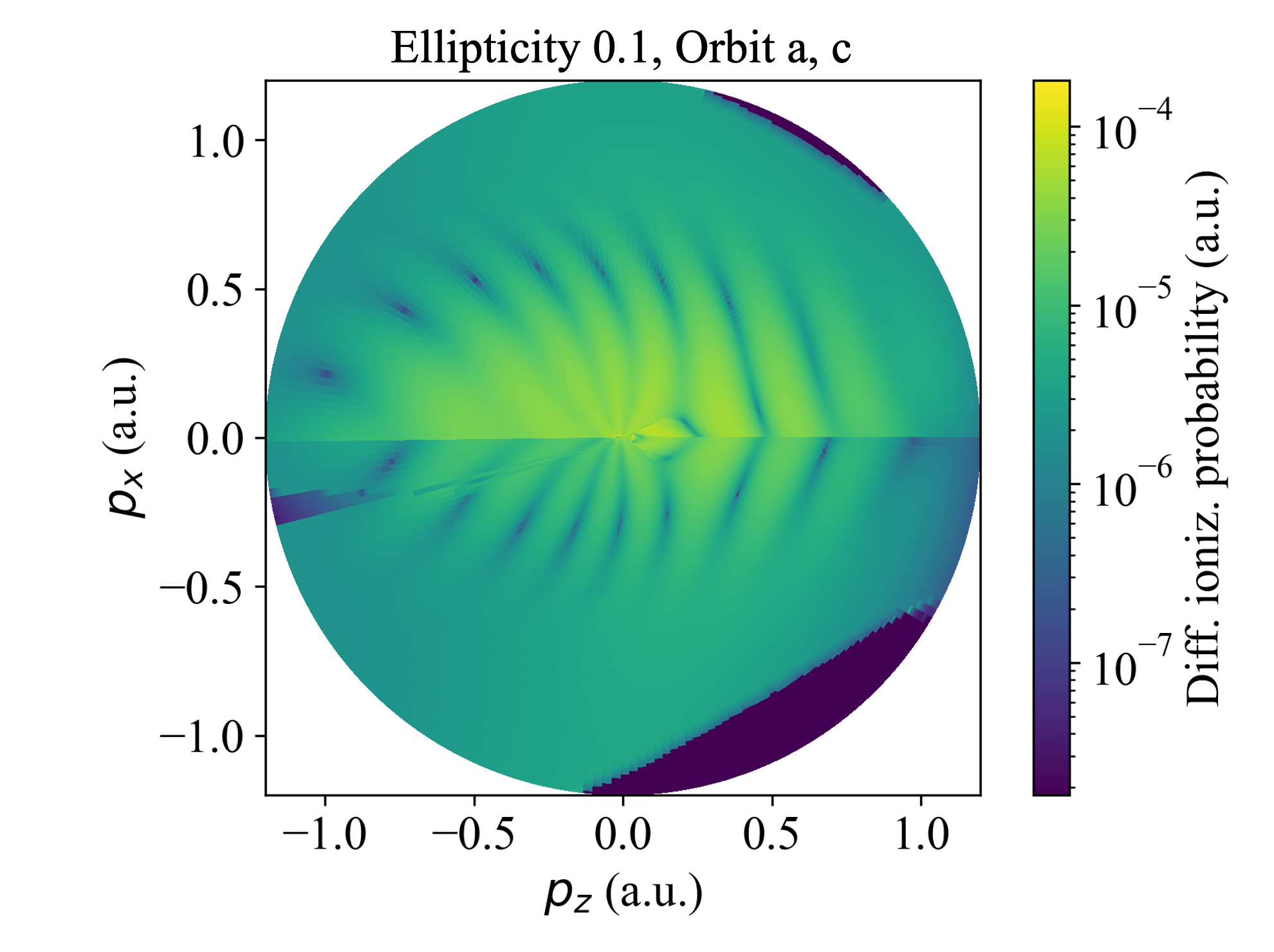}
  \\
   \includegraphics[width=0.32\textwidth] {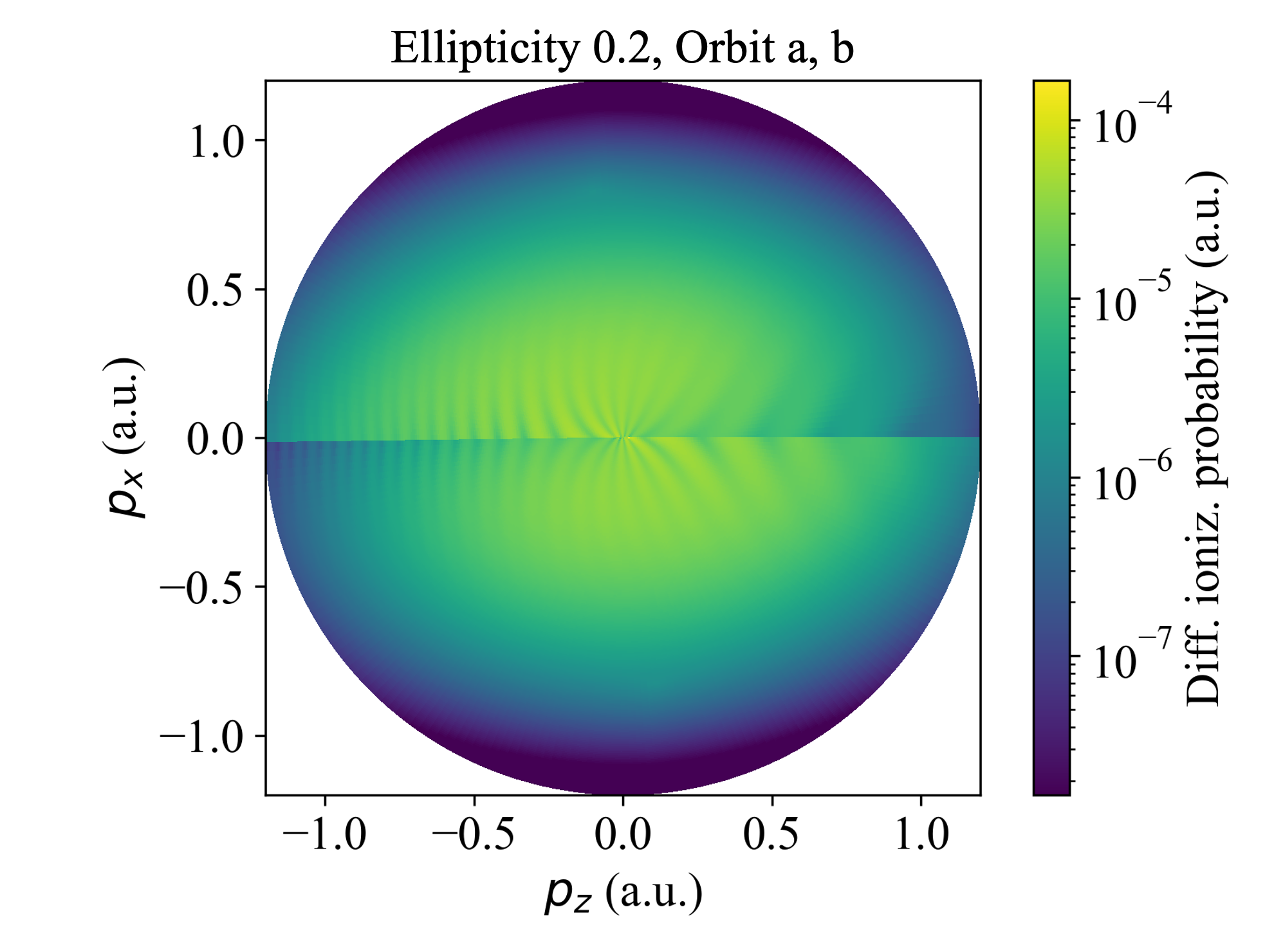}
   \includegraphics[width=0.32\textwidth] {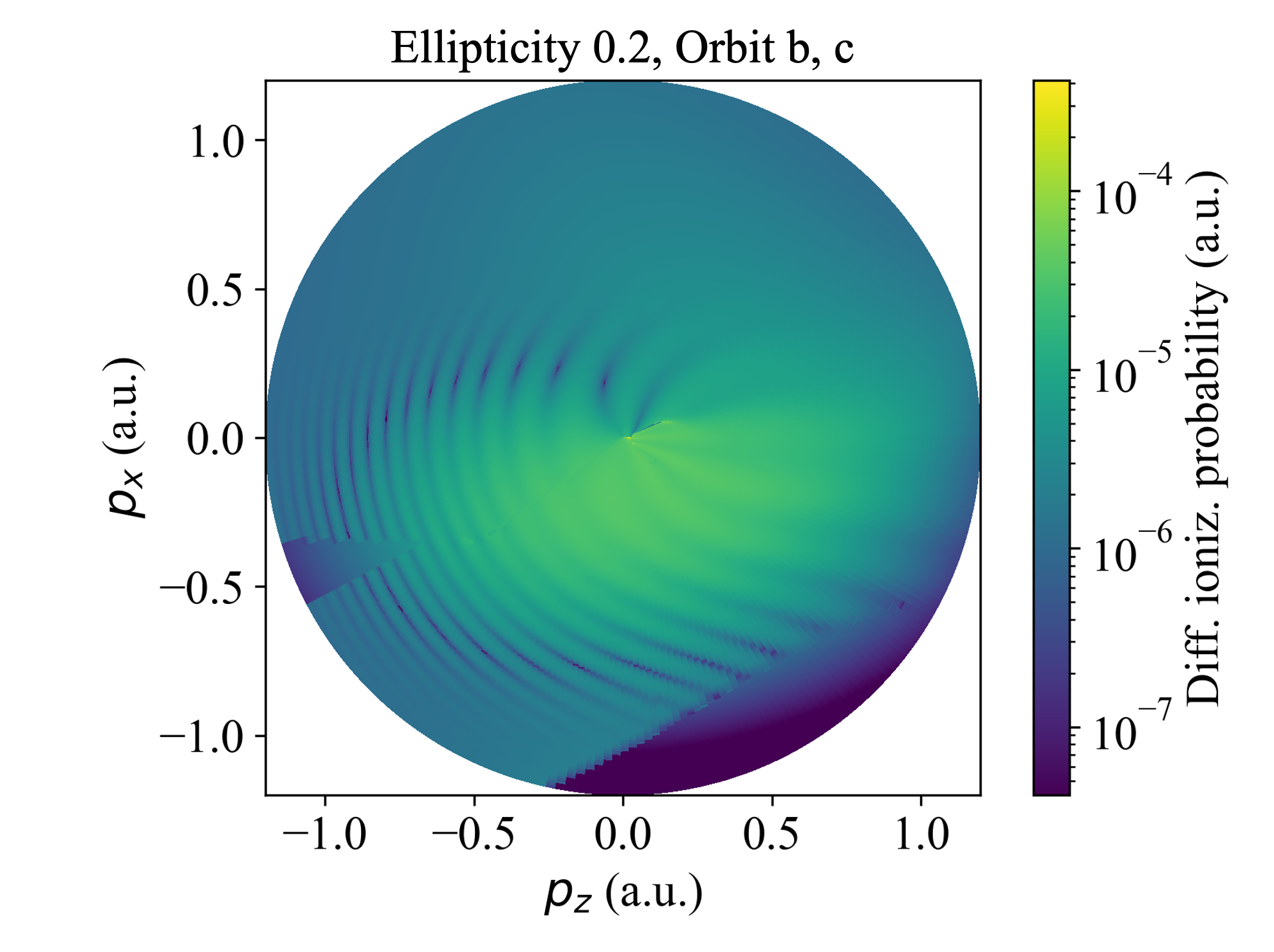} 
   \includegraphics[width=0.32\textwidth] {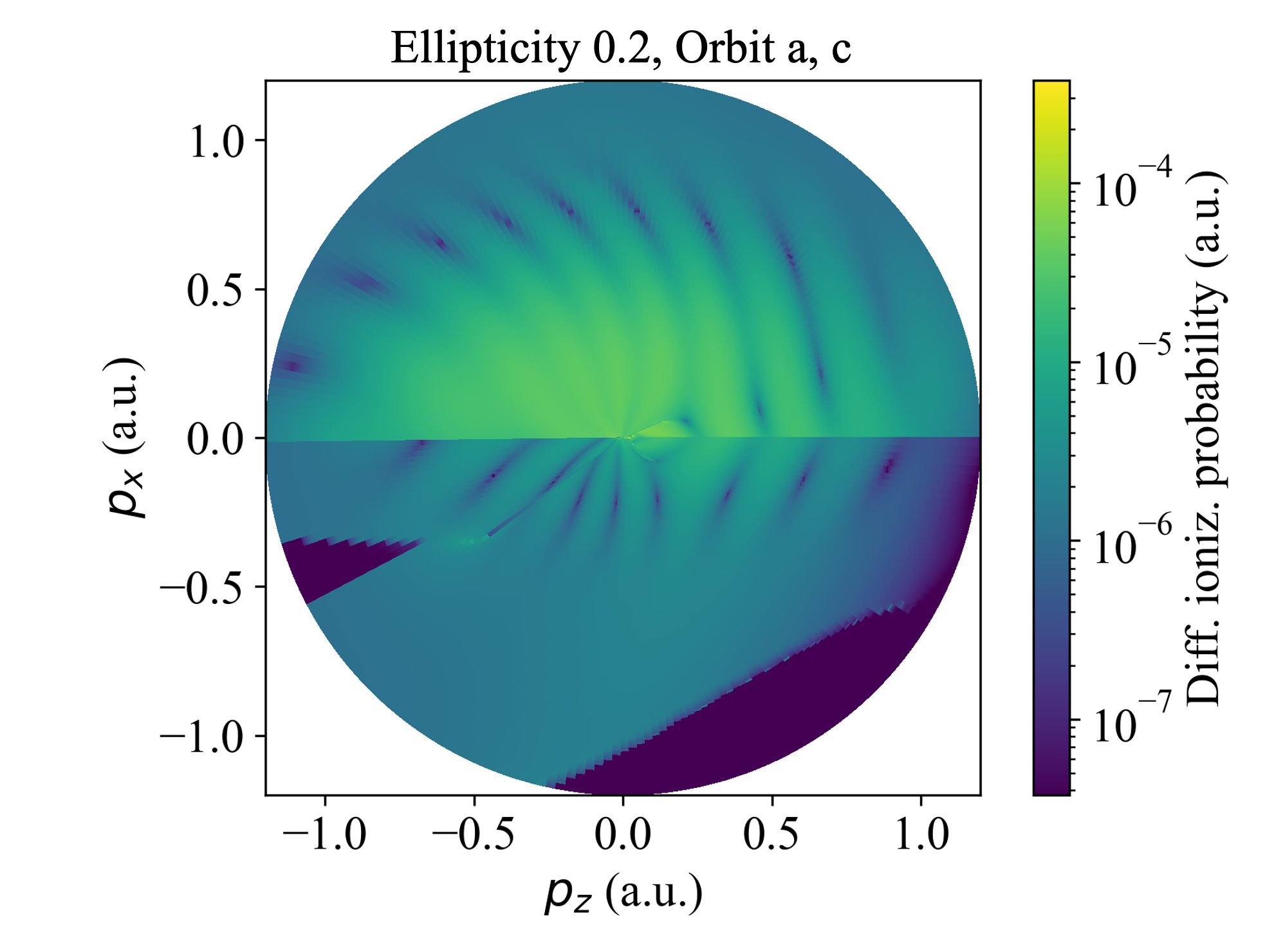}
  \\
   \includegraphics[width=0.32\textwidth] {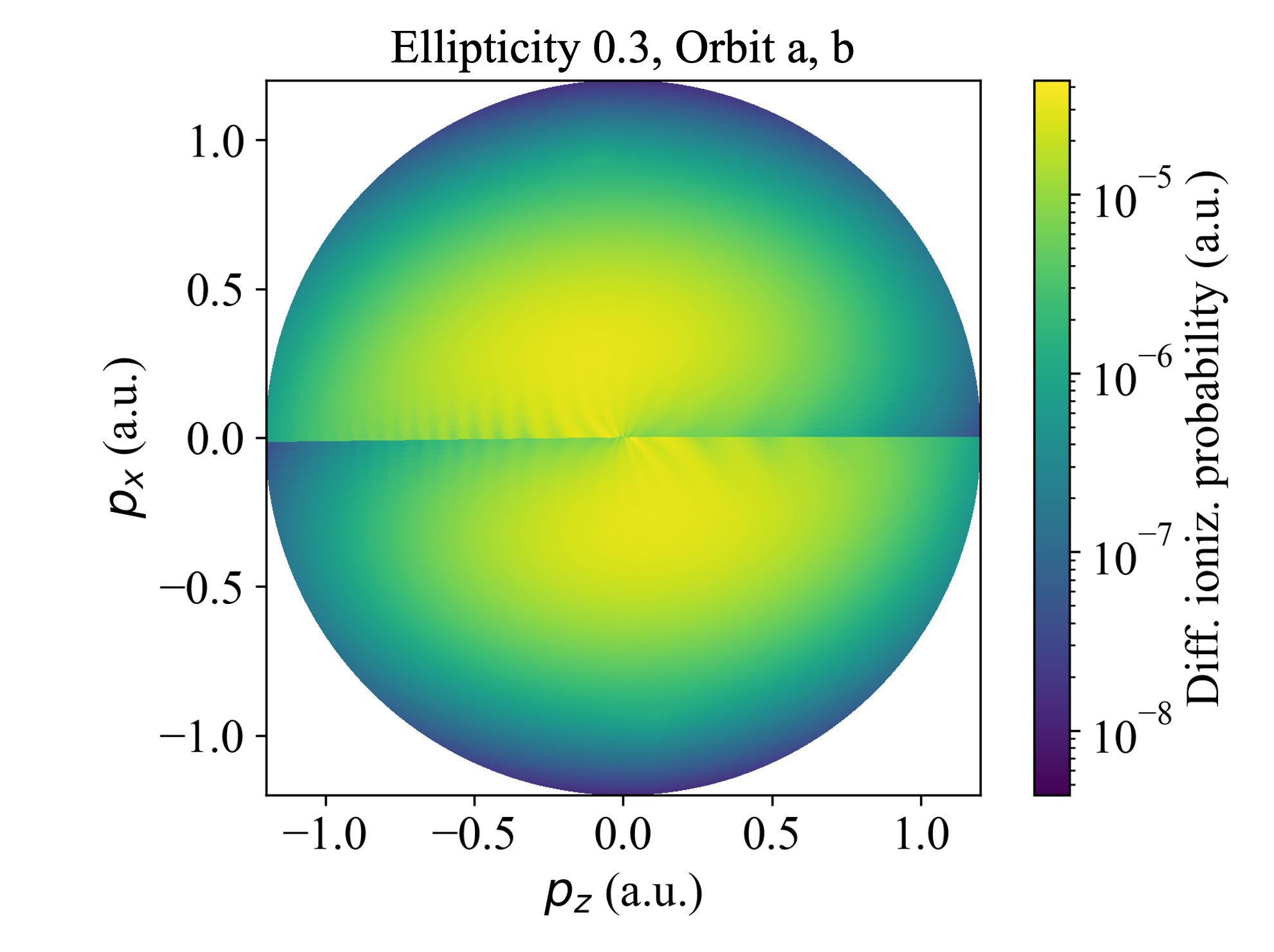}
   \includegraphics[width=0.32\textwidth] {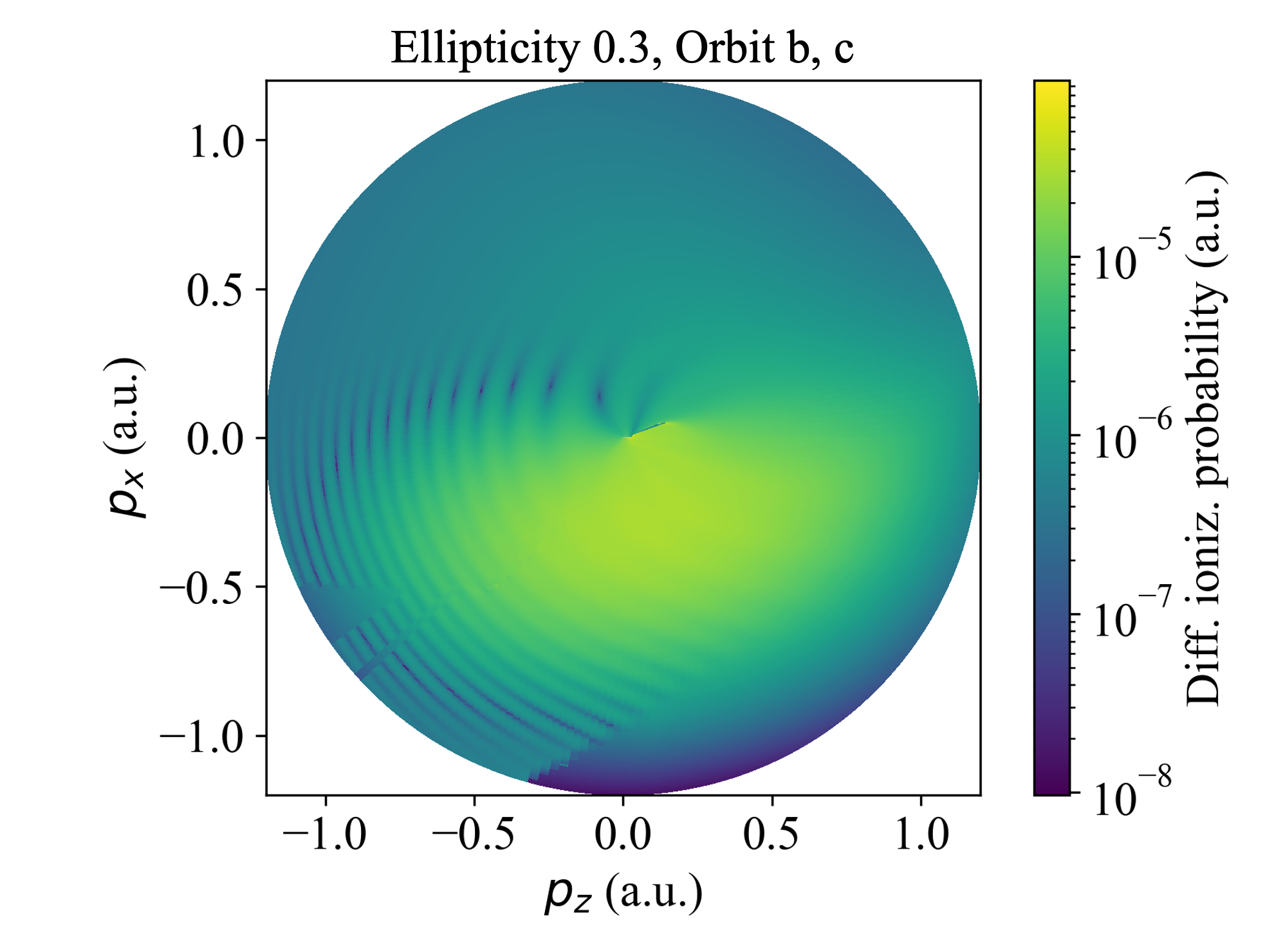}
   \includegraphics[width=0.32\textwidth] {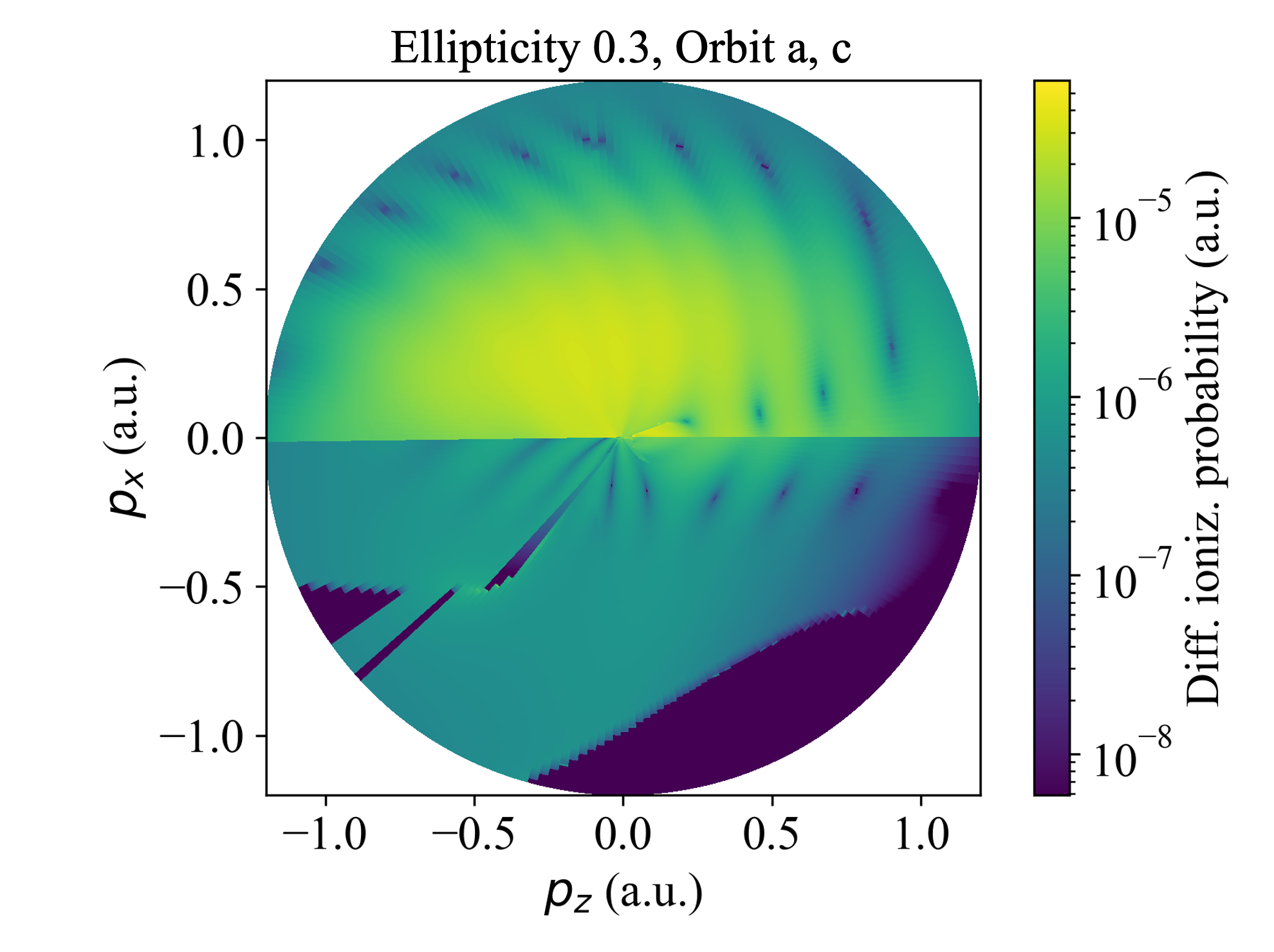}

    \caption{Photoelectron momentum distributions calculated for  helium in a field of intensity $2.5 \times 10 ^{14}$ W/cm$^2$ , wavelength $\lambda =$ 735 nm, whose ellipticity increases from $\epsilon=0$ to $\epsilon=0.3$, considering a single cycle and a unit cell with $\phi=0$. The left column shows the interference between orbits a+b, the middle column orbits b+c, and the right column orbits a+c. The orbit characterization is provided in Sec.~\ref{sec:classification}. All panels have been normalized to their maximum values and a logarithmic scale was employed. }
    \label{fig:PMDpairs}
\end{figure*}

In Fig.~\ref{fig:PMDpairs}, we analyze specific holographic structures by looking at how pairs of trajectories interfere.
Here, we employ the orbit classification in Sec.~\ref{sec:SaddlePointSolutions}, which keeps the distributions continuous along the minor axis. There may be, however, discontinuities along the major axis. 

In the left column of Fig.~\ref{fig:PMDpairs}, we see the PMDs resulting from the interference of orbits $a$ and $b$. For linear polarization, they give rise to a fan-shaped structure near the ionization threshold, displayed in the upper left corner of the figure. This is expected from our previous studies of holographic structures in linearly polarized fields \cite{Lai2017,maxwell2017coulomb}. Once the polarization increases, the fan starts to lose contrast until the interference pattern is ultimately washed out. The loss of contrast takes place away from the major polarization axis, with the peaks of the distributions moving further apart. This happens because the fan stems from interfering trajectories that start half a cycle apart, and whose momentum component $p_x$ parallel to the minor polarization axis does not change sign during the continuum propagation. Hence, the final momentum distributions resulting from such orbits will be peaked at opposite half planes and will overlap less and less as the ellipticity increases.  The remainder of the fan occurs where the overlap is still significant.

This behavior is very distinct from that of the spider, which is shown in the central column of Fig.~\ref{fig:PMDpairs} and results from the interference of orbits $b$ and $c$. For linear polarization, the spider is located in the region of positive $p_z$. According to the classification in Fig.~\ref{fig:Classification}, it is formed by the interference of orbits $D2$ and $C3$ ($D3$ and $C2$) in the upper (lower) half plane starting in the same half cycle. The structure forming in the region of negative $p_z$ also stems from the interference of orbits starting in the same half cycle, namely $D1$ interfering with $C4$ and $C1$ interfering with $D4$, although it is not known as ``the spider". 
For linear polarization, the spider is symmetric upon reflection with regard to the $p_z=0$ axis. For non-vanishing ellipticity, this symmetry is lost, with the whole structure undergoing an anticlockwise rotation and becoming more prominent in the lower momentum half plane. There is also a blurring in the spider fringes, initially close to the major polarization axis (second and third rows) and subsequently throughout (bottom row).

Finally, in the right column of Fig.~\ref{fig:PMDpairs} we plot the PMDs resulting from the interference of orbits $a$ and $c$. Those orbits start at different half cycles, but the final momenta will populate the same half plane. This is due to the momentum component $p_x$ along the minor polarization axis changing sign during the electron's continuum propagation. With increasing ellipticity, the fringes start to exhibit blurring in the vicinity of the $p_z$ axis, or, by inspecting the upper half plane, close to the maximum associated with orbit $a$. Contrast is retained for higher values of $p_x$. There is also a difference in strength in the upper and lower half plane.  
By inspection, one can see that many twisted patterns in Figs.~\ref{fig:QpropvsCQSFAlow} and \ref{fig:Allintracycle} in high photoelectron momentum regions can be attributed to the remnants of the spider, and of the fringes associated with the interference of orbits $a$ and $c$. 

Next, we will have a closer look at the blurring that occurs for the spider and the patterns due to the interference of orbits $a$ and $b$, among other effects. A loss of contrast may be due to changes in ionization probabilities, to orbit 3 being suppressed due to rescattering being hindered as the ellipticity increases, or to both effects.  

\begin{figure*}[h!tb]
    \centering
     \centering
   \includegraphics[width=0.45\textwidth] {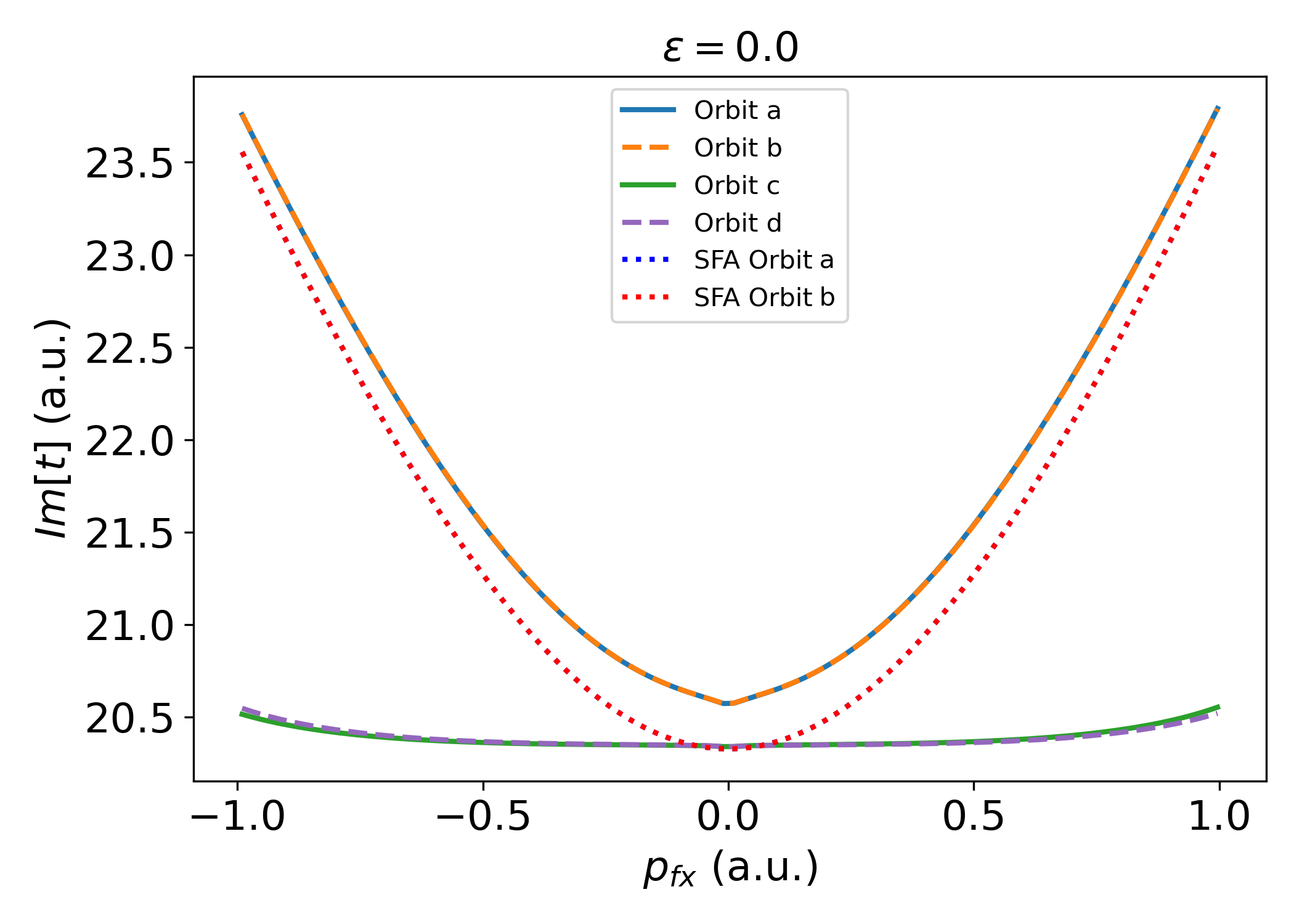}
   \includegraphics[width=0.45\textwidth] {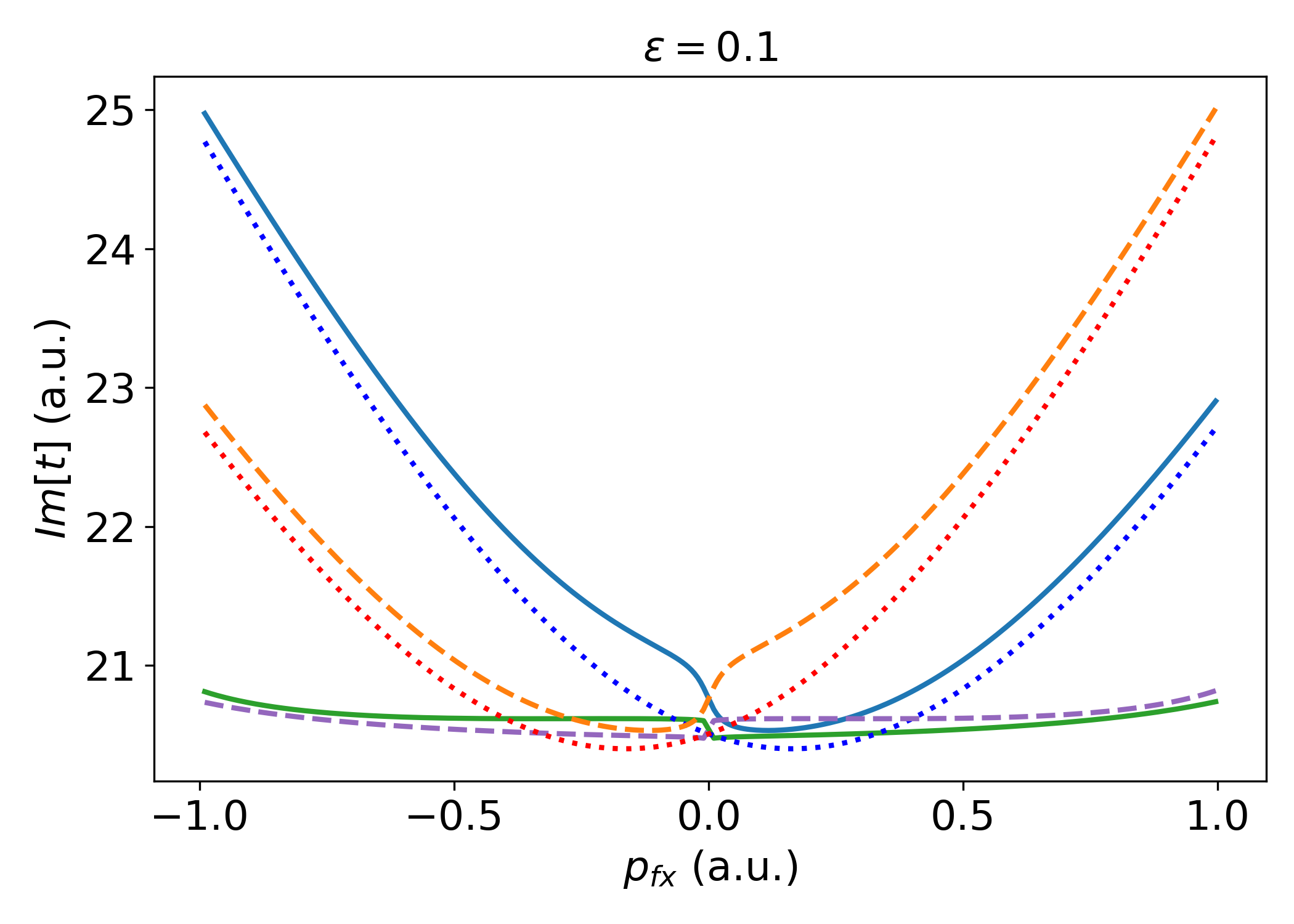}\\
      \includegraphics[width=0.45\textwidth] {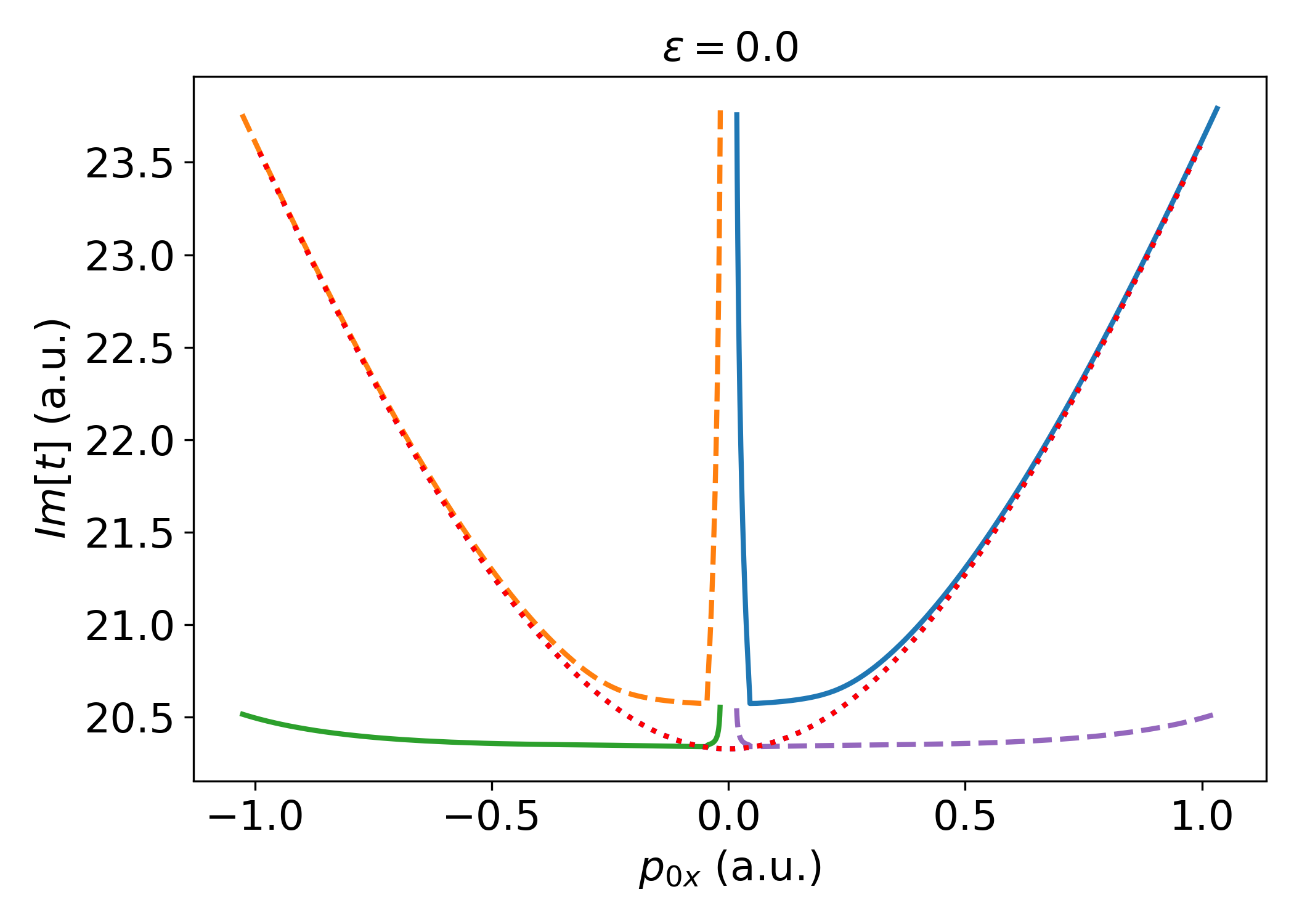}
   \includegraphics[width=0.45\textwidth] {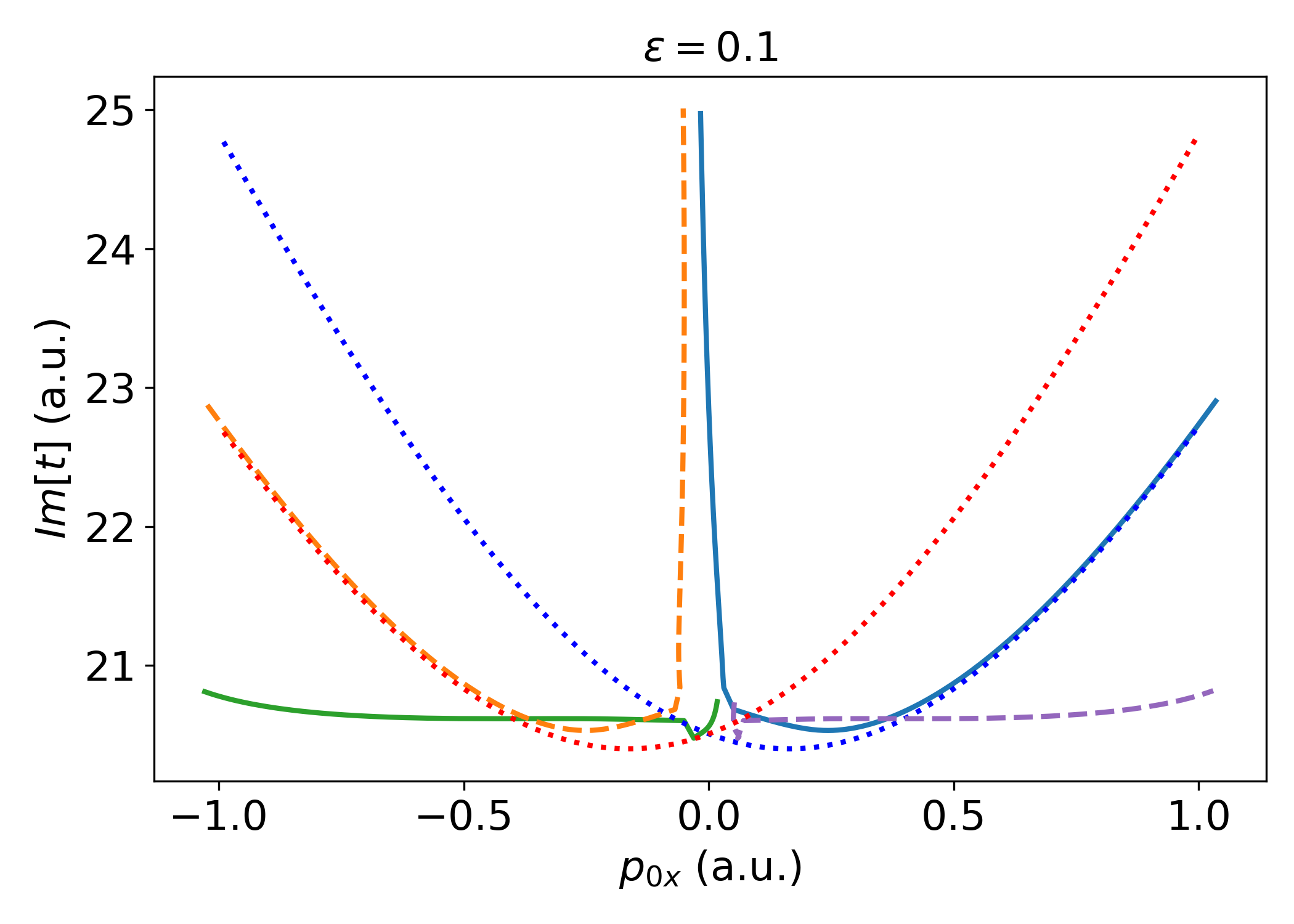}
\caption{Imaginary parts of the ionization times $t'$ for the CQSFA and the SFA as functions of the final and initial momentum components $p_{fx}$ and $p_{0x}$ taken along the fixed final $p_{fx}$ axis, with $p_{fz} = 0$ (upper and lower rows, respectively). Calculation done for helium in a field with intensity $2.5 \cross 10^{14} W/cm^2$, wavelength 735nm. The left and right column correspond to ellipticity $\epsilon=0$ and $\epsilon=0.1$, respectively. The remaining field and atomic parameters are the same as in the previous figures. }
    \label{fig:Imt}
\end{figure*}

The changes in the ionization probability can be inferred from the imaginary parts of the ionization times, as the ionization probabilities roughly scale as  $\exp[-2\mathrm{Im}[t']]$. Hence, the larger $\mathrm{Im}[t']$ is, the more suppressed a specific orbit will be. In Fig.~\ref{fig:Imt}, we plot $\mathrm{Im}[t']$ along the minor polarization axis $p_z=0$ for the ellipticities used in Fig.~\ref{fig:PMDpairs}. Our analysis will focus on the CQSFA orbits, but, in the upper panels of Fig.~\ref{fig:Imt}, we also provide $\mathrm{Im}[t']$  for the SFA orbits $a$ and $b$. 

Fig.~\ref{fig:Imt} displays $\mathrm{Im}[t']$ as functions of the final and initial electron momentum component $p_x$ along the minor polarization axis, that is, $p_{fx}$ and $p_{0x}$ (upper and lower row, respectively). In order to cover a larger range for the initial momentum, in the lower panels of Fig.~\ref{fig:Imt}, the orbits were selected such that all values of the parallel momenta are allowed. This is relevant as, strictly speaking, an electron cannot escape if $p_{0x}= p_{0z}=0$ and will cause the divergencies dicsussed below.  

For linearly polarized fields (left column in Fig.~\ref{fig:Imt}), $\mathrm{Im}[t']$ is symmetric with regard to the reflection $p_x\rightarrow-p_x$ for all orbits.  This behavior mirrors that observed for the PMDs in Figs.~\ref{fig:QpropvsCQSFAlow}, \ref{fig:Allintracycle} and  \ref{fig:PMDpairs}, which exhibit this symmetry for linear polarization. For the CQSFA orbits $a$ and $b$, $\mathrm{Im}[t']$ displays a behavior similar to its SFA counterparts, with a minimum at $p_{fx}=0$ (upper left corner of Fig.~\ref{fig:Imt}). This minimum indicates that, for an electron along orbits $a$ or $b$, the probability that the electron reaches the detector with final momentum component  $p_{fx}=0$ is largest.

For the SFA, this is related to the effective potential barrier being narrowest, as a non-vanishing $p_x$ will effectively raise the ionization potential (for a discussion of this shift see, e.g., \cite{Shaaran2010}). For the CQSFA, however, the interpretation is subtler, as suggested by $\mathrm{Im}[t']$  plotted against the initial momentum $p_{0x}$ (lower left corner of Fig.~\ref{fig:Imt}). For the CQSFA orbits $a$ and $b$, the figure shows that $\mathrm{Im}[t'] \rightarrow \infty$ for $p_{0x}=0$. This is due to the presence of the Coulomb potential and means that an electron along orbit $a$ and $b$ cannot escape with vanishing perpendicular momenta. This is clear as the Coulomb potential essentially decelerates an electron along orbit $a$, and a field dressed hyperbola starting half a cycle later, namely orbit $b$, requires  $p_{0x}\neq 0$. Therefore, there will be a minimal escape momentum for the electron, in order for it to reach the detector with final momentum $p_{fx}=0$.  This also holds for the other CQSFA orbits, as the maxima for $\mathrm{Im}[t']$  at $p_{0x}=0$ indicate. For large absolute values of $p_{0x}$, the imaginary parts $\mathrm{Im}[t']$ associated with the CQSFA orbits $a$ and $b$ tend to their SFA counterparts. This is expected, as, in this limit, they behave as SFA direct orbits and the Coulomb potential does not play a critical role \cite{maxwell2017coulomb}. For the CQSFA orbits $c$ and $d$ the curves are much flatter throughout. This flatter behavior of $\mathrm{Im}[t']$  stems from the real parts of the ionization times being restricted to narrower time ranges, closer to the peak of the field (for a recent discussion for linearly polarized fields see our preprint \cite{Werby2022}).

 Finally, Fig.~\ref{fig:Imt} is also a good indicator of the momentum regions for which the holographic fringes will show high contrast. Similar $\mathrm{Im}[t']$ for different trajectories at a specific final momentum means that their contributions to the whole transition amplitude are comparable, so that their interference will exhibit sharp fringes. According to the figure, this would happen for orbits $a$ and $b$, or orbits $c$ and $d$ for a wide range of perpendicular momenta $p_x$. An inspection of the PMDs along the $p_z=0$ axis shows, indeed, that the fan, caused by the interference orbits $a$ and $b$, and the carpet, caused by the interference of orbits $c$ and $d$, exhibit high contrast for the linearly polarized case regardless of $p_x$. On the other hand, the spider, coming from the interference of orbits $b$ and $c$, is only expected to be prominent near $p_x=0$, that is, the field-polarization axis. 

This overall behavior changes even for a small ellipticity (see right column of Fig.~\ref{fig:Imt}), for which the $p_x \rightarrow -p_x$ reflection symmetry is broken.
Non-vanishing ellipticity leads to a tilting of $\mathrm{Im}[t']$ with regard to $p_x=0$ for orbits $a$ and $b$, both for the SFA and CQSFA. For all CQSFA orbits, there is a step-wise behavior for $\mathrm{Im}[t']$ around the origin, if plotted as a function of the final momentum $p_{fx}$ (see upper right panel of Fig.~\ref{fig:Imt}). This feature is absent for the SFA. 

The tilting in $\mathrm{Im}[t']$ is caused by the field components parallel to the minor polarization axis, which either help or hinder the electron ionization along orbits $a$ and $b$. For instance, for orbit $a$ the `tilt' to the right indicates that the field component along the minor polarization axis helps ionization for positive momenta, but hinders it for negative momentum. A similar line of reasoning can be used for orbit $b$ `tilting' to the left, with the difference that in this case the curve will be the mirror image of that observed for orbit $a$. 

The step-wise feature is caused by the Coulomb potential, and can be understood by inspecting how $\mathrm{Im}(t')$ behaves as a function of $p_{0x}$ (see lower right panel of Fig.~\ref{fig:Imt}). Similarly to what happens for linearly polarized field, the electron cannot escape if its perpendicular momentum component $p_{0x}$ is vanishing and there is a minimum momentum value for which it may escape. Nonetheless, due to the field's non-vanishing ellipticity the escape momenta will be different for the positive and negative momentum half plane. This will lead to the step in $\mathrm{Im}(t')$ near the origin, if plotted as a function of the final momenta $p_{fx}$. Orbits $c$ and $d$ also exhibit the step-wise behavior mentioned above, and for the very same reasons.  
 
An inspection of the right upper panel of Fig.~\ref{fig:Imt} also provides valuable insight into the momentum regions for which specific holographic patterns are blurred or sharp.  For instance, the structure stemming from interference of orbits $a$ and $c$ becomes sharper away from the major polarization axis because the imaginary parts of the times cross each other for $p_{fx}=0.5$. This can be confirmed by looking at the corresponding PMD in Fig.~\ref{fig:PMDpairs} (see third column, second upper row therein). Similarly, one expects the spider to be sharper in the lower momentum half plane as $\mathrm{Im}[t']$ for orbits $b$ and $c$ are much closer for $p_{fx}<0$. The corresponding PMD, located at the second row and second column in Fig.~\ref{fig:PMDpairs}, shows that this is indeed the case.  

\begin{figure*}[h!tb]
    \centering
     \centering
   \includegraphics[width=0.45\textwidth] {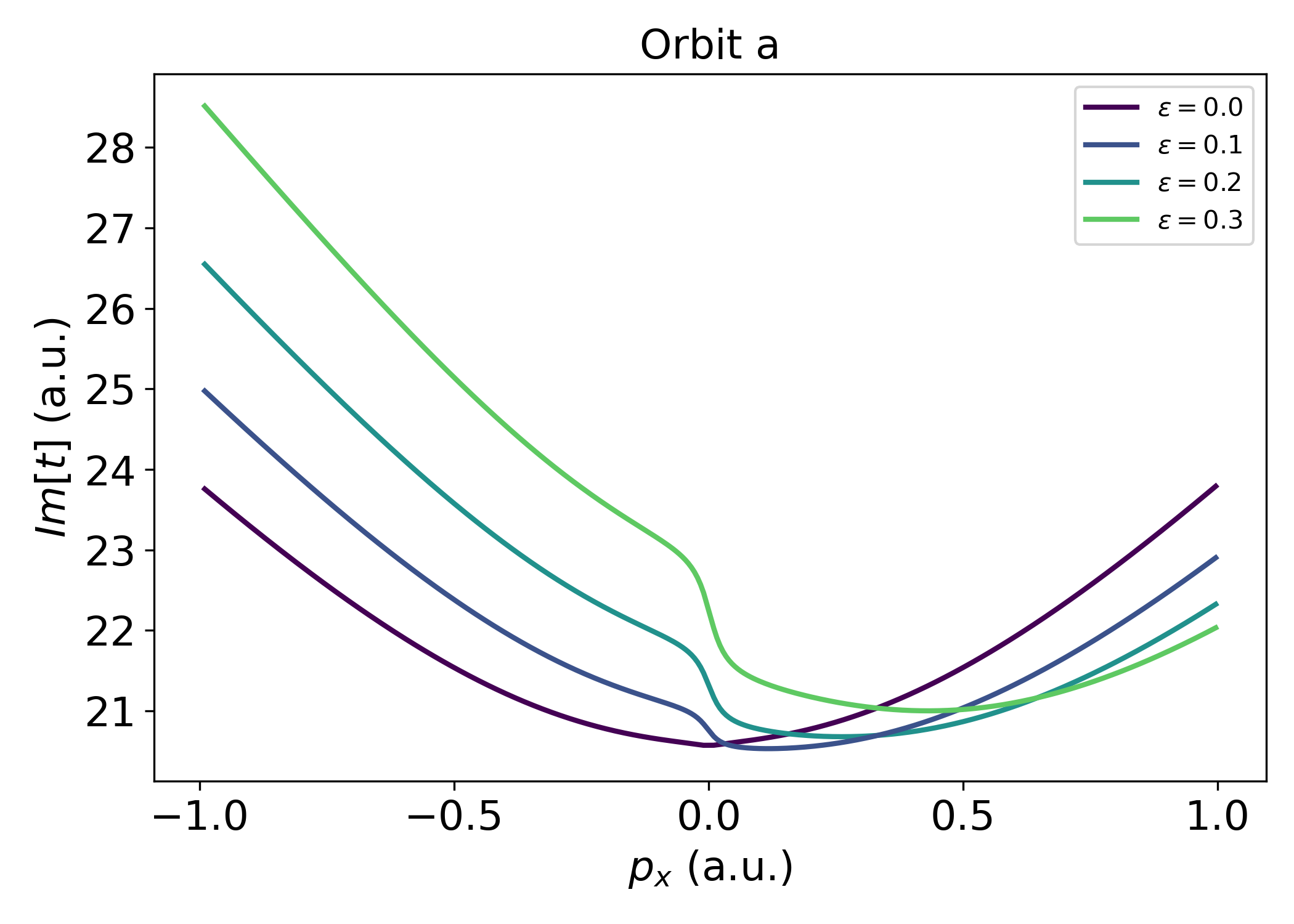}
   \includegraphics[width=0.45\textwidth] {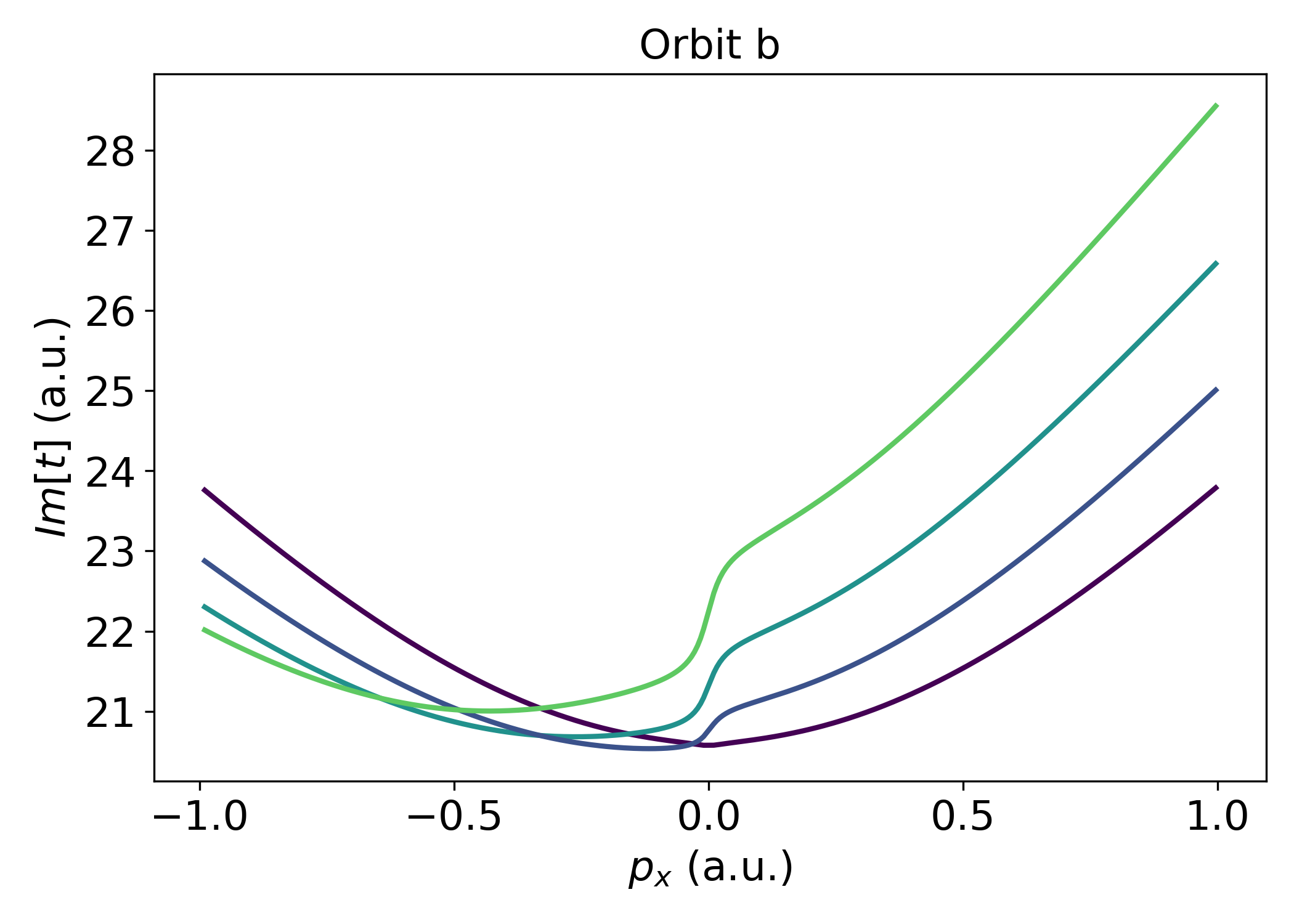}\\
\includegraphics[width=0.45\textwidth] {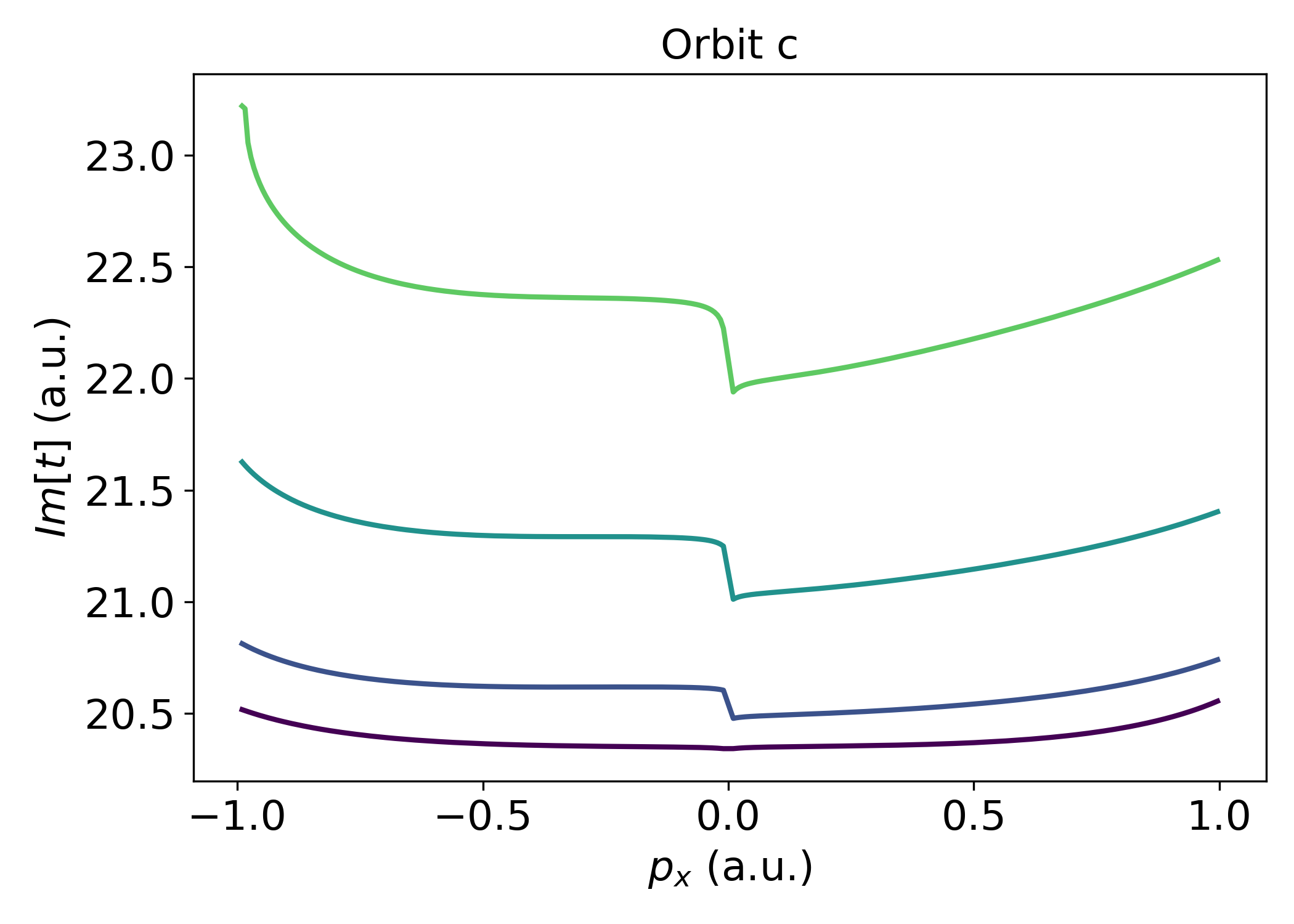}
\includegraphics[width=0.45\textwidth] {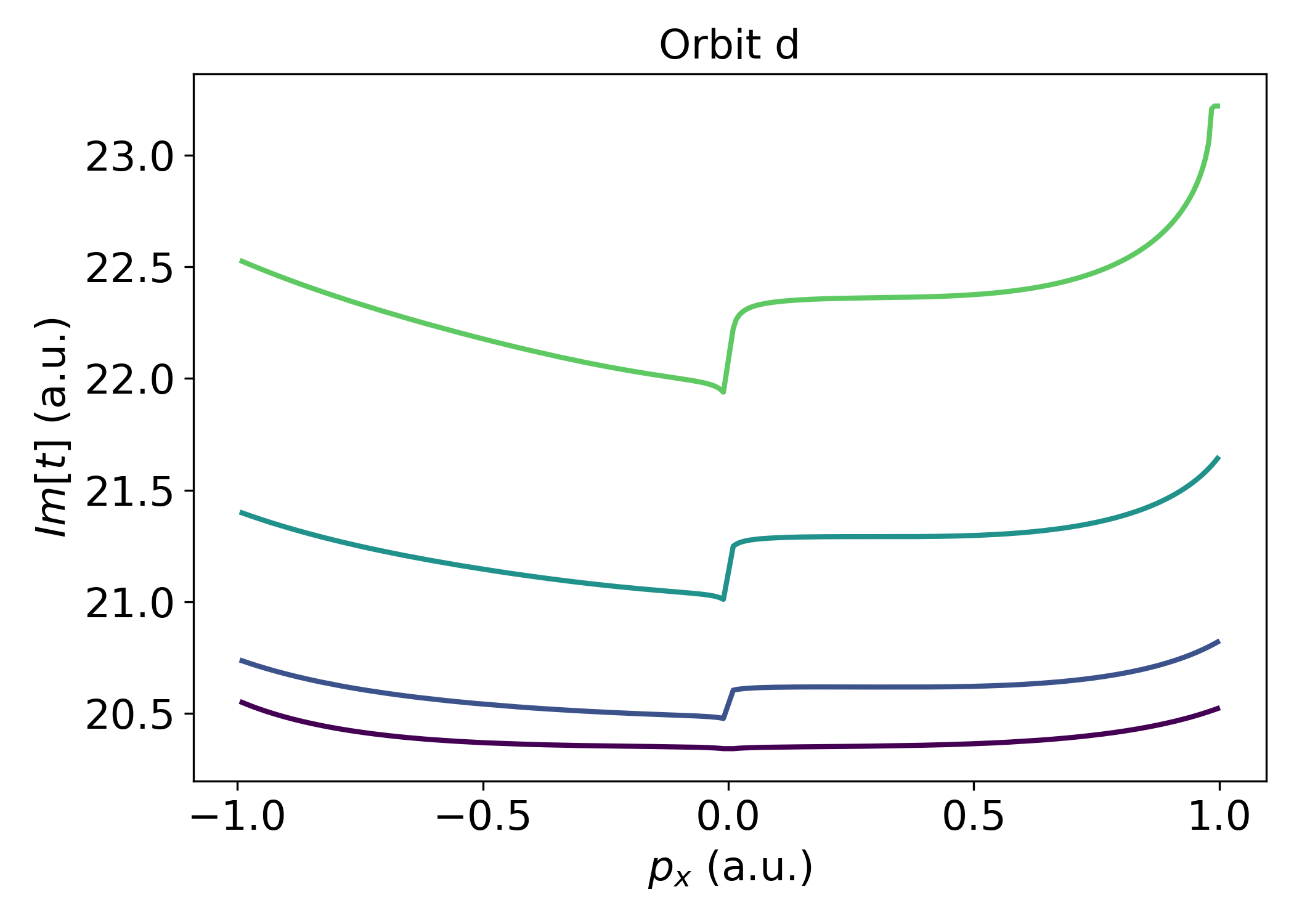}
\caption{Imaginary parts of the ionization times $t'$ for the CQSFA orbits $a$, $b$, $c$ and $d$ as functions of the electron's final momentum $p_x$ along the minor polarization axis, for increasing driving-field ellipticities. The orbits are classified according to Sec.~\ref{sec:classification} and the remaining parameters are the same as in the previous figures.}
    \label{fig:Imtvareps}
\end{figure*}

In Fig.~\ref{fig:Imtvareps}, we illustrate more thoroughly how $\mathrm{Im}[t']$ behaves for increasing field ellipticity. We consider the final electron momentum component $p_x$ along the minor polarization axis and plot each CQSFA orbit separately.  The tilting for Orbits $a$ and $b$ becomes more extreme for increasing ellipticity. This sheds light in the blurring of the fan-shaped fringes, caused by discrepant imaginary parts in the same momentum half plane, and on the shift of the maxima in the PMDs towards larger momentum values, caused by the changes in the minima of $\mathrm{Im}[t']$. For orbits $c$ and $d$, instead of a `tilt', we see a marked increase in $\mathrm{Im}[t']$ with the ellipticity. This hints at both orbits $c$ and $d$ becoming suppressed for larger values of $\epsilon$, which is not surprising, given that these orbits are associated with rescattering and will become rarer in the high-ellipticity regime.   
\clearpage
\section{Conclusions}
\label{sec:conclusions}

In this work, we investigate quantum interference in strong-field ionization in elliptically polarized fields, with emphasis on holographic patterns. We interpret the features encountered using the Coulomb-quantum orbit strong-field approximation (CQSFA), which is compared to the numerical solution of the time-dependent Schr\"odinger equation (TDSE) and the standard strong-field approximation (SFA), for which the binding potential is absent in the continuum propagation. The CQSFA is an
orbit-based method that accounts for tunneling, quantum
interference and the presence of the binding potential
in the continuum \cite{lai2015influence,Lai2017,maxwell2017coulomb}. So far, it had only been applied to photoelectron holography in linearly polarized fields. We focus on the low and intermediate ellipticity regime, for which intra-cycle holographic interference is present. This differs from typical studies of photoelectron emission in elliptical fields, whose main objective is to map a single ionization time to an offset angle. This mapping requires a high ellipticity, so that intra-cycle interference is strongly suppressed \cite{Hofmann2019}. In the low-ellipticity regime, there are many possible ionization times, which can be associated with electron orbits. 

We find that a non-vanishing ellipticity leads to twists in the holographic patterns. The twists are absent in the plain strong-field approximation, which neglects the residual binding potential, but have been identified in the CQSFA and in TDSE computations. This suggests that they are caused by the interplay of the elliptical field and the central potential. Further support to this is provided by a TDSE computation which truncates the tail of the Coulomb potential, but leaves the effective potential barrier intact. There is a decrease in the twists due to the removal of the Coulomb tail. However, a residual twist is present, which means that there is also a contribution from the barrier. Twists have been observed experimentally in the spider \cite{Xie2018} and angular shifts for ATI peaks of increasing order have been reported in \cite{Xiao2022}. However, most studies in the low to intermediate ellipticity regime concentrate on Coulomb focusing \cite{Shafir2013,Li2013}, or the maxima and width or the photoelectron momentum distributions \cite{Geng2014,Han2017b}. 

As the ellipticity increases, the contrast of the holographic fringes fades and the maxima of the PMDs move further apart. This is due to the transverse components of the momenta upon ionization and during continuum propagation. Estimates for the ellipticity range in which quantum interference is relevant has been provided in this work, and agree with the the outcomes of the TDSE and CQSFA computations.  

The twisting and the blurring are then understood in terms of interfering electron orbits, whose ionization times are first derived analytically in the SFA framework, in a generalization of the expressions in \cite{Paulus1998,Javsarevic2020} to a broader parameter range. These SFA expressions are then used as first guesses for the CQSFA ionization times. A noteworthy issue is that the orbit classification used in the CQSFA is highly dependent on the driving-field shape and existing symmetries. In fact, because the reflection symmetry with regard to the major polarization axis is broken, we have altered the classification in terms of orbits 1, 2, 3 and 4 \cite{Yan2010,lai2015influence}, with regard to the linearly polarized case. Other examples of modified CQSFA orbits have been used in the study of two-color linearly polarized fields \cite{Rook2022}. A very useful tool to understand the loss of contrast in the holographic patterns is the imaginary part of the ionization time, which one may relate to the ionization probability associated with a specific type of orbit. Comparable $\mathrm{Im}[t']$ for a pair $(i,j)$ of orbits means that there will be sharp fringes, while $\mathrm{Im}[t_i'] \ll \mathrm{Im}[t_j']$ means that blurring will occur. For a specific orbit, non-vanishing ellipticity will break the reflection symmetry of $\mathrm{Im}[t']$  for the lower and upper half planes. 

Although the loss of contrast in the holographic patterns 
is an overall feature, it occurs for different reasons, depending on the type of orbit which create the patterns. 
If a specific holographic pattern results from the quantum interference of events starting at different half cycles  whose momentum components $p_x$ do not change signs,
with increasing ellipticity their contributions will mainly populate different half planes.
Therefore, their maxima will move further apart and quantum interference will only be significant close to the major polarization axis. This can be observed, for instance, in the fan-shaped fringes.
If, on the other hand, the holographic pattern results from events in the same half cycle or $p_x$ changes during propagation, the contribution of such orbits to the PMDs will move to the same momentum half plane. However, those orbits interacting more closely with the core, such as orbits $c$ and $d$, will become rarer as the ellipticity increases. Consequently, the transition amplitudes associated with those specific pathways will be suppressed, and the patterns will blur.  This is the case of the spider and of the patterns stemming from the interference of orbits $a$ and $c$. The widely studied high ellipticity regime is reached when some of the solutions cease to exist and some merge. Methodologically, this involves Stokes transitions, which, for the field parameters considered in this work, happen outside the parameter range of interest (see Appendices 2 and 3). 

In summary, the twisted patterns reported in this paper are another manifestation of the interplay of the Coulomb potential and the elliptically polarized field: instead of a single offset angle in the PMD, which can be modelled classically, the long-range potential leads to offsets in holographic patterns, which can be understood in terms of interfering orbits. The present  studies may be useful for a wide range of scenarios in which quantum interference is important, such as diatomic molecules in elliptically polarized fields \cite{Yang2014}.   

\section*{Acknowledgements} This work was funded by grant No.\ EP/J019143/1, from the UK Engineering and Physical Sciences Research Council (EPSRC). C.H. acknowledges support by a Swiss
National Science Foundation mobility fellowship.
A.S.M. acknowledges funding support from the European Union’s Horizon 2020 research and innovation programme under the Marie Sk\l odowska-Curie grant agreement, SSFI No.\ 887153.
G.K.  was funded  by the University College London (UCL) summer research programme sponsored by the World-leading University Fostering Program in Seoul National
University and would like to thank UCL for its kind hospitality.
\appendix
\section*{Appendix 1 - Analytic expressions for ionization times}
\label{app:iontimes}

In this appendix, we briefly sketch the procedure to obtain the analytic solutions for the ionization times obtained with the strong-field approximation, which are given in Sec.~\ref{sec:SaddlePointSolutions}. The saddle-point solutions giving the ionization times are obtained from Eq.~\eqref{eqn: expanded saddle point equation}. 
By substituting of $\xi = \cos(\omega t)$ and by replacing $\sin^2(\omega t) = 1-\xi^2$ in Eq.~\eqref{eqn: expanded saddle point equation}, we can derive a quartic equation for $\xi$ as Eq.~\eqref{eqn: the quartic equation}.
\begin{equation}
    \xi^4 + 4\bar{p_z} \xi^3 + (2\bar{U} + 4(1+\epsilon^2)\bar{p_z}^2)\xi^2 + 4\bar{p_z}\bar{U}\xi + (\bar{U}^2 - 4\bar{p_x}^2\epsilon^2) = 0
	\label{eqn: the quartic equation}
\end{equation}

where, $\bar{p_z}$, $\bar{p_x}$, and $\bar{U}$ are defined as Eq.~\eqref{eqn: bar notation}.

\begin{equation}
    \begin{split}
        \bar{p_z} &= \frac{\sqrt{1+\epsilon^2}}{2(1-\epsilon^2)\sqrt{U_p}} p_z, \\
        \bar{p_z} &= \frac{\sqrt{1+\epsilon^2}}{2(1-\epsilon^2)\sqrt{U_p}} p_z, \\
        \bar{U} &= \frac{1}{(1-\epsilon^2)}\left(\frac{1+\epsilon^2}{4U_p}(2I_p + p_z^2 + p_x^2) + \epsilon^2\right)
    \end{split}
	\label{eqn: bar notation}
\end{equation}

Since the analytic form of the solutions of the quartic equation exists, we can obtain four solutions of the quartic equation as Eq.~\eqref{eqn: four solutions}. The explicit form of $\zeta$ and $\eta$ is given as
\begin{equation}
    \begin{split}
        \Delta_0 &= A_2^2 - 3A_3A_1+12A_0, \\
        \Delta_1 &= 2A_2^3 - 9A_3A_2A_1 + 27A_3^2A_0 + 27A_1^2 - 72A_2A_0, \\
        Q &= \left(\frac{\Delta_1 + i\sqrt{4\Delta_0^3 - \Delta_1^2}}{2}\right)^{\frac{1}{3}}, \\ 
        \zeta &= \frac{1}{2}\sqrt{-\frac{2}{3}\left(2\bar{U} + 4\epsilon^2\bar{p_z}^2-2\bar{p_z}^2\right) + \frac{1}{3}\left(Q + \frac{\Delta_0}{Q}\right)}, \\
        \eta &= -\frac{8\bar{p_z}^3\epsilon^2}{\zeta}, 
    \end{split}
	\label{eqn: zeta and eta}
\end{equation}
where $A_i$ are the coefficients of $\xi_i$ in the quartic equation Eq.~\eqref{eqn: the quartic equation}. Note that not all four solutions of the quartic equation are the solution of the original saddle point equation Eq.~\eqref{eqn: expanded saddle point equation}. However, by comparing the sign of Eq.~\eqref{eqn: expanded saddle point equation}, we can find the two valid solution sets as Eq.~\eqref{eqn: grouped ionization time}. Furthermore, we have checked that both $\zeta$ and $\eta$ go to zero at zero ellipticity, which makes our solutions consistent with the linearly polarized solution at zero ellipticity.

\section*{Appendix 2 - Stokes transitions and divergencies}

Within the present formalism, obstacles towards computing PMDs for fields of arbitrary ellipticity are coalescent saddles and Stokes transitions. Coalescent saddles mean that uniform asymptotic expansions that treat them collectively will be required \cite{Faria2002}. Stokes transitions lead to the asymptotic expansion becoming inaccurate due to a change of contour. This will result in divergent contributions, which must be discarded (for a detailed discussion and regularization methods see \cite{Berry1989}). 
In this appendix, we will highlight how coalescing saddles and Stokes transitions lead to divergencies in the PMDs, and explore in what momentum ranges this happens. This will be illustrated with the SFA. We will go beyond the studies in \cite{Paulus1998,Javsarevic2020}, which have been performed along the major polarization axis $p_z$, and look at how the PMDs are affected as a whole. The CQSFA will bring further challenges, such as branch cuts associated with rescattering. Preliminary studies in this direction already exist \cite{Pisanty2016,Maxwell2018}, but its full solution is beyond the scope of this article.

\begin{figure*}[h!tb]
    \centering
     \centering
   \includegraphics[width=0.325\textwidth] {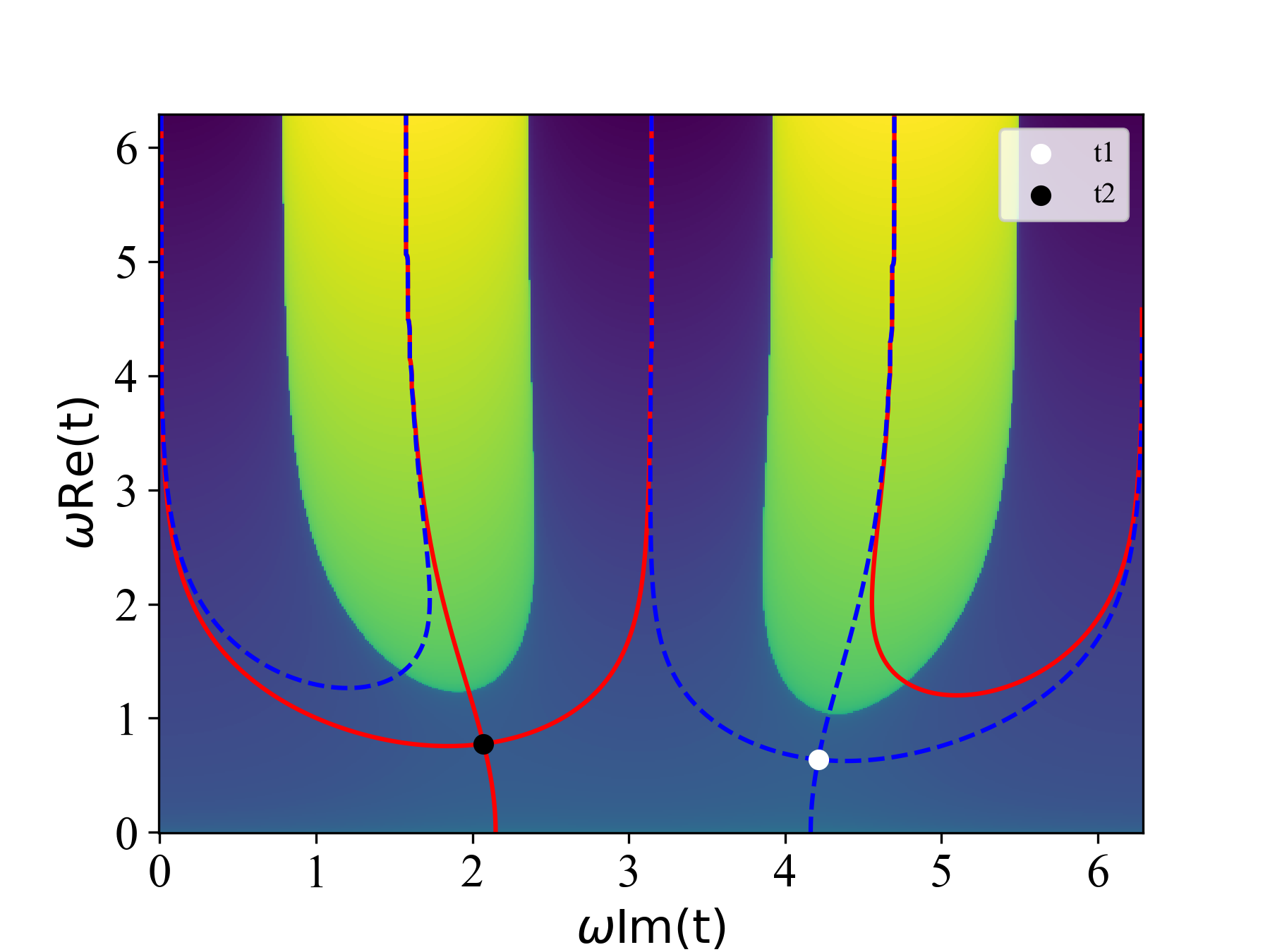}
   \includegraphics[width=0.325\textwidth] {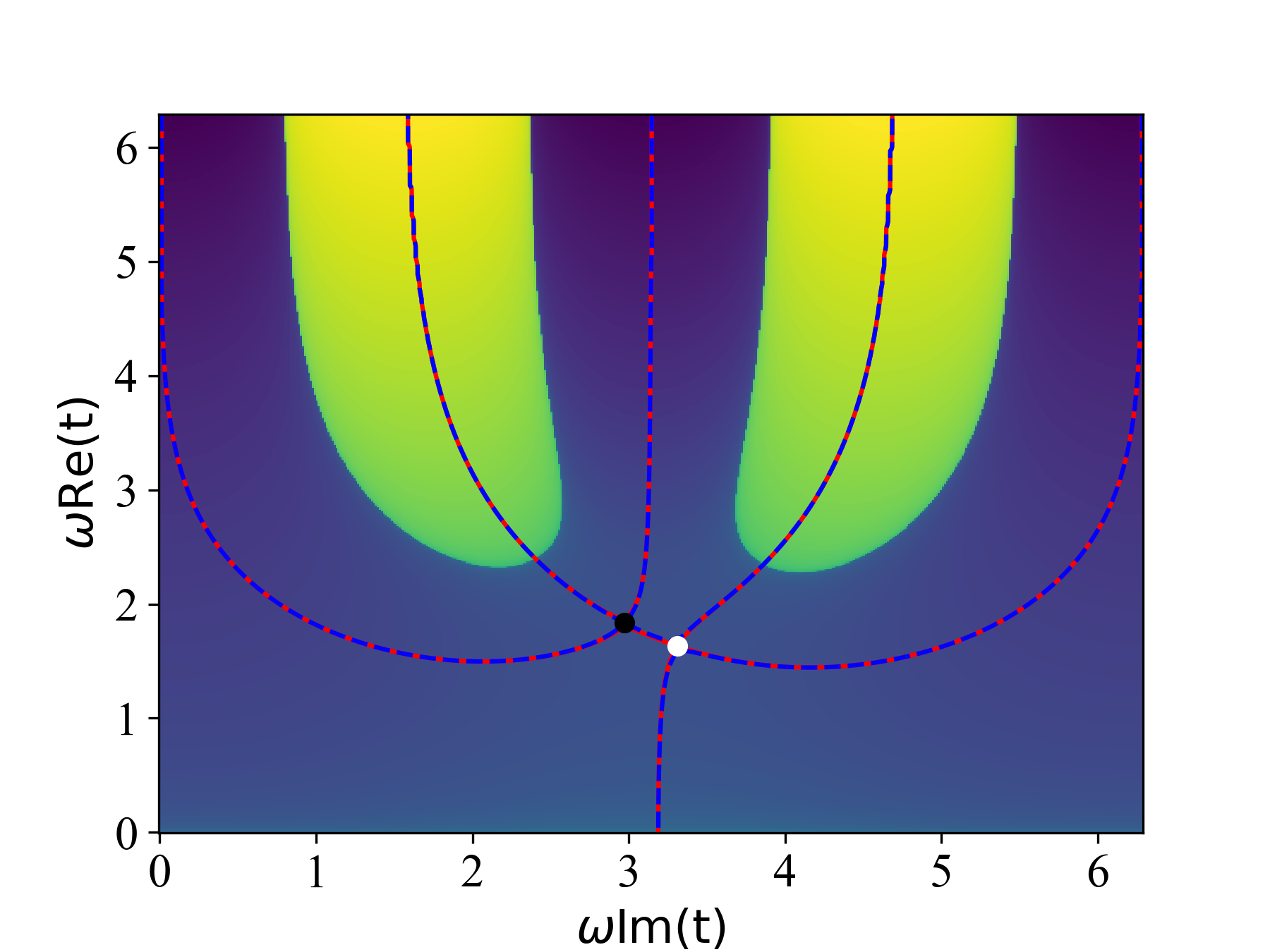}
  \includegraphics[width=0.325\textwidth] {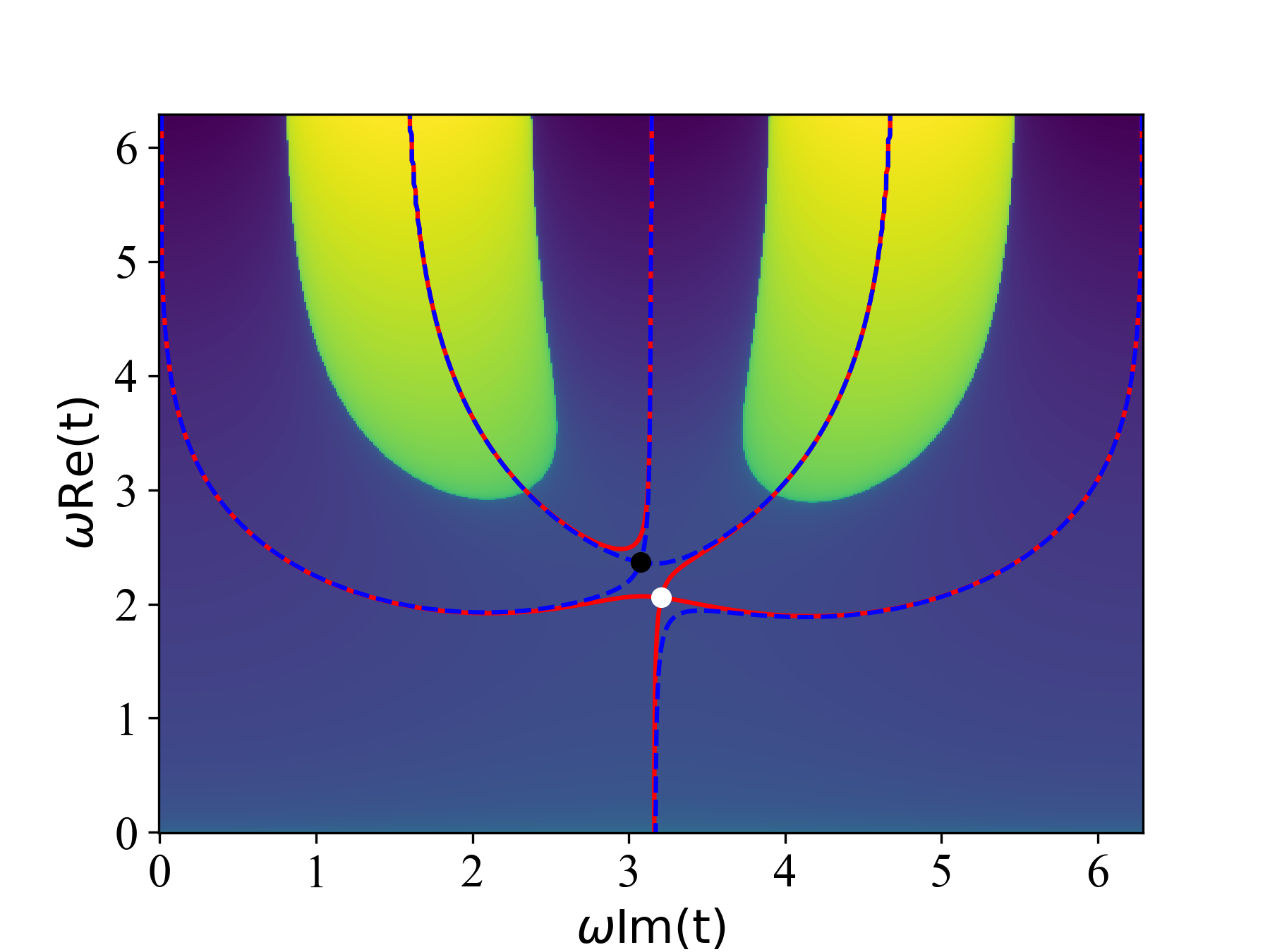}\\
   \includegraphics[width=0.325\textwidth] {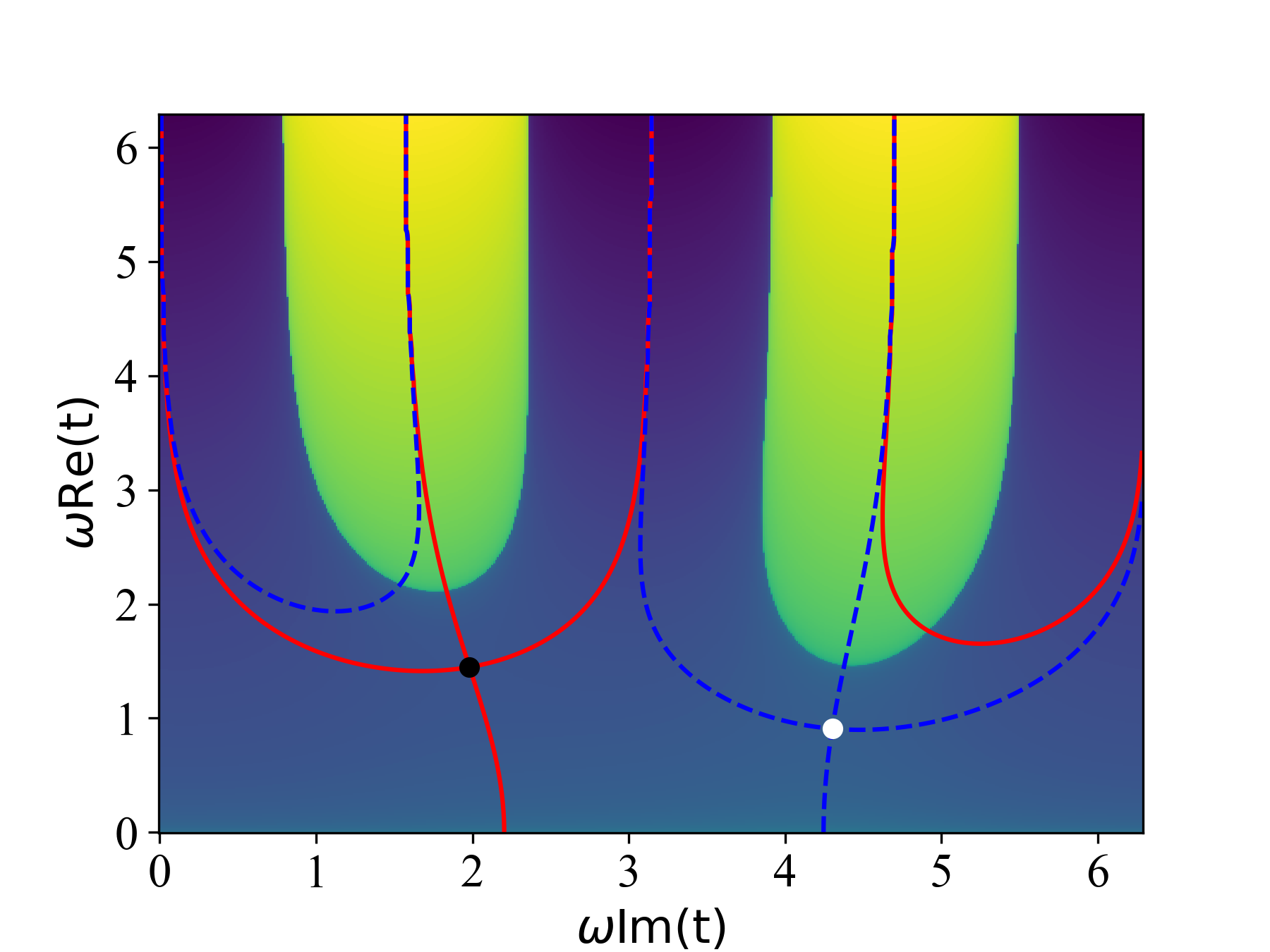}
   \includegraphics[width=0.325\textwidth] {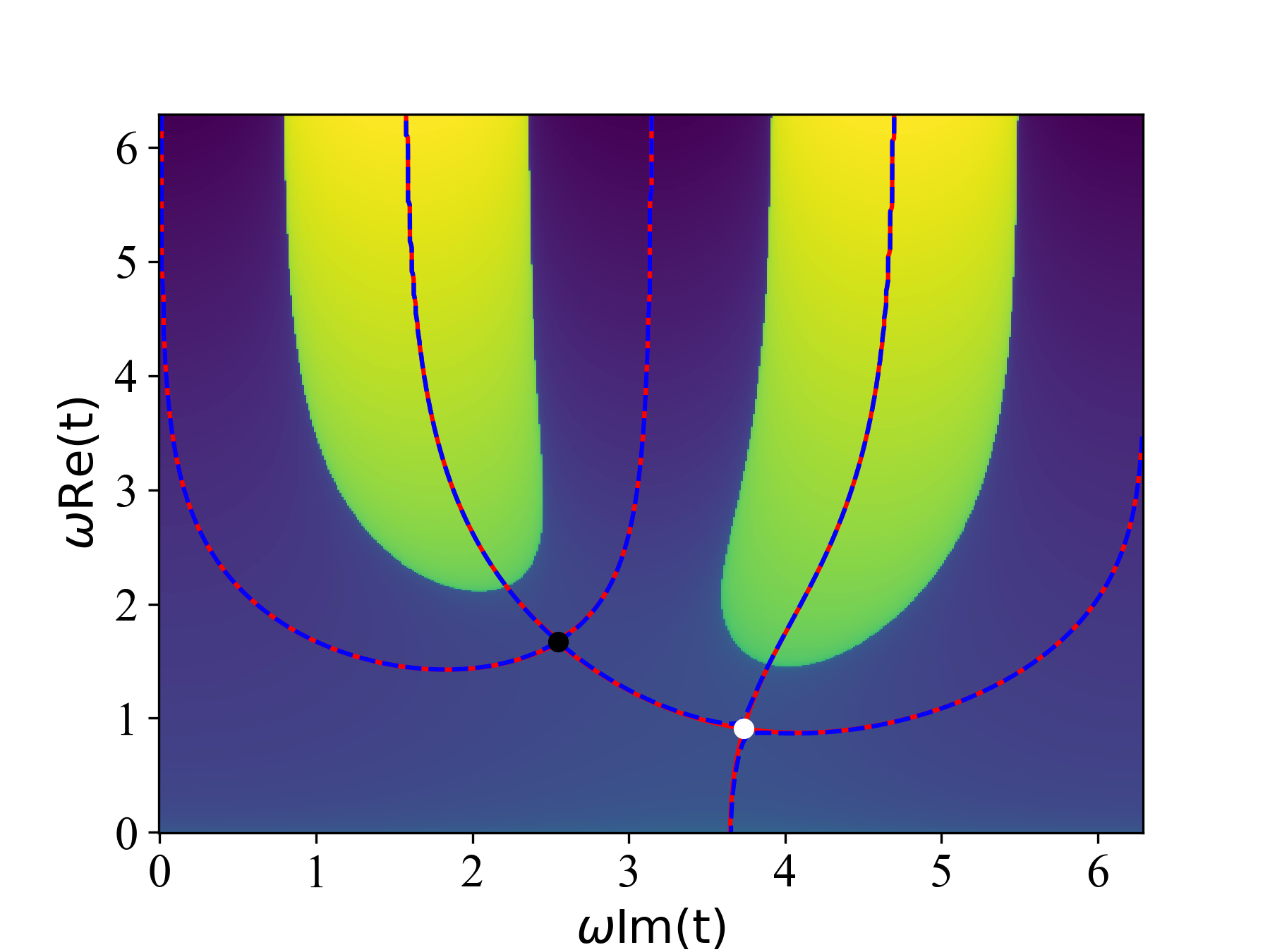}
  \includegraphics[width=0.325\textwidth] {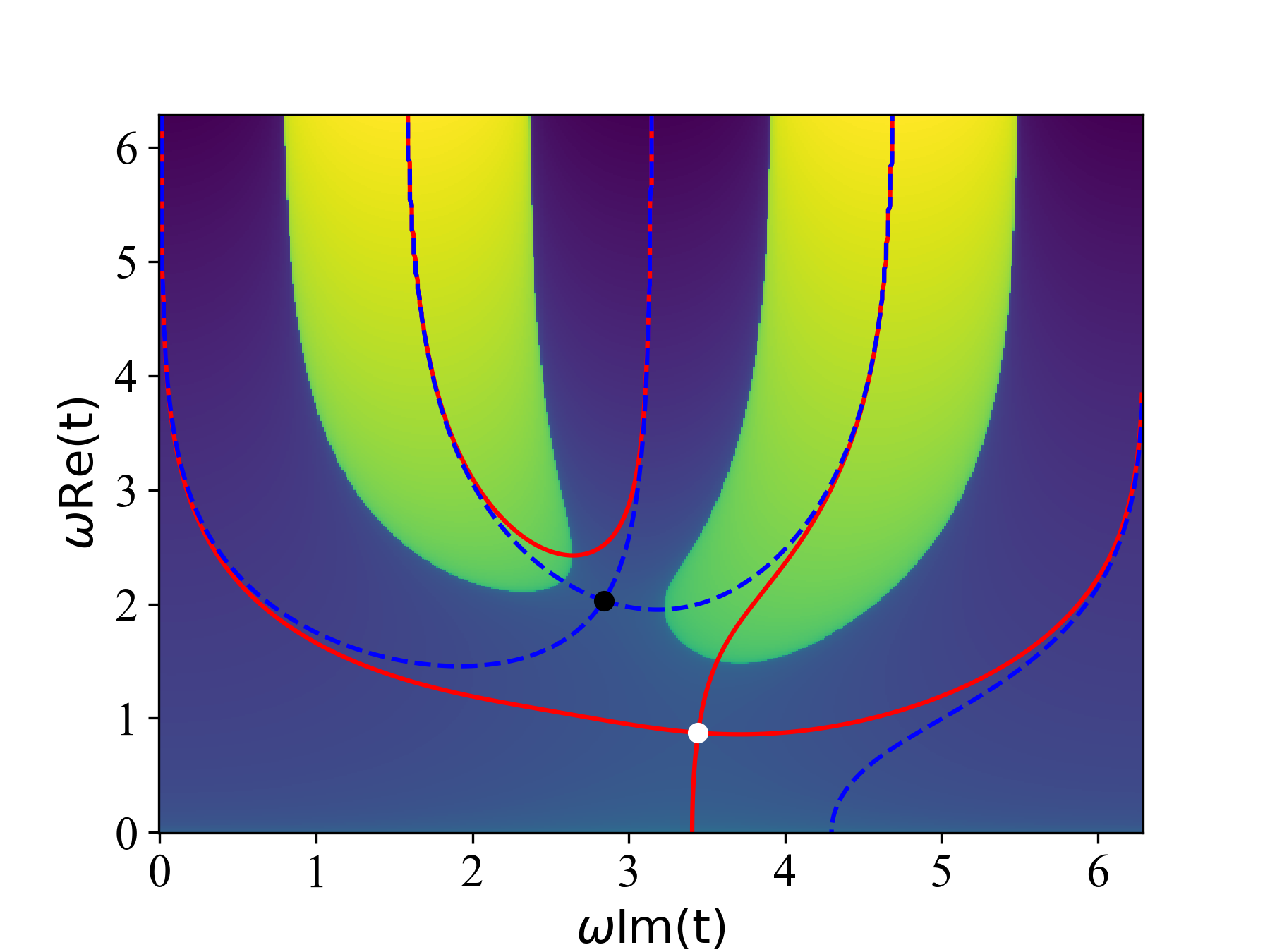}\\
    \caption{\textbf{Contours with the same real part of the action with saddle point solutions and imaginary part of the action in the background, computed for the SFA.} Blue (red) lines are contours with the same real part of the action with solution $t_1$ ($t_2$), and the yellow (blue) background represents the large negative (positive) imaginary part of the action. To calculate transition amplitude with Eq. 1, the integration should be done along these contours, and integrand is proportional to the $\exp{i\mathrm{Im}(t)}$. Therefore, while calculating the transition amplitude, only the contours which do not pass the yellow area should be selected. upper (lower) panels represents the ellipticity $0.2$ ($0.7$). left (middle, right) panels represents the perpendicular momentum $p_z$ is smaller than(same with, larger than) the critical momentum $p_{z, \rm{crit}}$. From the left panel to the right panel, the topology of the contour changes. This cause the Stokes transition. The remaining field and atomic parameters are the same as in the previous figures. }
    \label{fig:contours1}
\end{figure*}

In Fig.~\ref{fig:contours1}, we illustrate the change in the contour that occurs around a critical value of $p_{z}$, called here $p_{z, \rm{crit}}$, for which its topology changes. For simplicity, we keep the momentum component $p_{y}$ parallel to the minor axis fixed. To calculate the transition amplitude one must integrate from $t=-\infty$ to $t=+\infty$ along some contours in the figure, in the upper complex time half plane \footnote{The lower half plane would lead to unphysical results associated to diverging contributions inside a potential barrier \cite{Pisanty2016,Maxwell2018}; for a review on saddle-point methods in strong fields see \cite{Nayak2019}}. Since the action is periodic, it suffices to reduce our problem to a single field cycle. One should note that the contributions from the contours integrating from $0$ to $i\infty$ and from $2\pi + i\infty$ to $2\pi$ cancel each other.

The blue regions represent areas for which the imaginary part of the action causes the yield to vanish when the imaginary part of $\omega t$ tends to infinity, while the green areas depict regions for which it will diverge. The dots illustrate two saddle points, which will lead to key contributions to the PMDs. The contours passing through the saddle points are illustrated by the thick lines in the figure.

For $p_{z}<p_{z,crit}$ [left], the relevant contours encompass the two saddle points. Hence, there are two quantum trajectories  engaged in the ionization process, with the saddle $S_1$ being dominant, as seen from its closeness to the real time axis. At  $p_{z}=p_{z,crit}$ [center], the real parts of the action associated with saddles $S_1$ and $S_2$ become equal, which characterizes a Stokes transition \cite{Berry1989}. For $p_{z}>p_{z,crit}$ [right], the contour passing through $S_2$ will lead to divergencies, so that the saddle must be discarded. 

\begin{figure*}[h!tb]
    \centering
    \includegraphics{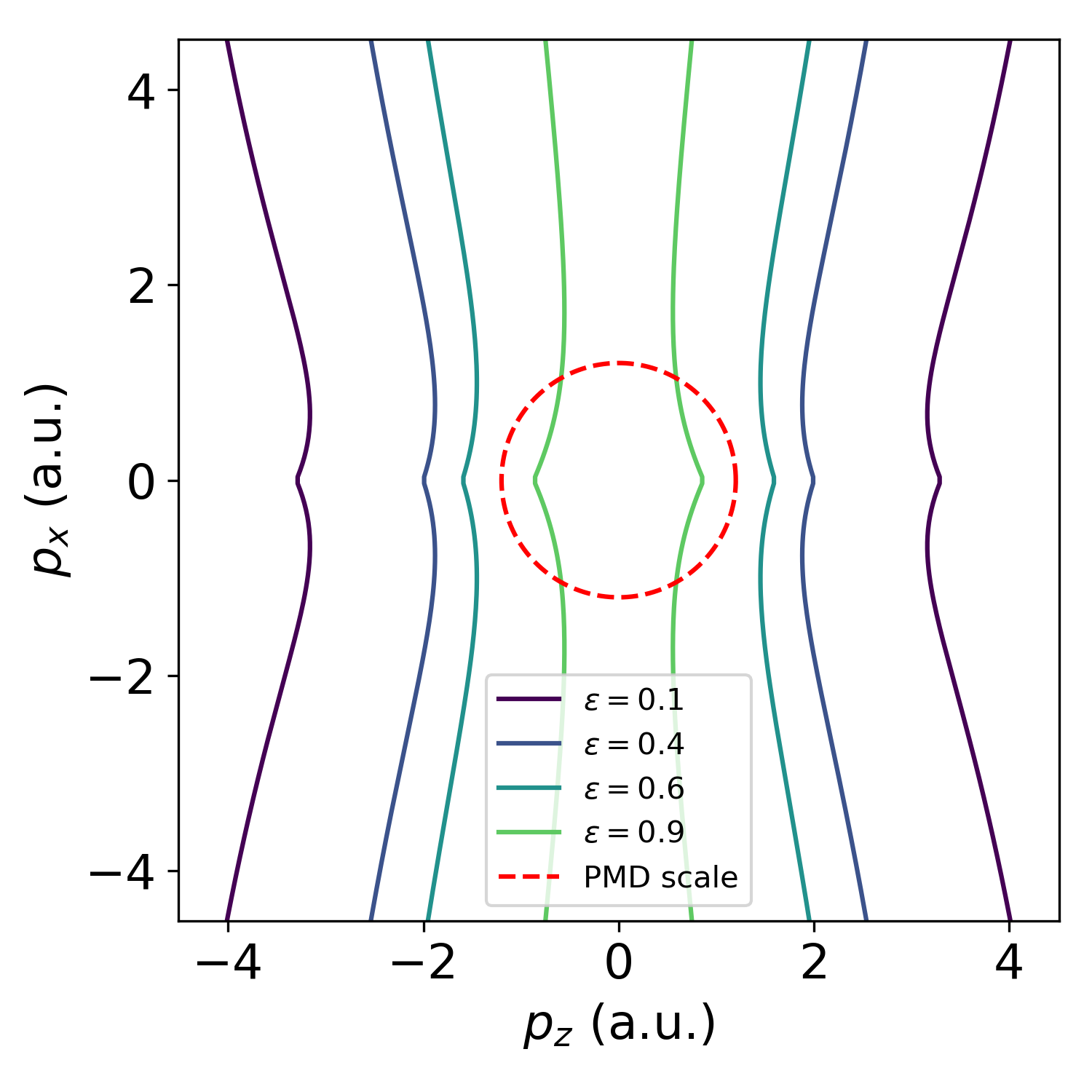}
    \caption{\textbf{Critical momenta for different field ellipticities, computed with the SFA.} If the perpendicular momentum is larger than the critical momentum indicated by solid lines, Stokes transition happens. The red dashed circled area shows the relevant momentum range of the PMDs used in this paper. The remaining field and atomic parameters are the same as in the previous figures.  }
    \label{fig:DivPMD}
\end{figure*}

Fig.~\ref{fig:DivPMD} displays the critical momenta for different field ellipticities, together with the region for which the PMDs are physically relevant (red dashed circle). This is the scale used in the remaining figures of this article. The figure shows that there is always a Stokes transition. However, the absolute value of $p_{z, \rm{crit}}$ decreases for increasing ellipticity. For small and moderate ellipticity, the Stokes transitions occur for momentum ranges far away from the regions of interest, and thus can be ignored, while for large ellipticities they encroach more and more into the physically relevant momentum regions. Nonetheless, the Stokes transitions always seem to occur in the half plane opposite to the physically relevant region.  Thus, matching the solutions in the physically relevant momentum ranges leads to sickle-shaped distributions, as expected. Including the Coulomb potential will lead to angular offsets, which are absent in the plain SFA (see, for instance,  Figs.~\ref{fig:QpropvsCQSFAlow} to \ref{fig:QpropvsCQSFAhigh} in the main body of the paper).

\section*{Appendix 3 - Circular polarization limit for ionization time}

For circularly polarized fields, there will be a single ionization time, which can be associated with a specific angle in the PMDs. This is the key idea upon which the attosecond angular streaking, also known as `the attoclock', is based.  Below we show how this time can be inferred analytically, using the SFA solution for the tunnel ionization time. 

For circular polarization ($\epsilon=1$), the saddle point equation  for $t'$  becomes 
\begin{equation}
\left[p_z+\sqrt{2U_p}\cos\omega t'\right]^2 +\left[p_x+\sqrt{2U_p}\sin\omega t'\right]^2=-2I_p.
\label{eq:circle}
\end{equation}
This is the equation of a circle with complex radius centered at 
\begin{equation}
(p_x,p_z)=\left(-\sqrt{2U_p}\sin\omega t',-\sqrt{2U_p}\cos\omega t'\right),
\label{eq:centercircle}
\end{equation}
which will lead to a ring-shaped distribution, due to the rotational symmetry. One may also show that, in this limit, $\mathrm{Re}[t']$ will be step functions. Thus, for each angle, the electron will escape at the time for which the field will have an extremum. The `step' happens when switching to different momentum half planes. 

Below we show that this holds for specific angles, but due to the symmetry of the problem this can be extended to an arbitrary axis using a rotation matrix. If we choose a momentum along the $p_z$ axis ($p_x=0$), Eq.~(\ref{eq:circle}) becomes 
\begin{equation}
p^2_z+2U_p+2p_z\sqrt{U_p}\cos \omega t'=-2I_p
\end{equation}
so that
\begin{equation}
\cos \omega t'=\frac{-1}{p_z\sqrt{2U_p}}\left[I_p+U_p+\frac{p^2_z}{2}\right].
\label{eq:timepz}
\end{equation}

Setting $t'=t'_r+it'_i$ in Eq.~(\ref{eq:timepz}) gives
\begin{equation}
\cos\omega t'_r\cosh\omega t'_i=\frac{-1}{p_z\sqrt{2U_p}}\left[I_p+U_p+\frac{p^2_z}{2}\right]
\end{equation}
and
\begin{equation}
\sin\omega t'_r\sinh\omega t'_i=0.
\end{equation}

Since $t_i'$ cannot vanish because tunneling is classically forbidden,
 $\sin\omega t'_r = 0$, which means that $\omega t'_r = n\pi$.

 Similarly, along the $p_x$ axis we find
 \begin{equation}
 \sin \omega t'=\frac{-1}{p_x\sqrt{2U_p}}\left[I_p+U_p+\frac{p^2_x}{2}\right], 
 \label{eq:timepy}
 \end{equation}
so that
\begin{equation}
\omega t'_r=(2n+1)\pi/2,
\end{equation}
which corresponds to an extremum for the other component of the field. 

\clearpage

\bibliography{attoclock.bib}

\end{document}